\documentclass{article}

\usepackage[a4paper, total={6in, 8in}]{geometry}

\usepackage{lipsum}
\usepackage{endnotes}
\usepackage{natbib}
\usepackage{graphicx}
\usepackage{moreverb,url}
\usepackage{xcolor}
\usepackage[bottom]{footmisc}
\usepackage{enumitem}
\usepackage{float}
\usepackage{amsmath}
\usepackage{amssymb}
\usepackage{authblk}
\usepackage{tabularx}
\usepackage[latin1]{inputenc}
\usepackage{booktabs}
\usepackage{url}
\usepackage{musicography} 

\newcommand{\citeauthorpos}[1]{{\citeauthor{#1}'s}}

\newcommand{\citepos}[1]{{\citeauthorpos{#1}~\citeyearpar{#1}}}

\usepackage[most]{tcolorbox}

\newtcolorbox{greybox}{
  colback=gray!10,
  colframe=gray!50,
  boxrule=0.4pt,
  arc=1mm,
  left=7pt,
  right=7pt,
  top=7pt,
  bottom=7pt
}

\usepackage[explicit]{titlesec}
\definecolor{sectionbg}{gray}{0.92}
\titleformat{\section}
  {\normalfont\Large\bfseries\setlength{\fboxsep}{9pt}}{}{0pt}
  {\fcolorbox{black}{sectionbg}{\parbox{\dimexpr\columnwidth-2\fboxsep-2\fboxrule\relax}{\thesection\quad#1}}}
\titleformat{name=\section,numberless}
  {\normalfont\Large\bfseries\setlength{\fboxsep}{9pt}}{}{0pt}
  {\fcolorbox{black}{sectionbg}{\parbox{\dimexpr\columnwidth-2\fboxsep-2\fboxrule\relax}{#1}}}

\usepackage{caption}
\newsavebox{\figframebox}
\newenvironment{framedfigure}[1][htbp]
  {\begin{figure}[#1]\centering\setlength{\fboxsep}{6pt}%
   \begin{lrbox}{\figframebox}\begin{minipage}{0.96\columnwidth}\centering}
  {\end{minipage}\end{lrbox}\fbox{\usebox{\figframebox}}\end{figure}}

\date{}

\title{Methods for pitch analysis in contemporary popular music: phenomenological analysis of Primaal's commercial works}

\author[1,3]{Emmanuel Deruty\thanks{CONTACT Emmanuel Deruty. Email: emmanuel.deruty@sony.com}}
\author[2]{Luc Leroy}
\author[2]{Yann Mac\'{e}}
\author[3]{David Meredith}
\affil[1]{Sony Computer Science Laboratories, Paris, France}
\affil[2]{Neodrome Entertainment, Paris, France}
\affil[3]{Department of Architecture, Design and Media Technology, Aalborg University, Denmark}

\begin{document}
\maketitle

\begin{greybox}
\small
This is the extended version of: Deruty, E., Leroy, L., Mac\'{e}, Y., \& Meredith, D. (2026). Methods for pitch analysis in contemporary popular music: phenomenological analysis of Primaal's commercial works. \emph{Journal of New Music Research}, July 2026, 1--19. \\\url{https://doi.org/10.1080/09298215.2026.2706790}
\end{greybox}

\vspace{.3cm}

\begin{abstract}\noindent\sloppypar

\noindent This article uses phenomenological analysis to examine pitch-related elements in commercial popular music, focusing on Primaal, a successful electronic music brand whose works have been licensed by major international companies. Working in collaboration with the producers, we identify musical parameters related to pitch and document how their treatment differs radically from Western classical music conventions. Central to Primaal's aesthetic is the deliberate cultivation of \emph{pitch uncertainty}, achieved through multiple techniques: tone inharmonicity, quasi-harmonic tones designed to evoke multiple simultaneous pitches, strategic boosting of upper partials, and continuous frequency trajectories inspired by the Roland TR-808 bass drum. Rather than discrete notes, pitches are distributed around scale degrees (`poles'), with the distribution width serving as an expressive parameter. The music is organised modally rather than tonally, typically focusing on a few degrees with one functioning as a `final'. We show how these pitch-related parameters articulate both large-scale and small-scale musical structure. Signal analysis, supported by psychoacoustic weighting, serves as our transcription method, avoiding the biases inherent in score-based notation. The findings contribute to music analysis, music information retrieval, computational creativity, and psychoacoustics, suggesting that pitch in contemporary popular music often operates as a continuous, multidimensional phenomenon that resists reduction to discrete note representations.

\end{abstract}

\vspace{5mm}


\newpage

\section{Introduction}

\vspace{.3cm}

This article analyses pitch-related musical parameters in ten commercial tracks produced by \emph{Primaal}, an electronic music brand of the Hyper Music company whose works have been licensed by major international brands. The study was conducted in direct collaboration with the producers, Luc Leroy and Yann Mac\'e, who validated all observations and conclusions. Our analytical method applies phenomenological reduction \citep{husserl1983ideas,ferrara1984phenomenology}: we deliberately \emph{bracket} established theoretical frameworks---including score-based notation and conventional pitch models---and treat signal analysis as our primary transcription tool, weighted by equal-loudness contours to reflect perception \citep{iso2262023}. This approach allows pitch to be examined as it presents itself in the music, rather than through the lens of Western classical conventions.

\paragraph{Pitch uncertainty.} A key goal for the producers is the manipulation of \emph{pitch uncertainty}, with the aim of creating an innovative and distinctive sound, enhancing perceptual richness, and boosting listener appeal. Comparable concerns have been documented in the music of the electronic producer Vitalic and others: single complex tones may be designed to evoke multiple pitches \citep{deruty2025multiple}, tones may be consistently slightly inharmonic, with the partials' positions determining a \emph{resultant tuning} \citep{deruty2025vitalictemperament,deruty2025emerging}, and frequencies often follow continuous trajectories \citep{deruty2025vitalictemperament}. This stands in sharp contrast to Western classical music, where tonal organisation and instrument design aim to \emph{disambiguate} pitch.

\paragraph{Hyper Music and Primaal.} Primaal is a project of the Hyper Music production company \citep{hypermusic}, referred to in \citet{deruty2022development} and \citet{deruty2022melatonin}. The producers recognise Trent Reznor, Daft Punk, Kanye West, Skrillex, and Rosalia as influences. Roland TR-808 bass tracks, common in popular music, especially hip-hop \citep{lavoie2020}, are key to Primaal's music. We therefore assume that their production techniques reflect shared practices in the music industry. We analyse ten songs: `Boom', `Cardinal', `Danger', `Elevate', `\textexclamdown{}Fire!', `R U Ready', `Silver', `Sweet Money', `Whomp', and `Yada Yada', selected on the basis of stem availability. Some of these songs were used in the Netflix series `Supacell', in a worldwide Chanel ad for the `Chance' perfume, in a UK commercial for O2 telecommunications, in the Canadian film `The Wedding Banquet', in the French film `KO', in the US Major League Soccer stadium PA music, and in the US Qualcomm Summit Keynote 2025 PA music, ensuring that the music analysed here is widely heard.

\paragraph{Terminology.} Throughout this article, terminology follows \citet{deruty2025multiple}. A \textit{partial} is the spectral representation of a single periodic sine wave. A \textit{harmonic complex tone} is an ensemble of partials whose frequencies are integer multiples of a fundamental frequency ($f_0$). An \emph{overtone} or \emph{upper partial} is any partial other than $f_0$, and the $n^{\text{th}}$ \textit{harmonic} has a frequency equal to $n$ times the fundamental. \textit{Inharmonicity} refers to a partial configuration in which overtone ratios deviate from integer multiples of a base frequency. No tone in this study is strictly harmonic, but degrees of inharmonicity vary, allowing the identification of \emph{quasi-harmonic tones}, in which each partial of the least-deviating harmonic series can still be associated with a partial of the original tone; in such cases, $f_0$ may be defined as the fundamental of this least-deviating series \citep{rasch1982perception}.

\newpage

\paragraph{Roadmap.} The paper is structured as follows. Section~\ref{sec:methods-of-analysis} describes the analytical process. Section~\ref{ref:signalanalysis} compiles the signal analyses of the studied tracks, which serve as transcriptions and constitute the empirical material on which the rest of the article builds. Section~\ref{sec:parameters} lists the identified expressive musical parameters related to pitch in Primaal's music, along with their organisation. Section~\ref{sec:pitchuncertainty} examines how the producers build pitch uncertainty. Section~\ref{sec:historical-context} places Primaal's pitch-related techniques in a historical context. Section~\ref{sec:analyses-of-works} focuses on specific pitch-related aspects in more detail. Finally, Section~\ref{sec:conclusion} summarises the paper's contributions. Appendix~\ref{ref:instruments} documents the tools and techniques used by the producers to manipulate pitch. Further supplementary material---including audio extracts and sources for audio production terminology---is available online at: \newline \\\url{https://hmp-suppl-mat.s3.eu-west-3.amazonaws.com/index_ext.html}

\vspace{1cm}

\section{Analytical methods and process}
\label{sec:methods-of-analysis}

This section describes the analytical process used in this study. After summarising the practical procedure (Section~\ref{sec:practicalprocess}), it reviews issues identified in previous work on popular music analysis (Section~\ref{ref:analysisofpopularmusic}), then details how phenomenology is applied to analyse the music with as few cultural, \textit{a priori} assumptions as possible (Section~\ref{sec:phenomenologicalreduction}). This approach proceeds from open listening to signal analysis as transcription, followed by syntactical listening---the identification of syntax, understood here as the organisation of expressive parameters. The procedure involves the producers, whose knowledge is treated as empirical clarification of intentional production choices rather than an imposed external analytical template. Section~\ref{ref:mircommunity} examines how `pitch' is commonly represented in the machine learning and music information retrieval communities, and why these representations were also bracketed. We then explain how signal analysis incorporates psychoacoustic weighting, how pitch estimates are informed by standard spectral and temporal models (Section~\ref{sec:signal-analysis-methods}), and how the temporal evolution of expressive parameters is described in terms of large- and small-scale structure (Section~\ref{sec:structurebackground}). Overall, the procedure proposes a method for analysing pitch in the `primary text' \citep[p.~1]{moore2018rock} of recent popular music.

\subsection{Workflow}\label{sec:practicalprocess}

In practice, the procedure consisted of seven steps.

\begin{enumerate}[leftmargin=*,itemsep=3pt,topsep=4pt]
  \item \textbf{Song selection.} Songs were selected on the basis of stem availability and the producers' assessment that they were representative.
  \item \textbf{Open listening.} We performed open listening using phenomenological reduction, as detailed in Section~\ref{sec:phenomenologicalreduction}; these listening sessions were conducted both with and without the producers.
  \item \textbf{Example selection.} In consultation with the producers, we selected sections for signal analysis and jointly decided to place greater emphasis on the `bass' tracks, which are central to this music.
  \item \textbf{Signal analysis.} The selected stems were subjected to signal analysis.
  \item \textbf{Syntactical listening and discussion.} We combined syntactical listening with discussions of the signal analysis results with the producers, in order to assess observations and hypotheses raised during listening, their relevance to the music's syntax, and their links to production methods.
  \item \textbf{Compilation of results.} The results were compiled in written and graphical form, providing the basis for the manuscript.
  \item \textbf{Producer validation.} The producers reviewed and confirmed the observations and conclusions.
\end{enumerate}

\vspace{.3cm}

The stems used for phenomenological listening and signal analysis always include drums and bass. Some songs feature lead vocals as rap flows or edited loops, but none include melody-based lead singing. When vocal melodies occur, they take the form of short, recurring `hooks' \citep[p.~58]{delson1980dictionary}. Human shouts, vocal effects, and synthetic effects often act as `punctuation marks', according to the producers. Other stems may include more traditional elements, such as brass samples and arpeggios.

\subsection{Analysing `popular' music}\label{ref:analysisofpopularmusic}

The procedure summarised in Section~\ref{sec:practicalprocess} addresses several issues that have been identified in previous work on popular music analysis; see \citet{wicke2003popmusik} for comments on many of these concerns (in German). This section reviews these issues and relates them to the choices made in the present study.

\subsubsection{The `primary text'}\label{ref:primarytext}

\citet[p.~1]{moore2018rock} oppose the `primary text' (what is constituted by the sounds themselves) to the `secondary text' (the commentaries on the sounds), and consider that insufficient attention has been paid to the `primary text' in popular music analysis. If the analysis of popular music focuses on the `primary text' then, according to \citet[p.~338]{keil1966motion}, it may be `the analyst's primary obligation [...] to elucidate the syntax or grammatical rules of the musical system or style with which he is dealing', on the grounds that `[a]ll music has syntax or embodied meaning'. In this article, we focus on the `primary text' and attempt to elucidate the syntax and grammatical rules that relate to pitch in Primaal's music.

\subsubsection{Harmony as a bias}

The elucidation of the syntax or grammatical rules of popular music may be hindered by an intrinsic bias that originates from the practice of music analysis itself. According to \citet[p.~19]{brackett2023interpreting}, `we must recognize that the metalanguage of music analysis is not transparent, but that it is a medium that comes with its own ideological and aesthetic baggage which will affect what we can say'. As far as popular music analysis is concerned, according to \citet[pp.~104--105]{middleton1990studying}, `the bundle of methods, assumptions, and ideologies which came to constitute ``mainstream musicology'' in the later nineteenth and the twentieth centuries renders it a less than useful resource in many ways'.

\newpage

To be more specific, still according to \citeauthor{middleton1990studying}, `[t]here is a rich vocabulary for certain areas, important in musicology's typical corpus [amongst which] harmony (chord types, chord functions, and relations), tonality [but] an impoverished vocabulary for others [such as] pitch nuance and gradation outside the steps of the diatonic/chromatic system'. In this article, we highlight the importance of pitch nuance and gradation outside the steps of the diatonic/chromatic system. In the process, we have indeed found that `harmony may not be the most important parameter; rhythm, pitch gradation, timbre and the whole ensemble of performance articulation techniques are often more important; ``dissonance'' and ``resolution'' may be produced by non-harmonic means' \citep[pp.~104--105]{middleton1990studying}.

\vspace{.3cm}

\subsubsection{The score as a bias}\label{ref:scoreasbias}

According to \citeauthor{middleton1990studying}, one important obstacle that prevents the analysis of the `primary text' in popular music is the \emph{score}. Musicologists tend to have a `notation-centric' approach, which `encourages reification: the score comes to be seen as ``the music'', or perhaps the music in an ideal form'. The score `induces particular forms of \emph{listening}, and these then tend to be applied to \emph{all} sorts of music, appropriately or not'. As a result, the musical analysis methodology is often `slanted by the characteristics of notation': `[m]usicological methods tend to foreground those musical parameters which can be easily notated', and `tend to neglect or have difficulty with parameters which are not easily notated: non-standard pitch and non-discrete pitch movement (slides, slurs, blue notes, microtones, and so on)' \citep[pp.~104--105]{middleton1990studying}. According to \citet[p.~16]{mellers1974twilight}, the score `may be not only inadequate but also misleading: for written notation can represent neither the improvised elements nor the immediate distortions of pitch and flexibilities of rhythm which are the essence (not a decoration) of a music orally and aurally conceived'. In this article, so as not to become notation-centric, we prefer representations stemming from signal analysis to scores (Section~\ref{ref:signalanalysis}).

\citet{tagg1982analysing,middleton1990studying,moore2016song,moore2018rock} provide examples of popular music transcribed into Western-style scores. As an illustration of the bias brought by both tonality and the score, \citet[p.~11]{moore2018rock} cite the example of \citet{everett1986fantastic}, who uses Schenkerian foreground graphs (\emph{Urlinie-Tafeln}) \citep{schenker1969five} of The Beatles' `Strawberry Fields Forever' and `If I Fell'.

\vspace{.3cm}

\subsubsection{Pertinent approaches}

According to \citet[p.~10]{moore2018rock}, one goal of the musicologist who wants to analyse popular music is to `elucidate theoretical approaches pertinent to the music [...] This activity is best considered pre-analytical, since any analysis must be based on theoretical preconceptions'. We propose that collaborative critical listening with the music's producers, supported by signal analysis (Section~\ref{ref:signalanalysis}), may constitute a suitable `pre-analytical' theoretical approach. The analyses reported in this article are conducted using this approach.

\newpage

\subsubsection{The producers' point of view}\label{ref:producerspov}

A final documented obstacle for popular music analysis is the absence of communication with the music's producers. \citet[p.~271]{everett2000expression} writes: `I'll demonstrate expressive manipulations of formal construction, vocal and instrumental colorings, rhythmic relationships, melodic device, and tonal systems, without once ever considering whether the composers, arrangers, artists, or engineers might have been fully conscious of how I might hear what they were doing'. We hope that conducting the analysis and writing the article in collaboration with the producers themselves addresses this problem.

\subsection{Phenomenological reduction}\label{sec:phenomenologicalreduction}

\subsubsection{Husserl's phenomenology}

Husserl introduces the notion of the philosophical \emph{epoch\'{e}} as a suspension of judgment, in which the observer refrains from endorsing or rejecting the doctrinal claims of prior philosophies and instead confines analysis to what can be examined within this suspension \citep[p.~33]{husserl1983ideas}.\footnote{In \citepos{husserl1983ideas} formulation, the philosophical epoch\'{e} `shall consist of our completely abstaining from any judgment regarding the doctrinal content of any previous philosophy and effecting all of our demonstrations within the limits set by this abstention'. In the original German: `Die philosophische $\grave{\varepsilon} \pi o \chi \acute{\eta}$, die wir uns vornehmen, soll [...], darin bestehen, da{\ss} wir uns hinsichtlich des Lehrgehaltes aller vorgegebenen Philosophie vollkommen des Urteils enthalten und alle unsere Nachweisungen im Rahmen dieser Enthaltung vollziehen' \citep[p.~33]{husserl1913german}.} This abstention does not amount to a refutation but to a bracketing of presuppositions: they are rendered inactive, or `put between brackets' (\emph{eingeklammert}), such that existing theoretical frameworks are neither denied nor affirmed \citep[pp.~59--61]{husserl1983ideas}. Through this act of bracketing, the observer deliberately sets aside all disciplinary assumptions, regardless of their perceived validity\footnote{The observers `exclude all sciences relating to this natural world no matter how firmly they stand there for [them], no matter how much [they] admire them, no matter how little [they] think of making even the least objection to them' \citep[p.~61]{husserl1983ideas}.}, thereby enabling what \citet[p.~66]{husserl1983ideas} terms \emph{phenomenological reduction}.

In physics, phenomenology has likewise been contrasted with theory. Phenomenological laws describe observable appearances; theoretical laws explain indirectly inferred realities \citep{cartwright1984laws}. This distinction motivates our own methodological priority: descriptive adequacy to the heard signal comes before explanatory fit with an established music-theoretical framework. The score- and theory-related biases discussed above (Section~\ref{ref:scoreasbias}) are therefore bracketed rather than treated as false.

\subsubsection{Phenomenology and music analysis}

In the context of music analysis, `[t]he process of [phenomenological] epoch\'{e} [has been described as] disregard[ing] the influences and biases of everyday knowledge and [...] naively mak[ing] judgments on things from face value' \citep[p.~3]{serhan2022phenomenological}. As pointed out by \citet{kim2010critique} and \citet{serhan2022phenomenological}, an application of phenomenology to music analysis is Schaeffer's reduced listening (original French: \emph{\'{e}coute r\'{e}duite}). Reduced listening is `the listening attitude which consists in listening to the sound for its own sake, as a sound object, by removing its real or supposed source and the meaning it may convey'; `[t]he name reduced listening refers to the notion of phenomenological reduction [...] because [...] it consists of stripping the perception of sound of everything that is not ``in itself''' \citep[p.~33]{chion1983guide}.

\newpage

A documented difficulty in the application of phenomenology to music analysis is the existence of the score, which echoes the bias mentioned in Section~\ref{ref:scoreasbias}. \citet{benson2011phenomenology} remarks that in \citepos{ingarden1986work} ontology of the musical work, `the score ends up taking on a centrality which almost seems to eclipse that of the composition', even though \citeauthor{ingarden1986work} acknowledges that `[b]ecause of the imperfection of musical notation, the score is an incomplete, schematic prescription for performance'. According to \citeauthor{benson2011phenomenology}, the problem is intrinsic to the notation, and such indeterminacies `would likely not disappear [...] even with a far superior notational system'.

\subsubsection{Phenomenological reduction as the first analytical phase}\label{ref:reduction}

\citet[p.~359]{ferrara1984phenomenology} describes the first phase of phenomenological music analysis as \emph{open listening}, which orients the analyst to the work. It is termed `open' because the analyst may respond to any level of meaning in the work. At the start of our analysis, we chose to have minimal information about the music, bracketing cultural preconceptions: we did not know in advance whether pitch arose from harmonic tones, whether pitches were stable enough to form notes, whether their distribution formed a scale, or whether simultaneous pitches formed known chords. At this stage, we retained only \citepos{oxenham2012pitch} statement according to which `pitch is the perceptual correlate of the periodicity [...] of an acoustic waveform'. One major obstacle to open listening is the temptation to transcribe the music into \emph{scores}: score-based transcription foregrounds parameters that are easily notated while neglecting those that are not, and thus carries its own ideology (Section~\ref{ref:scoreasbias}). We therefore propose that signal analysis, supported by collaborative critical listening with the producers, serves as the transcription system---consistent with early 14\textsuperscript{th}-century theorists' recognition that some notation is necessary because sounds are volatile \citep[p.~103]{aluas1996quatour}. Within the seven-step workflow described in Section~\ref{sec:practicalprocess}, open listening, example selection, and signal analysis thus constitute the first analytical phase.

\subsubsection{Syntactical listening}

Ferrara's second analytical phase is the identification of syntactical meanings \citep[p.~359]{ferrara1984phenomenology}. This aligns with \citet[p.~338]{keil1966motion}, who argues that the analyst's main duty is to explain the syntax or grammatical rules of a musical system, and with \citet[p.~10]{moore2018rock}, who maintain that the musicologist should `elucidate theoretical approaches pertinent to the music' (Section~\ref{ref:analysisofpopularmusic}). In practice, a spectral transform of a short window of a stem's signal typically reveals partials forming what is heard as a single \emph{auditory stream} \citep{bregman1994auditory}. There is strong regularity in the distribution of these partials, but the relations between their frequencies are often not harmonic. Several pitches may be heard for a single set of partials, and the notes that are heard often do not correspond to the lowest partial. Short-term spectral transforms show that frequency values are almost never stable, and frequency distributions across stems seldom coincide, even when centred on similar values (Section~\ref{ref:signalanalysis}). Consultations with the producers allow us to relate tone morphology to production methods, confirm that evoking multiple pitches is deliberate, and assess the musical relevance of continuous frequency distributions. Within the practical workflow, this syntactical phase corresponds to the fifth step, in which the signal analysis results are interpreted with the producers. Together, these observations and consultations motivate the expressive parameters listed in Section~\ref{sec:parameters}.


\newpage

\subsection{`Pitch' as viewed by the machine learning and M.I.R.\ communities}\label{ref:mircommunity}

Because our approach brackets not only score-based representations but also the standard computational representations of pitch, this section reviews how `pitch' is commonly conceived in the machine learning and music information retrieval (M.I.R.) communities, and why these conceptions were also bracketed during the analysis.

\subsubsection{MIDI files and piano rolls as reification of the score}

The reification of the score mentioned by \citet[pp.~104--105]{middleton1990studying} and reported in Section~\ref{ref:scoreasbias} influences the M.I.R.\ community. \citet[sec.~4.2]{briot2020deep} `believe that the essence of music (as opposed to sound) is in the compositional process, which is exposed via symbolic representations (like musical scores or lead sheets)'. A classic music information retrieval task is audio-to-MIDI alignment \citep{raffel2016optimizing}, in which MIDI files are aligned with corresponding audio recordings. MIDI files have been referred to as symbolic `versions' of pieces of music \citep{ewert2012towards,raffel2016optimizing}, as well as `transcriptions' \citep{turetsky2003ground,raffel2016learning,benetos2018automatic} of the pitched content of the music.

\subsubsection{The AI Song Contest and Italian secular monodies with \emph{basso continuo}}

`The term basso continua [sic] designates a mode of accompaniment in use primarily between 1600 and 1750 [...] a continua [sic] part has two components: a bass line, which is provided by the composer and is generally to be performed as notated [...] and a set of harmonies, which may be specified by signs or implied by standard chord progressions within a given style' \citep{ashworth2012basso}. A seminal work for the codification of the \emph{basso continuo} is Giulio Caccini's \emph{Nuove Musiche}, published in Florence in 1602 \citep{caccini1978nuove}. \emph{Basso continuo} was extensively used in the context of Italian secular monodies \citep{fortune1953italian}, of which the \emph{Nuove Musiche} is an example. \citet{arnold1957alessandro,baron1968monody,caccini1978nuove,williams2001continuo} show several examples of 17\textsuperscript{th}-century Italian secular monodies with \emph{basso continuo}.

The AI Song Contest\footnote{\url{https://www.aisongcontest.com/}} is a flagship event of co-creative music composition, in which human musicians use AI-based technology to combine human and AI creativity. Its focus on song generation makes it a useful case study of computational representations employed in a popular-music context. \citet{AIsongcontest_long} provide an account of the 2020 AI Song Contest, showing that musicians distinguished `building blocks'---aspects of the musical discourse addressed by AI technology---such as `melody', `harmony', and `bass line'. Similar approaches can be found in the 2021 contest \citep{deguernel2022personalizing} and in taxonomies of generation systems \citep{herremans2017functional}. This decomposition into melody, harmony, and bass resembles the organisation of 17\textsuperscript{th}-century Italian secular monodies with \emph{basso continuo}. On the evidence presented in this paper, we argue that this representational model may describe those monodies more closely than it describes at least some recent popular music. One possible reason for the persistence of this Baroque-derived model is suggested by \citet{williams2001continuo}, who observes that `[t]he \emph{continuo} was fundamental to music in the 17\textsuperscript{th} and 18\textsuperscript{th} centuries to such an extent that its characteristic manner of notation, the figured bass [...] became the basis for teaching composition and analysis and has remained in use for theoretical purposes throughout the 19\textsuperscript{th} and 20\textsuperscript{th} centuries'. We therefore bracket this melody--harmony--bass decomposition, because open listening must first establish whether these discrete, score-like layers are pertinent to the music under analysis.

\newpage

\subsubsection{The harmonic complex tone as the sole origin for the impression of pitch}\label{ref:soleorigin}

According to \citet{oxenham2012pitch}, `[t]he most commonly considered form of pitch-evoking sound is a harmonic complex tone [...] two [harmonic complex] tones generally have the same pitch if they share the same $f_0$'. \citeauthor{oxenham2012pitch}'s view is widely reflected in the M.I.R.\ and machine learning communities. Pitch detection, also called `pitch tracking' or `pitch estimation', has been performed using `traditional machine learning' \citep{drugman2018traditional} and neural networks \citep{kim2018crepe,riou2023pesto}. Both approaches are often formulated as \emph{$f_0$ estimation tasks} \citep{signol2008evaluation,drugman2018traditional,kim2018crepe}.

Traditional machine learning approaches generally rely on the assumption according to which the pitched part of the signal is `periodic' or `quasiperiodic' in the time domain, with the spectrum consisting `of a series of impulses at the fundamental frequency and its harmonics' \citep[p.~400]{rabiner1976comparative}. Similarly, neural network-based pitch-tracking models are trained on harmonic or quasi-harmonic audio. For instance, \citet{riou2023pesto} use MIR-1K, a dataset of quasi-harmonic samples \citep{hsu2009improvement}, and MDB-stem-synth, a dataset of `exact multiples of the $f_0$' \citep{salamon2017analysis}. The underlying hypothesis according to which pitched content is produced by harmonic complex tones can be found in a variety of other music information retrieval tasks, such as harmonic-percussive source separation \citep{ono2008real}, automatic music transcription \citep{ewert2014score}, and multi-pitch detection \citep{christensen2008multi}. A rare occurrence of the inclusion of inharmonic models in pitch detection may be found in \citet{vincent2008harmonic}.

Equating `pitch' with `the $f_0$ of harmonic complex tones' contrasts with the wider consideration of inharmonicity observed in the field of psychoacoustics (see Section~\ref{sec:inharmonicity}). In the case of Primaal's music, it contrasts with the observations reported in Section~\ref{ref:signalanalysis}, according to which (1) most of the analysed tones are inharmonic to some degree, and (2) pitch may be carried by partials higher than the fundamental. In addition, \citet{deruty2024inharmonicity} show that the \textit{HarmonicRatio} \citep{peeters2004large} of popular music released after 1970 is significantly lower than that of orchestral and piano music. Although low \textit{HarmonicRatio} values may stem from both the inharmonicity of individual sounds and the relative position of sounds on the frequency axis, an example is shown in which the \textit{HarmonicRatio} of a single instrument from a well-known album is lower than that of church bells.

\subsection{Signal analysis methods}\label{sec:signal-analysis-methods}

Because our goal is to remain close to the perceived primary text, we apply equal-loudness-contour weighting, recognising that humans are not equally sensitive to all frequencies \citep{fletcher1933loudness}. Various models exist that weight the power spectrum to reflect perception \citep{skovenborg2004evaluation}; we use the ISO~226:2023 50-phon equal-loudness contour \citep{iso2262023} before spectral analysis. 

Our interpretation and implementation of the \textit{spectral} and \textit{temporal} models of pitch perception \citep{yost2009pitch} follow \citet{deruty2025multiple}. For individual partials, we use simplified spectral modelling. For complex tones, temporal modelling uses either autocorrelation or differences between adjacent partials \citep{meddis2006virtual,yost2009pitch}. These models are used as descriptive probes rather than as ground-truth accounts of perceived pitch.

\newpage
\subsection{Structure and form}\label{sec:structurebackground}

To study the temporal evolution of expressive parameters, we distinguish between large-scale structure, called `semiotic structure' or `sectional form' \citep{bimbot2012semiotic}, and small-scale structure, called `form' \citep{caplin1998classical} or `morpho-syntagmatic level' \citep{bimbot2014semiotic}. The \emph{large-scale structure} describes long-term regularities between musical parts \citep{bimbot2012semiotic}. Segments at this scale are `autonomous and comparable blocks' \citep{bimbot2010decomposition}, with a typical duration of around 11.5~s \citep{deruty2013methodological}. In Western classical music, such segments have been referred to as `periods' \citep[pp.~149--150]{monelle2014linguistics} or classified as `sentences' and `periods' \citep{schoenberg1967fundamentals,caplin1998classical}. The \emph{small-scale structure} refers to the internal organisation of large-scale segments. Documented functions at this level include \emph{contrast} \citep{bimbot2016system} and, in Western classical music, \emph{antecedent} and \emph{consequent} \citep{caplin1998classical}. A typical duration for the units forming the small-scale structure in Primaal's music is the beat. The relations between these units can be described in terms of \emph{implications}, a contrast being identified as a deviation from a system of implications derived from the relations between the units \citep{bimbot2016system}. These concepts are applied to Primaal's music in Sections~\ref{sec:largescalestructure} and~\ref{sec:smallscalestructure}.

\vspace{1.5cm}

\section{Signal analysis of Primaal tracks}\label{ref:signalanalysis}

\vspace{.4cm}

As mentioned in Sections~\ref{sec:practicalprocess} and~\ref{sec:phenomenologicalreduction}, the signal analyses serve as transcriptions and are part of the first analytical phase. These analyses are presented before any interpretation because the whole study proceeds from them: the musical parameters identified in Section~\ref{sec:parameters}, the forms of pitch uncertainty described in Section~\ref{sec:pitchuncertainty}, and the detailed observations of Section~\ref{sec:analyses-of-works} all refer back to individual analyses given here. Reading them first establishes concretely what the object of the study consists of---how these tones are actually constituted, and how their frequencies behave over time---before anything is claimed about them.

The analyses involve pitch estimations. These estimations are not meant to provide ground truth for the perceived pitch values; instead, pitch is used as a frequency unit that musicians can easily interpret. The mention `weighted' indicates prior weighting of the audio using the ISO~226:2023 50-phon equal-loudness contour \citep{iso2262023}---see Section~\ref{sec:signal-analysis-methods}. Audio extracts corresponding to the analysed excerpts are available in the online supplementary material.\footnote{\url{https://hmp-suppl-mat.s3.eu-west-3.amazonaws.com/index_ext.html}}

\vspace{.3cm}

\subsection{`Boom', bass}\label{ref:boombass}

The bass track for `Boom' was generated in Omnisphere using Seismic Shock's `808 Woofer Warfare' patch, mode 1, high amount. The sound was then processed using an equaliser highlighting high frequencies, a SansAmp PSA-1 distortion to generate new medium- and high-frequency partials, and a compressor.

\newpage

Figure~\ref{fig:boombass} (a) and (b) show the evolution of the bass's first partial in terms of pitch. The TR-808 bass drum-style initial decrease mentioned in Appendix~\ref{ref:bassgeneration} is clearly visible. In this track, it covers an unusually large ambitus of six semitones. Figure~\ref{fig:boombass} (c) shows the bass part's weighted STFT. The partials with increasing frequencies are the result of the SansAmp process. Figure~\ref{fig:boombass} (d) shows the power spectrum at the frame indicated by the blue line in (c). Harmonic 5 (major third) was significantly boosted and is the loudest harmonic. Harmonic 11 (out-of-tune augmented fourth) is also clearly audible. Figure~\ref{fig:boombass} (e) shows the distance between the partials in (d). The distance fluctuates around the $f_0$ value within an ambitus of $\pm 1$ semitone.

\begin{framedfigure}[h!]
  \centering
  \includegraphics[width=.84\columnwidth]{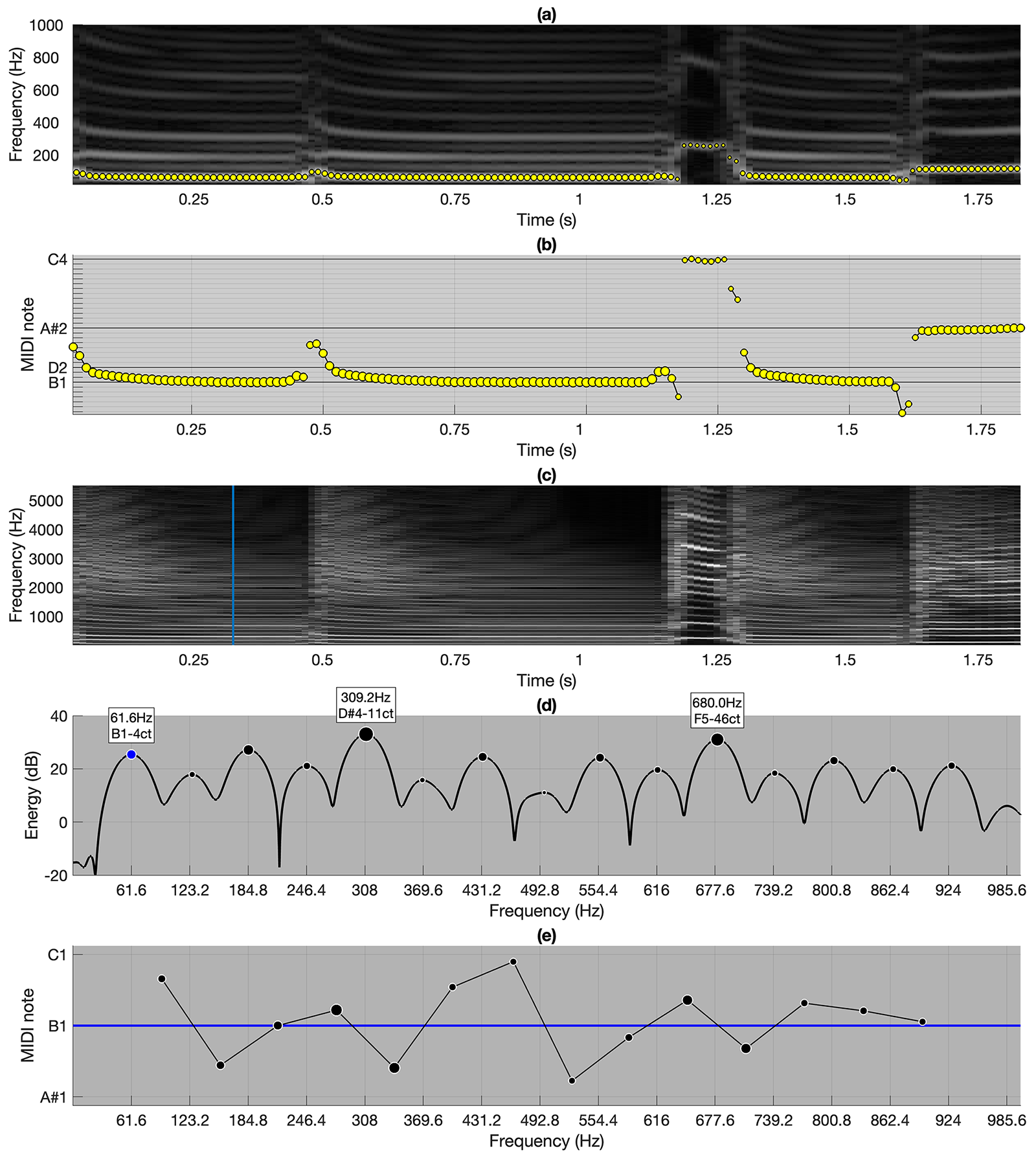}
  \caption{`Boom', bass track, 0'17 to 0'19. (a) STFT. The yellow scatter plot tracks the lowest partial. (b) Lowest partial's frequency, expressed as MIDI notes. As in all similar representations, the markers' area denotes the energy. (c) STFT, weighted. (d) FT, weighted, at frame indicated by blue line in (c). The x-axis grid is set on the harmonic positions for the lowest partial. (e) Frequency difference between consecutive partials from (d). Blue line, frequency of the lowest partial.}
\label{fig:boombass}
\end{framedfigure}

\clearpage
\subsection{`Boom', vocals}\label{ref:boomvocals}

The vocal track in `Boom' derives from a 2-second-long sample. The sample is looped, and the repeated occurrences are heavily edited. The pattern approximately follows `B3 B3 D4 B3 A3 G\#3'. Figure~\ref{fig:BoomVocals} (b) shows that the $f_0$ of each `note' is not stable. It is decreasing in a similar manner to the bass from `Elevate' (Section~\ref{ref:elevatebass}). The third `note' is produced using a bend. Its $f_0$ is between D4 and D\#4. Figure~\ref{fig:BoomVocals} (c) shows that the vocals are based on a quasi-harmonic complex tone. Figure~\ref{fig:BoomVocals} (d) can be used as a reference to which less harmonic complex tones can be compared (see, for instance, the similar representations in Figures~\ref{fig:DangerLowBassUnweighted}, 
\ref{fig:DangerHighBassUnweighted}, 
\ref{fig:SilverBass}, and \ref{fig:SilverBass2}).

\begin{framedfigure}[h!]
  \centering
  \includegraphics[width=.95\columnwidth]{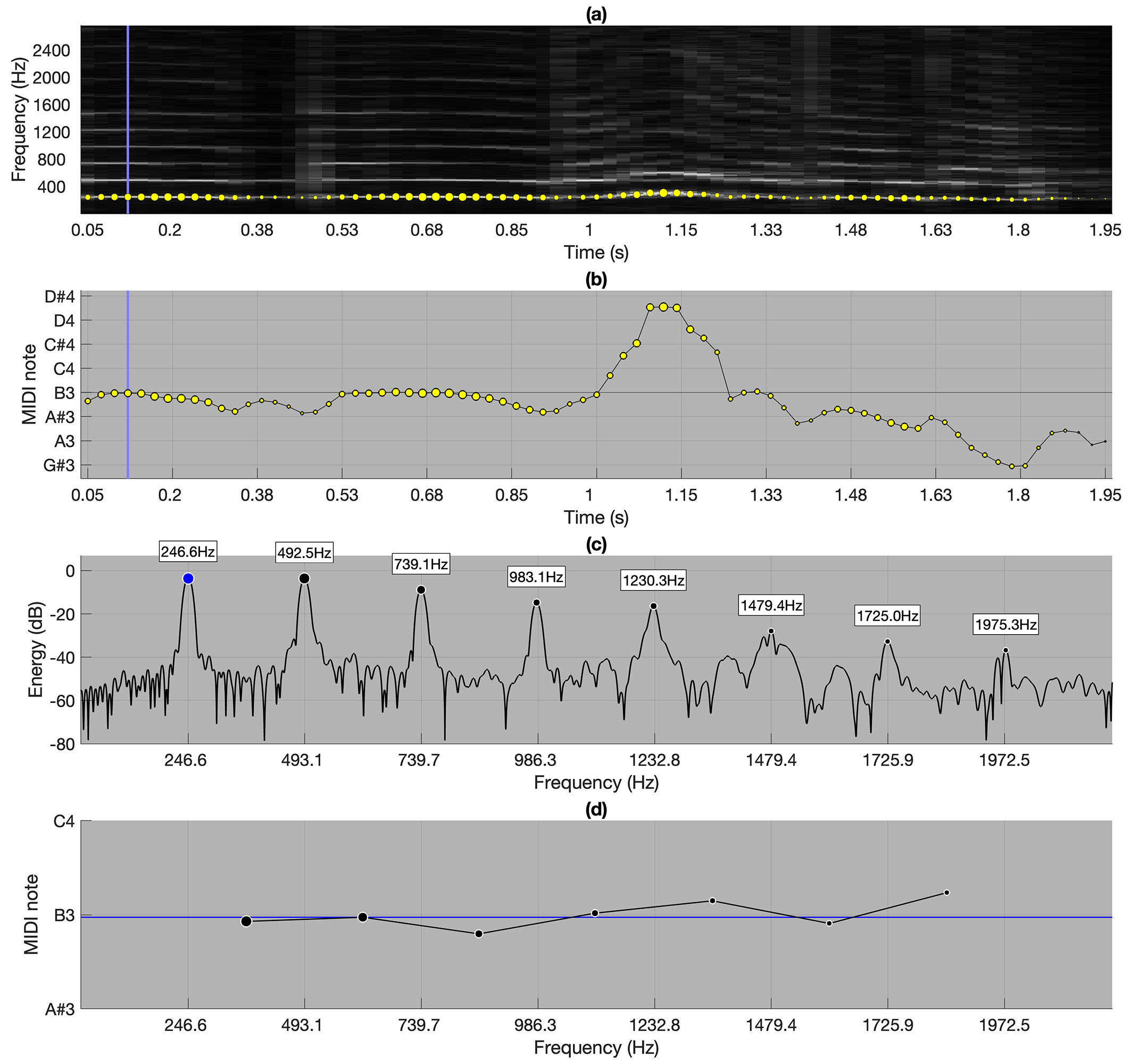}
  \caption{`Boom', vocal track, 0'17 to 0'19, unweighted. (a) STFT. The yellow scatter plot follows the lowest partial. (b) Lowest partial, frequency expressed as MIDI notes. (c) FT at the frame indicated by the blue line in (a) and (b). The x-axis grid is set on the harmonic positions for the lowest partial. (d) Frequency difference between consecutive partials from (c). The blue line shows the first partial's frequency.}
\label{fig:BoomVocals}
\end{framedfigure}

\clearpage
\subsection{`Boom', relations}\label{ref:boomrelations}

Figure~\ref{fig:BoomSpecanalysis} superimposes pitch values from the bass, the vocals, and the kick drum in `Boom'. Figure~\ref{fig:BoomSpecanalysis} (a) shows the pitch values derived from the $f_0$. The three tracks feature a continuous $f_0$ evolution. The kick drum's $f_0$ is almost identical to the bass's $f_0$ during the first two attacks. Figure~\ref{fig:BoomSpecanalysis} (b) is (a) folded into one octave. The vocals' $f_0$ is generally slightly lower than the bass's. Figure~\ref{fig:BoomSpecanalysis} (c) confirms that the folded $f_0$ pitch values are lower in the case of the vocals. Figure~\ref{fig:BoomSpecanalysis} (d) shows how differences in partial positions accentuate the detuning between the bass and vocals (see Section~\ref{sec:inharmonicityprimaal} for a discussion of this measure).

\begin{framedfigure}[h!]
  \centering
  \includegraphics[width=.75\columnwidth]{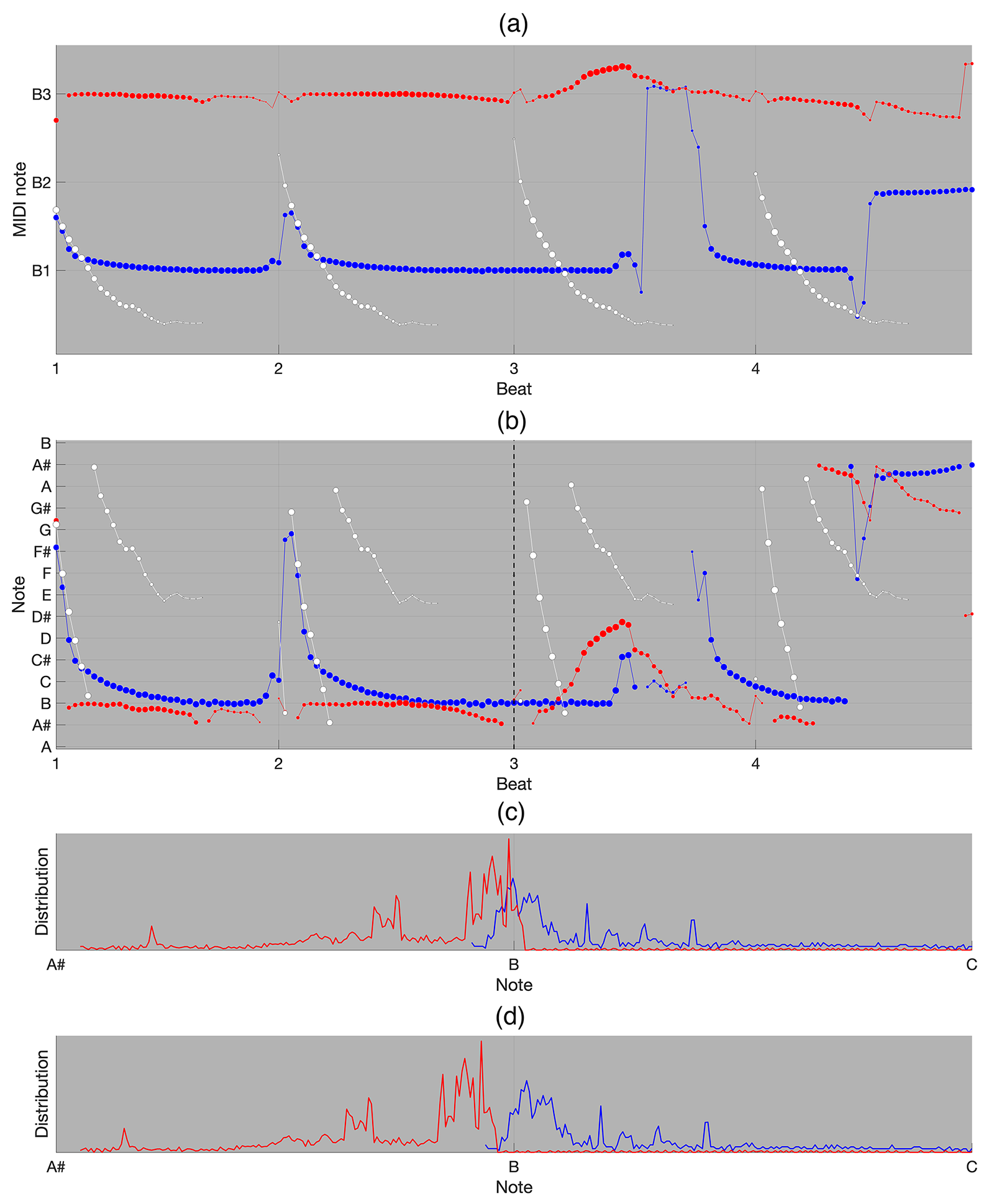}
  \caption{`Boom', 0'17 to 0'19, TR bass (blue), vocals (red), kick drum (white). (a) Pitches corresponding to the $f_0$. (b) Pitches corresponding to the $f_0$, folded inside one octave (A$\sharp$-A$\sharp$). (c) Bass and vocals, $f_0$ distributions, left of the vertical dashed line in (b). (d) Bass and vocals, $f_0$ distributions offset by the difference between $f_0$s and autocorrelation-based pitches.}
\label{fig:BoomSpecanalysis}
\end{framedfigure}

\clearpage
\subsection{`Cardinal', kick drum}\label{ref:cardinalkick}

Figure~\ref{fig:CardinalKick} shows (a) the STFT, (b) the weighted STFT, and (c) the MIDI notes corresponding to the first two partials ($f_0$ and harmonic 3) of the kick drum in `Cardinal'. The frequency values evolve continuously over approximately one octave. The kick drum is inharmonic and features only odd-numbered partials.

\vspace{.5cm}

\begin{framedfigure}[htbp]
  \centering
  \includegraphics[width=1\columnwidth]{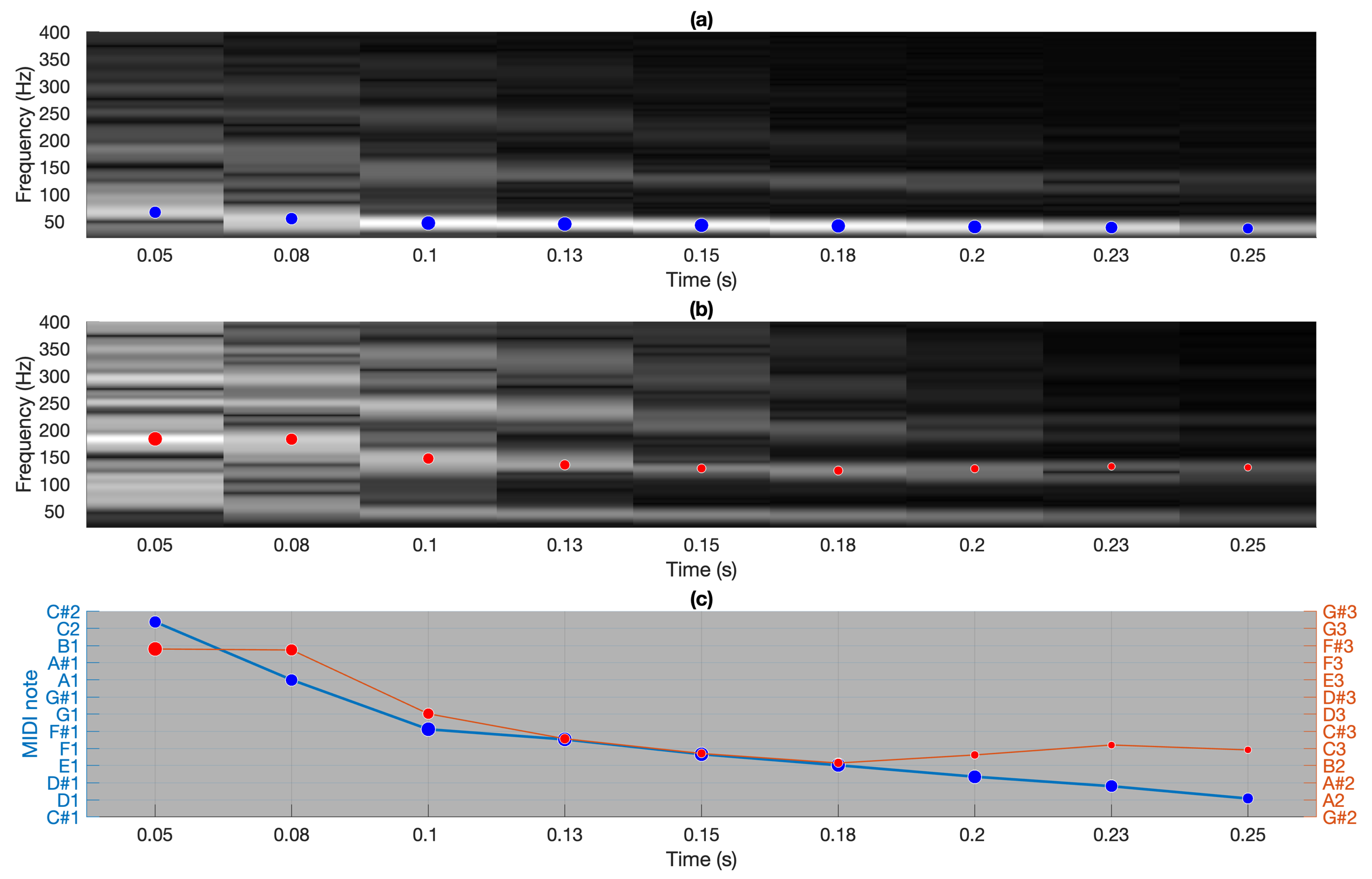}
  \caption{`Cardinal', kick drum. (a) STFT. The blue scatter plot tracks the lowest partial. (b) STFT, weighted. The red scatter plot tracks the loudest partial (close to harmonic 3, one octave and one fifth). (c) Comparison between the lowest and loudest partials.}
\label{fig:CardinalKick}
\end{framedfigure}

\clearpage
\subsection{`Cardinal', main vocals}\label{ref:cardinalvocals}

The vocal track in `Cardinal' derives from a 3-second-long sample. The sample revolves around a single pitch value. The sample is looped, and the repeated occurrences are heavily edited. Figure~\ref{fig:CardinalVocals} (a) and (b) suggest that the producers have filtered out the partials above 2~kHz. Figure~\ref{fig:CardinalVocals} (b) shows that the partials are harmonic. Figure~\ref{fig:CardinalVocals} (c) shows that the ambitus of the glides applied to this instance of the loop ranges from approximately a fourth to approximately a seventh. Figure~\ref{fig:CardinalVocals} (d) shows that outside the glides, the vocals' $f_0$ evolves within one semitone. Figure~\ref{fig:twodistributions} from Section~\ref{sec:continousfreqs} provides the distribution for the $f_0$.

\begin{framedfigure}[htbp]
  \centering
  \includegraphics[width=.87\columnwidth]{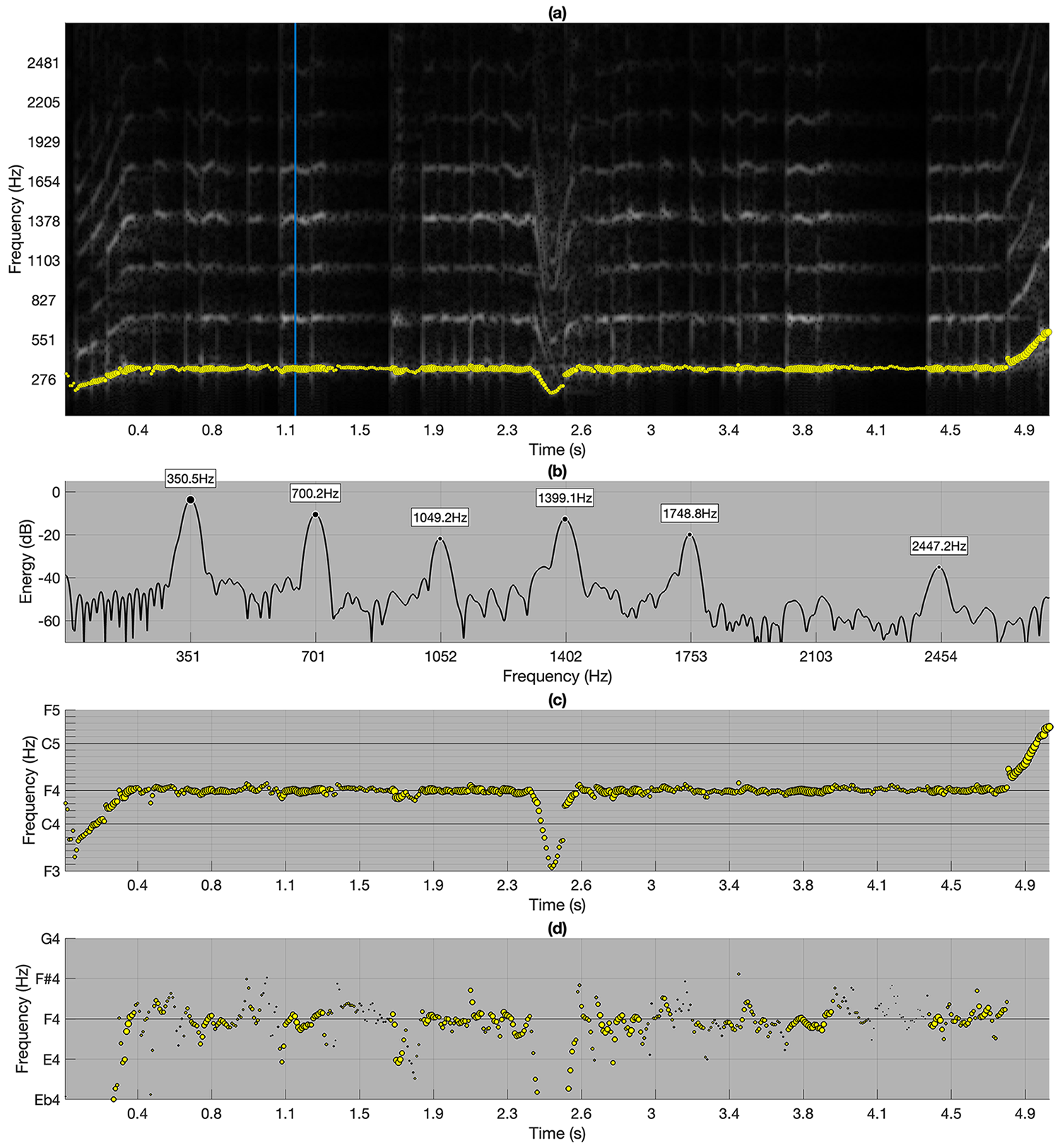}
  \caption{`Cardinal', vocals, unweighted audio. (a) STFT. The yellow scatter plot tracks the lowest partial. (b) FT at frame indicated by blue line in (a). (c) Pitch evolution of the lower partial in (a), expressed as MIDI pitch. (d) Same data as (c), zoomed in on a four-semitone range around F4.}
\label{fig:CardinalVocals}
\end{framedfigure}

\clearpage
\subsection{`Cardinal', synth break}\label{ref:cardinalsynthbreak}

The synth break (contrast) from `Cardinal', 2'03--2'05, was made using a Cyclone Analogic TT-303\footnote{\url{https://www.cyclone-analogic.fr/en/34-bass-bot-tt-303-0701980493430.html}}, a revival of the Roland TB-303\footnote{\url{https://www.roland.com/global/products/rc_tb-303/}}. The strong formants shown in Figure~\ref{fig:CardinalSynthBreak} originate from the `resonance' slider being set to a high value. The continuous pitch values of the formants derive from the use of the `slide time' feature.

Figure~\ref{fig:CardinalSynthBreak} (a) and (b) suggest that the set of partials related to the sound's $f_0$ is used as a scale. The audible elements in the scale are set by the resonance slider. Figure~\ref{fig:CardinalSynthBreak} (c) shows that both the scale and the elements audible within it change over time. The audible elements are set between one and four octaves above the $f_0$.

\vspace{.5cm}

\begin{framedfigure}[htbp]
  \centering
  \includegraphics[width=1\columnwidth]{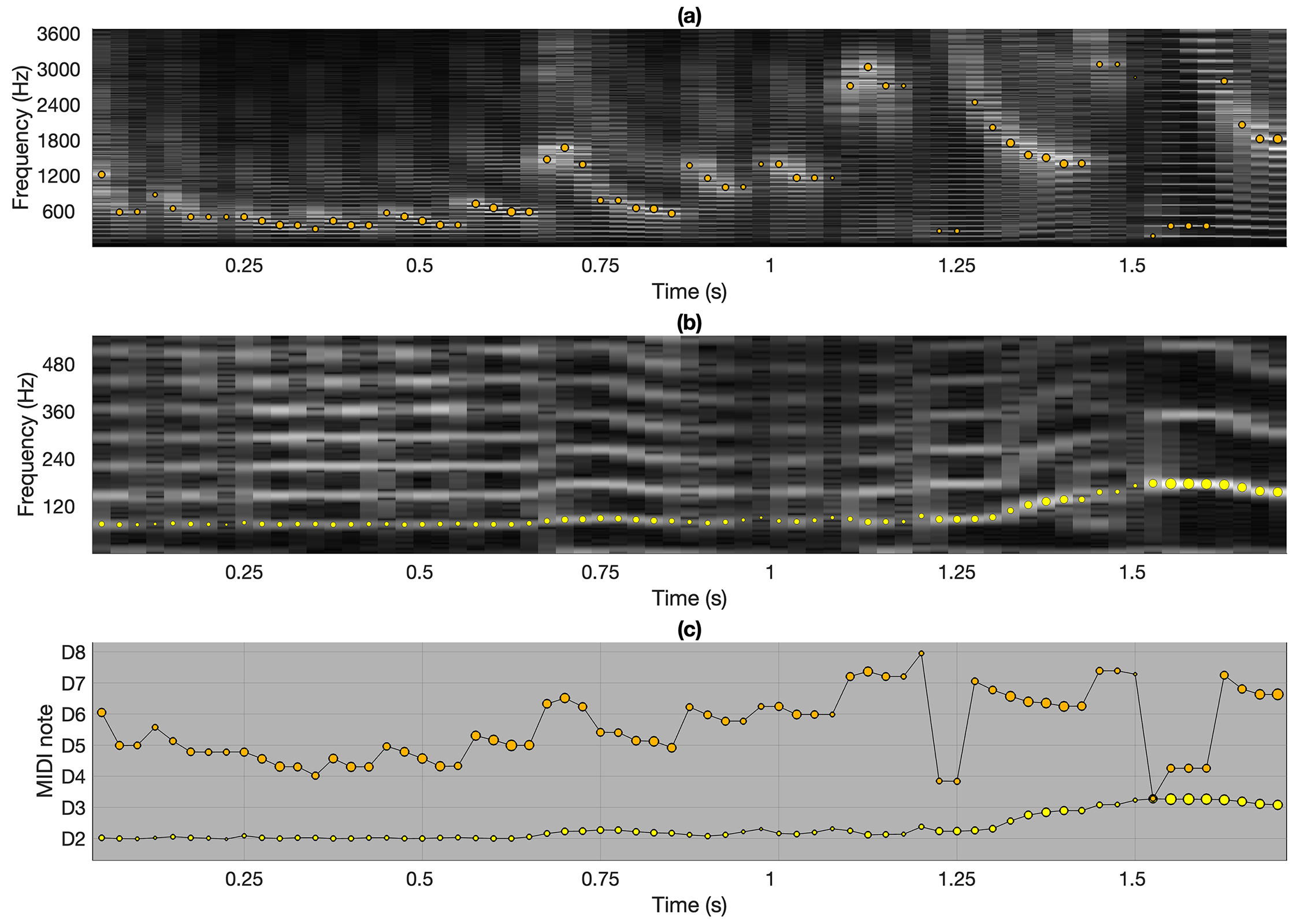}
  \caption{`Cardinal', synth break. (a) STFT, weighted. Orange scatter plot, partials with maximum loudness. (b) STFT, unweighted, zoom on the lower frequencies. Yellow scatter plot, tracking of the tone's lowest partial. (c) Yellow, evolution of the lowest partial in (b). Orange, evolution of the loudest partial in (a).}
\label{fig:CardinalSynthBreak}
\end{framedfigure}

\clearpage

\subsection{`Danger', bass tracks and kick drum}\label{ref:dangerbassandkick}

`Danger' features two bass tracks and a tuned kick track. The three parts are homorhythmic and loosely tuned between E$\flat$ and E, around the song's final of E$\flat$.

\subsubsection{Lower bass track}\label{ref:dangerlowbass}

The lower bass track for `Danger' was made in Omnisphere using Seismic Shock's `808 Woofer Warfare' patch. No effects were applied. Figure~\ref{fig:DangerLowBassUnweighted} (a) and (c) show that the bass mainly occupies the low-frequency register. Figure~\ref{fig:DangerLowBassUnweighted} (b) illustrates the 808-bass drum-like frequency profile. The ambitus of the initial glide ranges from one to more than three semitones. Figure~\ref{fig:DangerLowBassUnweighted} (c) shows that the bass's sound relies on odd harmonics. Figure~\ref{fig:DangerLowBassUnweighted} (d) shows that the partials are generally inharmonic. Figure~\ref{fig:DangerLowBassUnweighted} (e) and (f) show that partials 3 and 5 are the loudest. The main perceived pitch value for this `note' is E, close to the fundamental.

\subsubsection{Higher bass track}\label{ref:dangerhighbass}

The higher bass part for `Danger' was made in Omnisphere using Seismic Shock's `Subsonic Fracking' patch. It was heavily processed: Seismic Reverb, Tape Slammer, waveshaper, ring modulator, pitch bend, and dynamic filtering. The comparison between Figure~\ref{fig:DangerHighBassUnweighted} (a) and Figure~\ref{fig:DangerLowBassUnweighted} (a) suggests that the higher bass in `Danger' complements the lower bass in higher frequencies. Figure~\ref{fig:DangerHighBassUnweighted} (b) and (d) show that the higher bass's sound relies on odd partials. Figure~\ref{fig:DangerHighBassUnweighted} (c) shows that the sound is highly inharmonic, with frequency differences between consecutive partials corresponding, modulo the octave, to intervals up to three semitones below the fundamental. Figure~\ref{fig:DangerHighBassUnweighted} (e) suggests that several higher partials are much louder (ca. 30~LU) than the fundamental. One perceived pitch lies near E$\flat$, though its octave is uncertain. The perceived pitch is difficult to link with the partials.

\subsubsection{Kick drum}\label{ref:dangerkick}

The kick drum from `Danger' originates from an audio library. Figure~\ref{fig:DangerKick} (a) shows that the sample contains periodicity, with four clear partials forming a slightly inharmonic complex tone. Figure~\ref{fig:DangerKick} (b) shows that the pitch value derived from temporal modelling of pitch perception is close to the fundamental.

\clearpage

\begin{framedfigure}[h!]
  \centering
  \includegraphics[width=1\columnwidth]{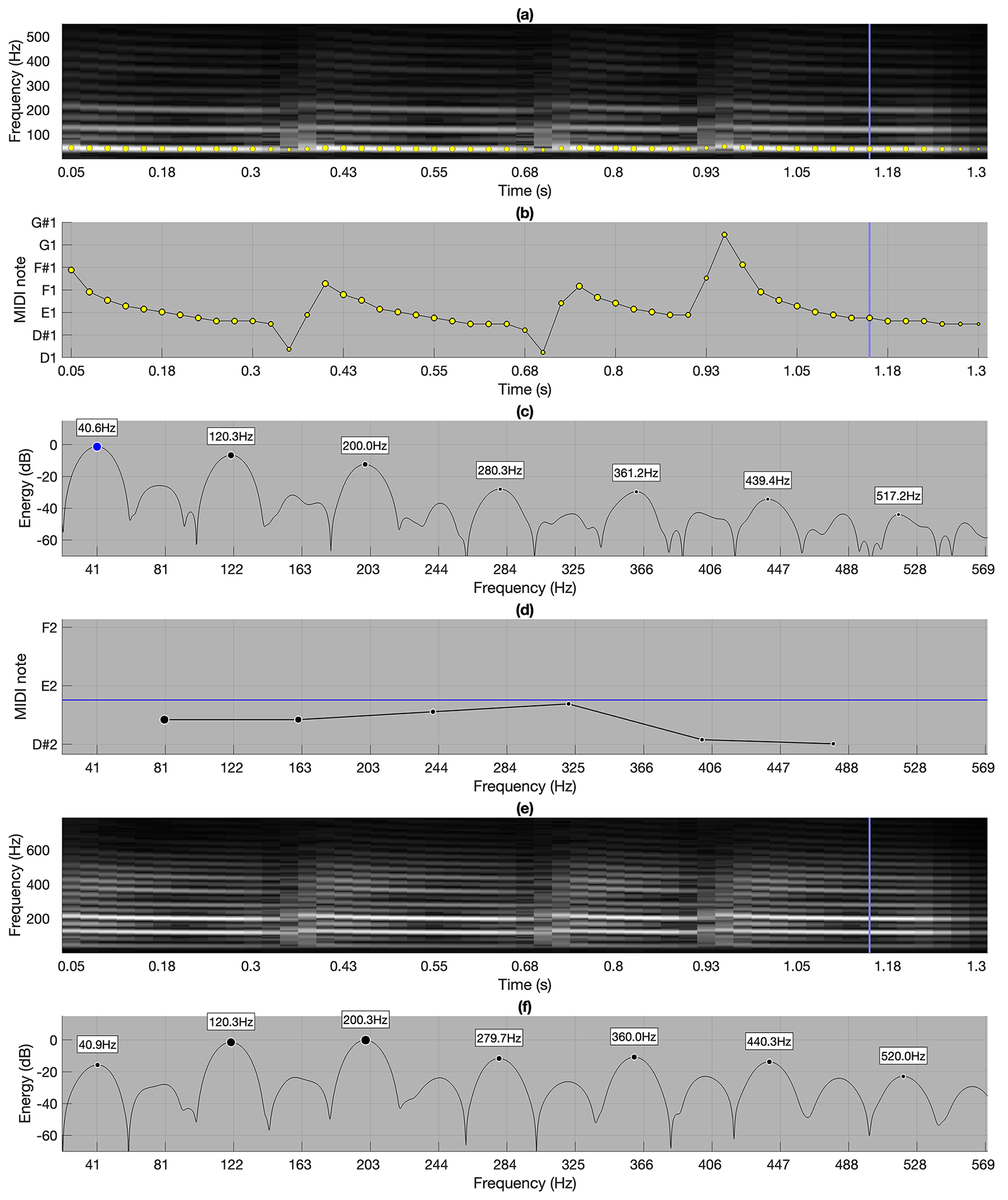}
  \caption{`Danger', lower bass track, 0'49 to 0'50. (a) to (d): unweighted. (e) and (f): weighted. (a) STFT. Yellow scatter plot, tracking of the tone's lowest partial. (b) Pitch evolution of the lower partial in (a). (c) FT at frame indicated by blue line in (a) and (b). The x-axis grid is set on the harmonic positions for the lowest partial. (d) Frequency difference between consecutive partials from (c). The horizontal blue line indicates twice the $f_0$ value from (b). (e) STFT. (f) FT at frame indicated by blue line in (e).}
\label{fig:DangerLowBassUnweighted}
\end{framedfigure}

\clearpage

\begin{framedfigure}[h!]
  \centering
  \includegraphics[width=1\columnwidth]{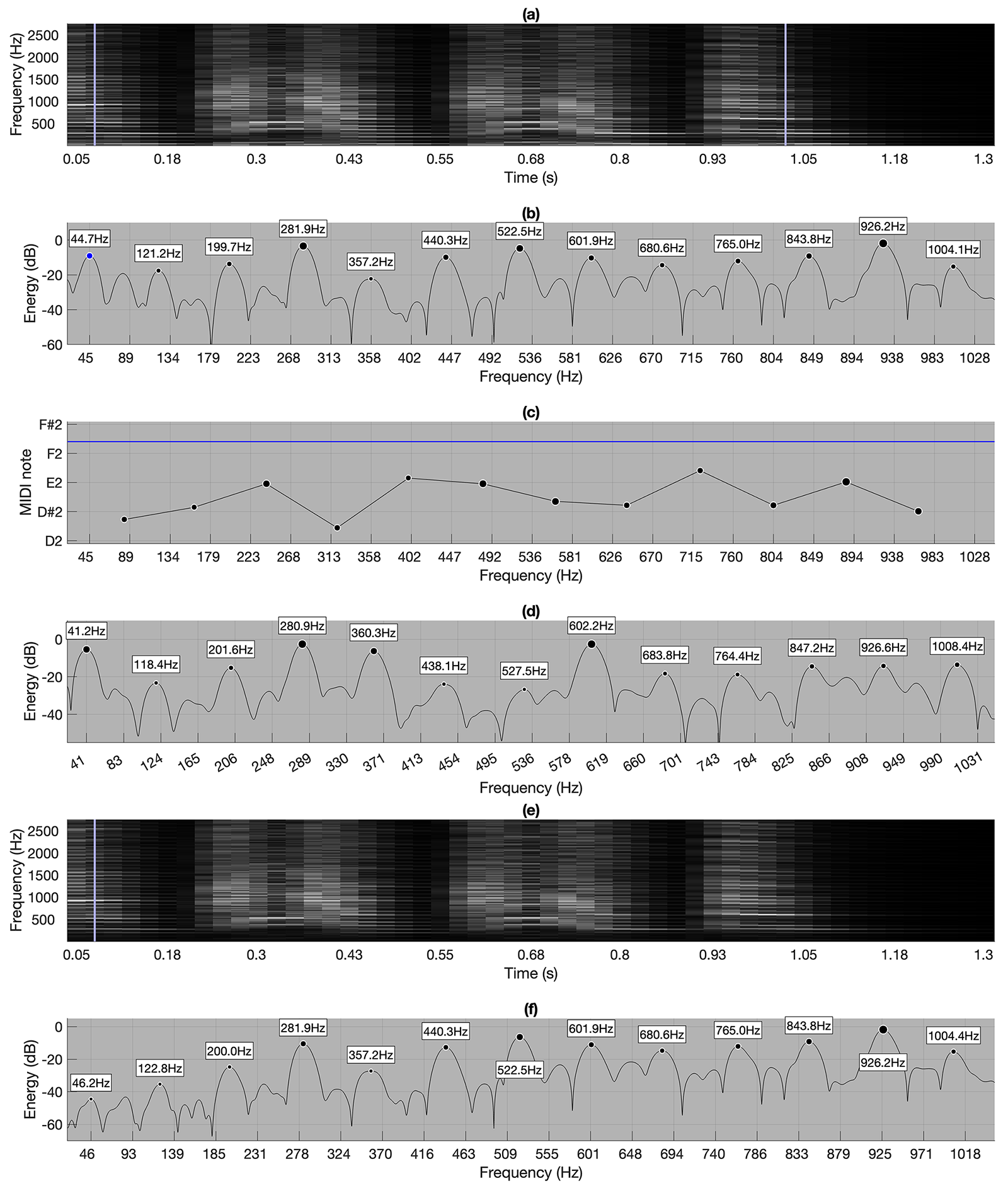}
  \caption{`Danger', higher bass track, 0'49 to 0'50. (a) to (d): unweighted. (e) and (f): weighted. (a) STFT. (b) FT at frame indicated by leftmost blue line in (a). The x-axis grid is set on the harmonic positions for the lowest partial. (c) Frequency difference between consecutive partials from (b). The horizontal blue line indicates twice the $f_0$ value from (b). (d) FT at frame indicated by rightmost blue line in (a). (e) STFT. (f) FT at frame indicated by blue line in (e).}
\label{fig:DangerHighBassUnweighted}
\end{framedfigure}

\clearpage

\begin{framedfigure}[htbp]
  \centering
  \includegraphics[width=1\columnwidth]{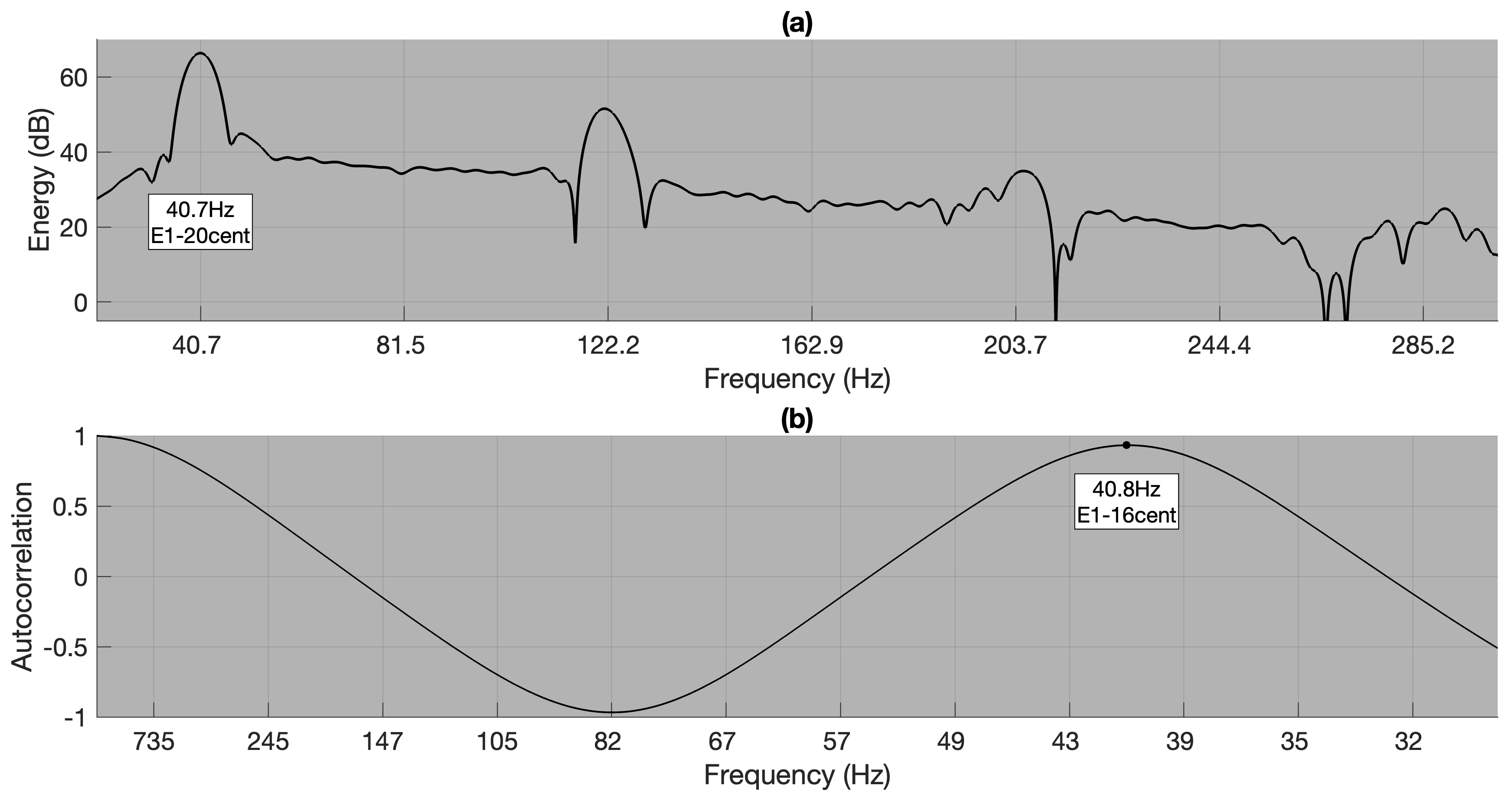}
  \caption{`Danger', kick drum sample, unweighted. (a) FT. The x-axis grid is set on the harmonic positions for the lowest partial. (b) Autocorrelation.}
\label{fig:DangerKick}
\end{framedfigure}

\vspace{1cm}

\subsection{`Elevate', bass}\label{ref:elevatebass}

Figure~\ref{fig:Elevate_bass} corresponds to the bass track from `Elevate', 35.7 to 37.5~s. It was generated using the `808 Woofer Warfare' patch and processed with Omnisphere's WaveShaper, Unison, and Seismic Pump plug-ins. A pitch bender was used during the recording of the part.

\begin{framedfigure}[h]
  \centering
  \includegraphics[width=1\columnwidth]{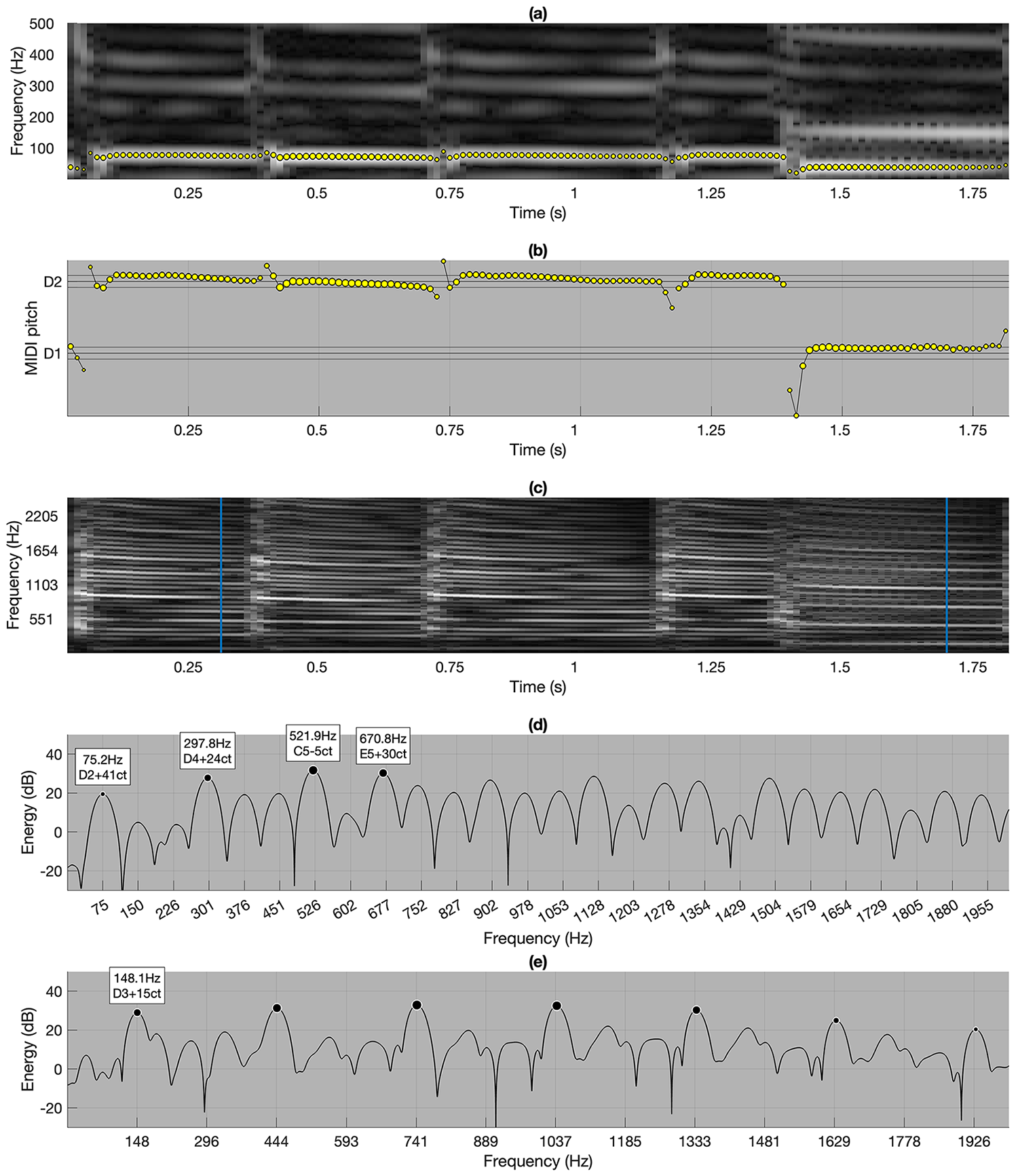}
  \caption{`Elevate', bass track, 35.7 to 37.5~s. (a) STFT. The yellow scatter plot tracks the lowest partial. (b) Lowest partial as MIDI notes. (c) STFT, weighted. (d) and (e), FT for the two frames corresponding to the blue lines in (c). The x-axis grid is set on the harmonic positions for the lowest partial.}
\label{fig:Elevate_bass}
\end{framedfigure}

Figure~\ref{fig:Elevate_bass} (a) and (b) show that frequency values are not stable. They typically evolve within one semitone. At least two lines can be heard simultaneously: (1) `D2 C\#2 D2 D2 D1' and (2) `D4 C\#4 D4 D4 D3'. Line (1) approximately corresponds to the physical maxima seen in Figure~\ref{fig:Elevate_bass} (a). Line (2) derives from upper partials shown in Figure~\ref{fig:Elevate_bass} (c). The partials giving rise to the perception of line (2) depend on the `note' in the sequence. Figure~\ref{fig:Elevate_bass} (d) suggests that the first D4 from sequence (2) can be heard as harmonics 4, 12, and 20 of D2, which become harmonics 1, 3, and 5 of D4. Figure~\ref{fig:Elevate_bass} (e) suggests that D3 from sequence (2) can be heard as a result of harmonics 4, 6, 8, 10, \ldots{} of D1.

\clearpage
\subsection{`Elevate', rising keyboard}\label{ref:elevatekey}

The rising keyboard from `Elevate' was made using a Super Jupiter MKS-80\footnote{\url{https://en.wikipedia.org/wiki/Roland_MKS-80}} waveform processed in Omnisphere with Waveshaper, Unison, a delay, and a compressor. The producers used a monophonic playback engine with slow glide settings and played two notes simultaneously, a major third apart, causing the engine to alternate unpredictably between upward and downward glides. According to the producers, this part was influenced by Missy Elliott's `Lose Control'\footnote{\url{https://www.youtube.com/watch?v=na7lIb09898}}.

\vspace{1cm}

\begin{framedfigure}[h]
  \centering
  \includegraphics[width=1\columnwidth]{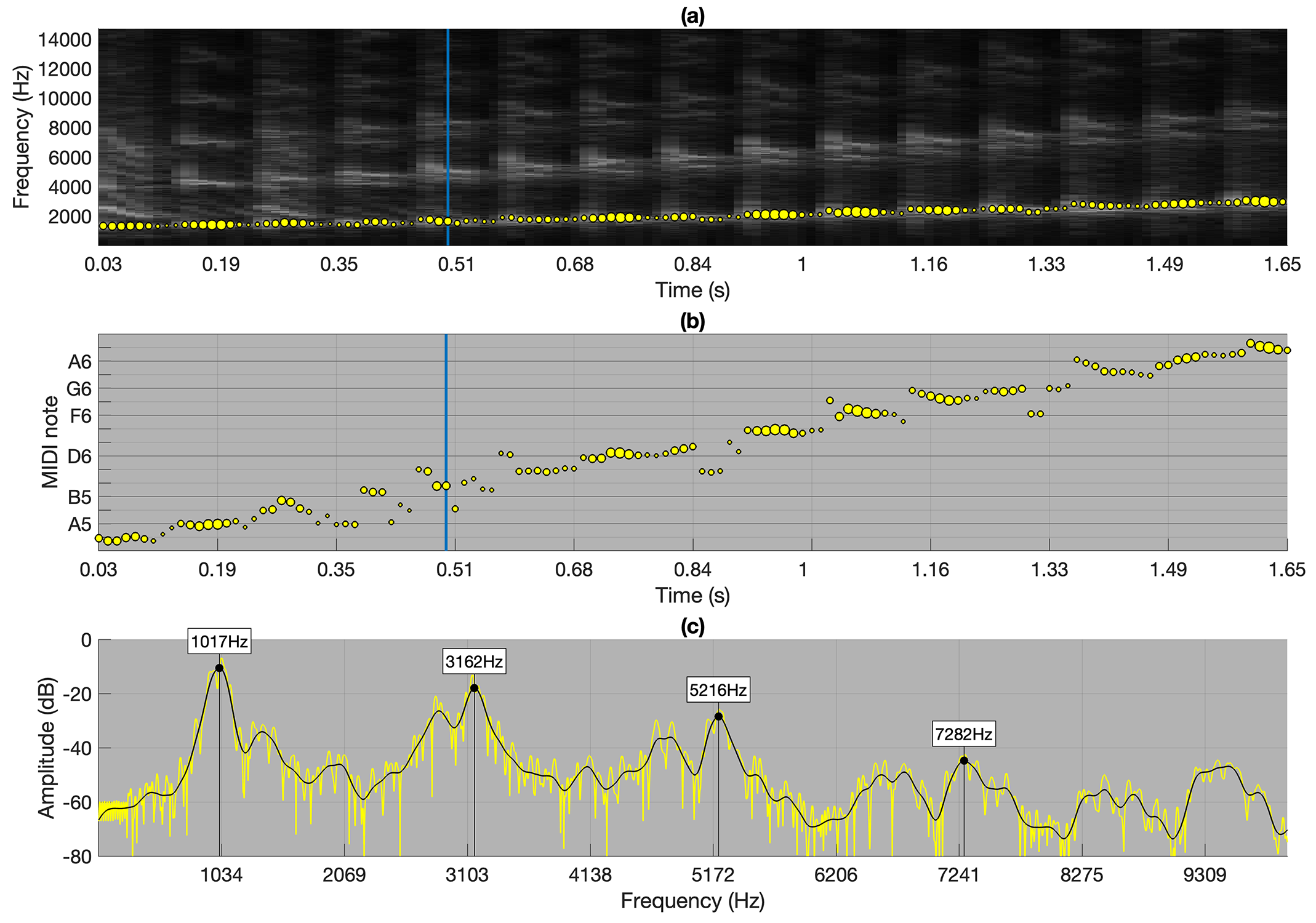}
  \caption{`Elevate', keyboard track, 0'14 to 0'16. (a) STFT, weighted. The yellow scatter plot tracks the lowest partial. (b) Lowest partial, MIDI pitches. (c) FT for the two frames corresponding to the blue lines in (a) and (b). Yellow, original values. Black, smoothed spectrum.}
\label{fig:Elevate_keyboard}
\end{framedfigure}

\clearpage

\vspace{2cm}

\subsection{`\textexclamdown{}Fire!', bass}\label{ref:firebass}

The bass track in `\textexclamdown{}Fire!' combines two sources: (1) a low-frequency sine wave processed with Omnisphere's Waveshaper, and (2) an Industrial Music Electronics Piston Honda processed with the Mutable Instruments Rings resonator\footnote{\url{https://pichenettes.github.io/mutable-instruments-documentation/modules/rings/}}. The Waveshaper is used to generate upper partials. The Rings unit uses resonators to modify the spectrum's envelope.

\subsubsection{Unweighted}

Figure~\ref{fig:FireBass} (a) shows the STFT of the unweighted audio for the bass track, focusing on 20 to 2600~Hz. One can observe a low-frequency partial (37~Hz) originating from the sine wave. Figure~\ref{fig:FireBass} (b) shows the evolution of the lower partial's frequency over time. The grid indicates the position of the `notes'. Their frequency slowly increases. The last `notes' are subject to octaviation. Figure~\ref{fig:FireBass} (c) shows the FT corresponding to the blue line in Figure~\ref{fig:FireBass} (a). The partials that are near multiples of the 37~Hz partial derive from both sources: source (1) (up to ca. 600~Hz) and source (2) (from ca. 400~Hz). Source (2) is set so that its partials approximately coincide with source (1)'s. Figure~\ref{fig:FireBass} (d) highlights the inharmonicity of the resulting tone.

\subsubsection{Weighted}

Figure~\ref{fig:FireBassWeighted} (a) shows the STFT of the weighted audio for the bass track. The weighting attenuates the contribution of source (1). Figure~\ref{fig:FireBassWeighted} (c) shows the FT corresponding to the blue line in Figure~\ref{fig:FireBassWeighted} (a). According to Figure~\ref{fig:FireBassWeighted} (c), the envelope's peaks have no harmonic relations. The 37~Hz fundamental is very weak, suggesting that any corresponding pitch is recovered primarily through temporal modelling of pitch perception \citep{yost2009pitch}. Figure~\ref{fig:FireBassWeighted} (b) shows that the partial around 500~Hz, generated by source (2), evolves upwards continuously except near the end of the extract, where it is subject to octaviation.

\clearpage

\begin{framedfigure}[h!]
  \centering
  \includegraphics[width=1\columnwidth]{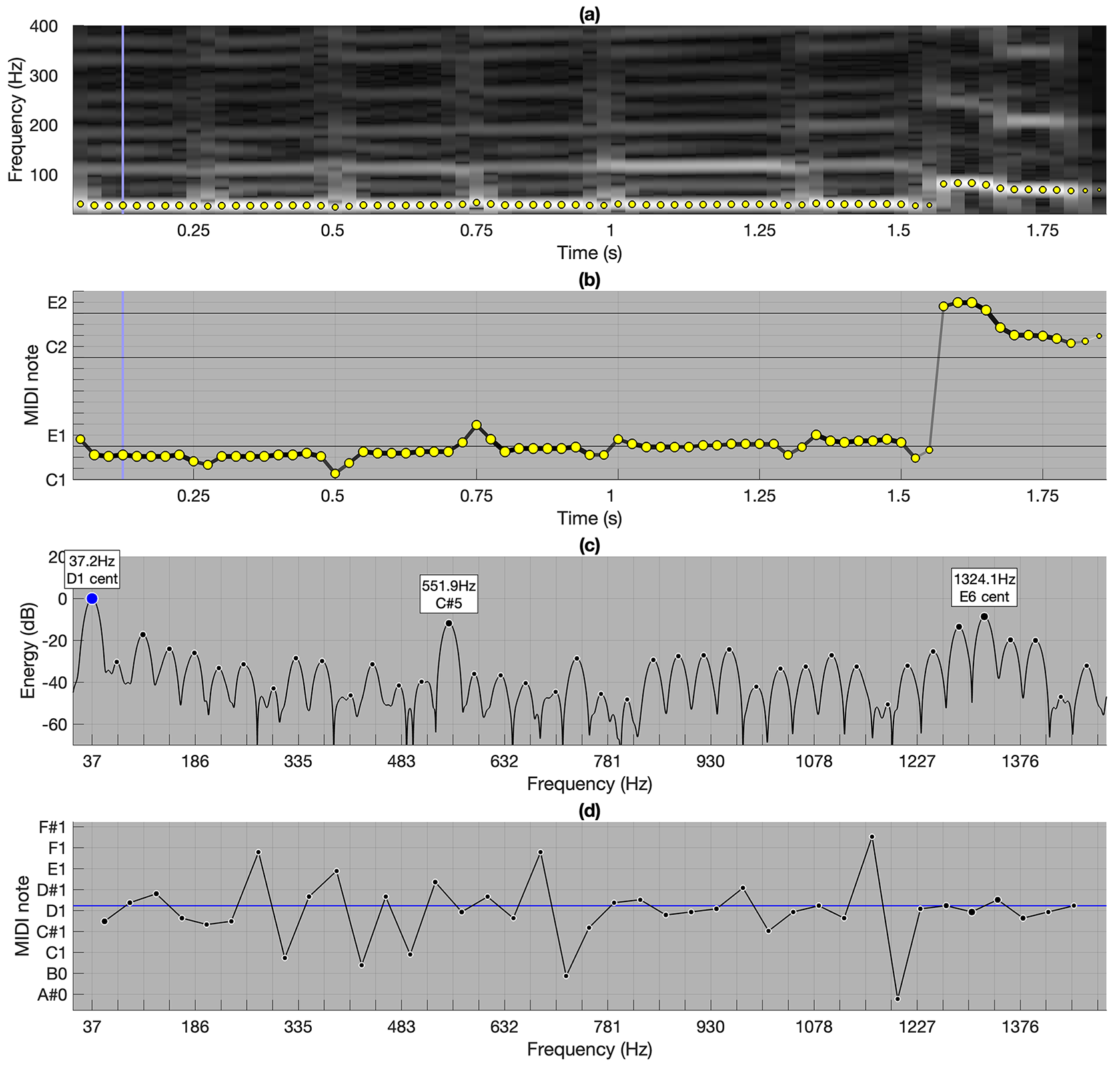}
  \caption{`\textexclamdown{}Fire!', bass track. (a) STFT, unweighted. (b) MIDI note corresponding to the $f_0$ value from (a). (c) FT at frame indicated by blue line in (a) and (b). The x-axis grid is set on the harmonic positions for the lowest partial. (d) Frequency difference between consecutive partials from (c). The horizontal blue line indicates the $f_0$ value from (b).}
\label{fig:FireBass}
\end{framedfigure}

 \clearpage

\begin{framedfigure}[h]
  \centering
  \includegraphics[width=1\columnwidth]{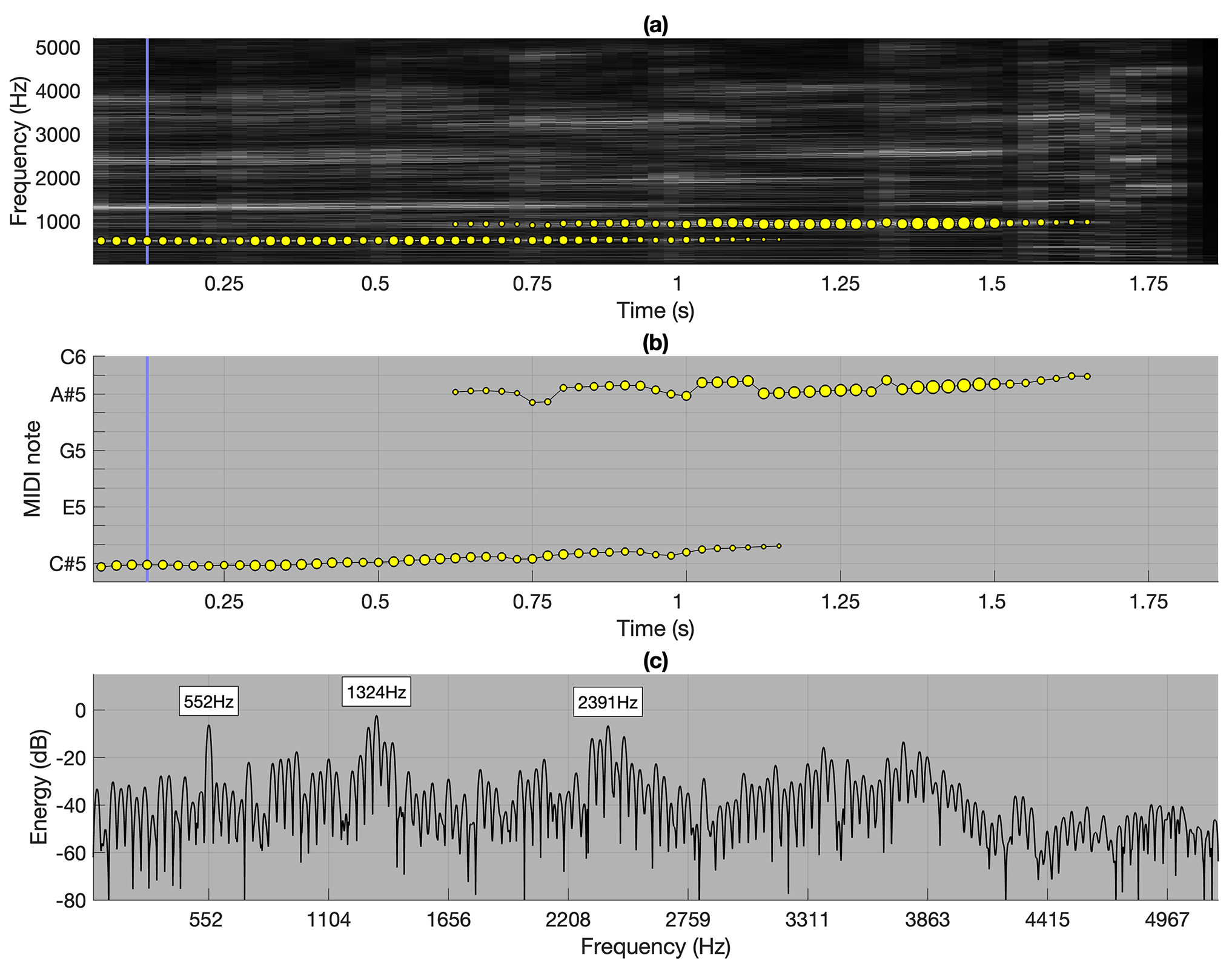}
  \caption{`\textexclamdown{}Fire!', bass track. (a) STFT, weighted. The yellow scatter plots track the maximum energy in the lower register. (b) Yellow scatter plots from (a), frequencies as MIDI notes. (c) Power spectrum for the frame indicated by the blue lines in (a) and (b). The x-axis grid is set on the harmonic positions of the 552~Hz partial.}
\label{fig:FireBassWeighted}
\end{framedfigure}

\clearpage

\subsection{`R U Ready', bass}\label{ref:areureadybass}

\subsubsection{Synth bass}\label{ref:areureadysynthbass}

The bass track for `R U Ready' was generated in Omnisphere using Seismic Shock's `Subsonic Fracking' preset. The sound was then processed with a variety of effects, all in Omnisphere: Waveshaper, Ring Modulation, Frequency Modulation, Unison, Seismic Verb, and Stomp Modeler's distortion\footnote{\url{https://support.spectrasonics.net/manual/Omnisphere2/25/en/topic/stompbox-modeler}}.

Figure~\ref{fig:areyoureadybass} (b) shows that the $f_0$ values for the bass are not stable. Inside a single `note', they evolve continuously within an ambitus of one to two semitones. Figure~\ref{fig:areyoureadybass} (c) shows that the bass relies on odd harmonics only. Figure~\ref{fig:areyoureadybass} (c) and (d) show that the bass sound is highly inharmonic, with pitch values derived from temporal modelling that vary within a four-semitone ambitus and are generally lower than the pitch corresponding to the $f_0$.

Figure~\ref{fig:areyoureadybass} (e) suggests that the partials higher than ca. 300~Hz are louder than the fundamental. During the contrast (see Section~\ref{sec:smallscalestructure}), harmonic 3 is much louder than the others. Figure~\ref{fig:areyoureadybass} (f) suggests that the strongest partials of the first `note' are B4 and D\#5, without an obvious connection to the low D1 fundamental.

\subsubsection{Electric bass}\label{ref:areureadyelecbass}

The electric bass in `R U Ready' was generated in Native Instruments Kontakt\footnote{\url{https://www.native-instruments.com/en/products/komplete/samplers/kontakt-7/}} using the Scarbee Rickenbacker bass instrument\footnote{\url{https://scarbee.com/collections/basses/products/scarbee-rickenbacker-bass}}, preset `So What'. The instrument's output was then processed in Guitar Rig using a compressor, an EQ, and an amp simulator. Figure~\ref{fig:areyoureadyelecbass} (a) and (c) show that harmonic 2 has more energy than the fundamental. Figure~\ref{fig:areyoureadyelecbass}~(b) suggests that the $f_0$ values of the electric bass remain closer to the chromatic scale than those of the synth bass. Figure~\ref{fig:areyoureadyelecbass} (d) shows that partials 2 to 6 are harmonic, whereas upper partials are not. According to \citet{carlyon1994comparing}, lower harmonic partials result in robust pitch even if higher partials are inharmonic.

\clearpage

\begin{framedfigure}[h!]
  \centering
  \includegraphics[width=1\columnwidth]{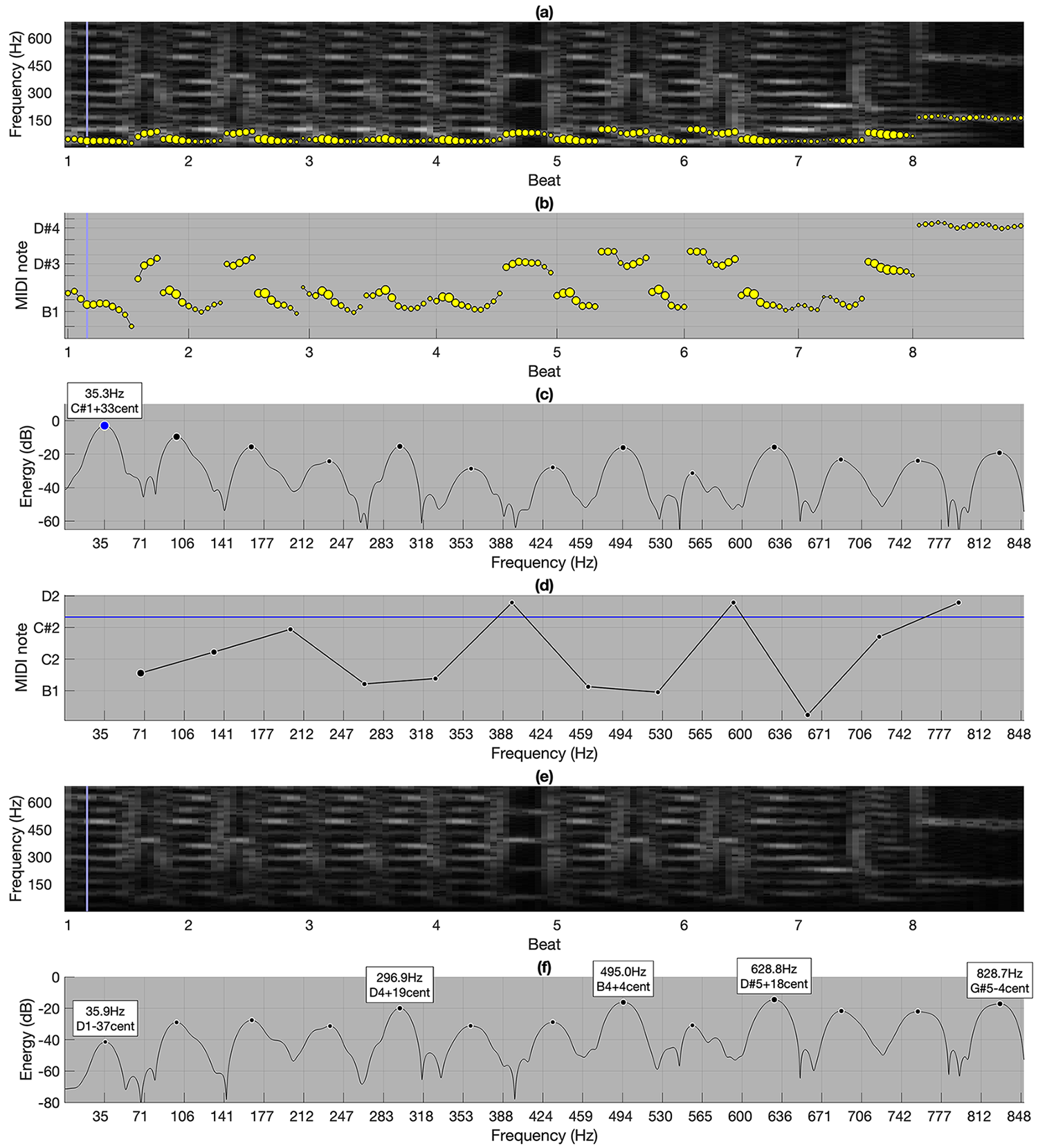}
  \caption{`R U Ready', synth bass track, 0'40 to 0'44. (a) to (d), unweighted. (e) and (f), weighted. (a) STFT. The yellow scatter plot tracks the lower partial close to the lower perceived pitch sequence. (b) Lower partial as MIDI pitch. (c) FT at frame indicated by blue line in (a) and (b). The x-axis grid is set on the harmonic positions for the lowest partial. (d) Frequency differences between consecutive partials from (c). The horizontal blue line indicates twice the $f_0$ value from (b) (odd-numbered harmonics). (e) STFT. (f) FT at frame indicated by blue line in (e).}
\label{fig:areyoureadybass}
\end{framedfigure}

\clearpage

\begin{framedfigure}[h!]
  \centering
  \includegraphics[width=1\columnwidth]{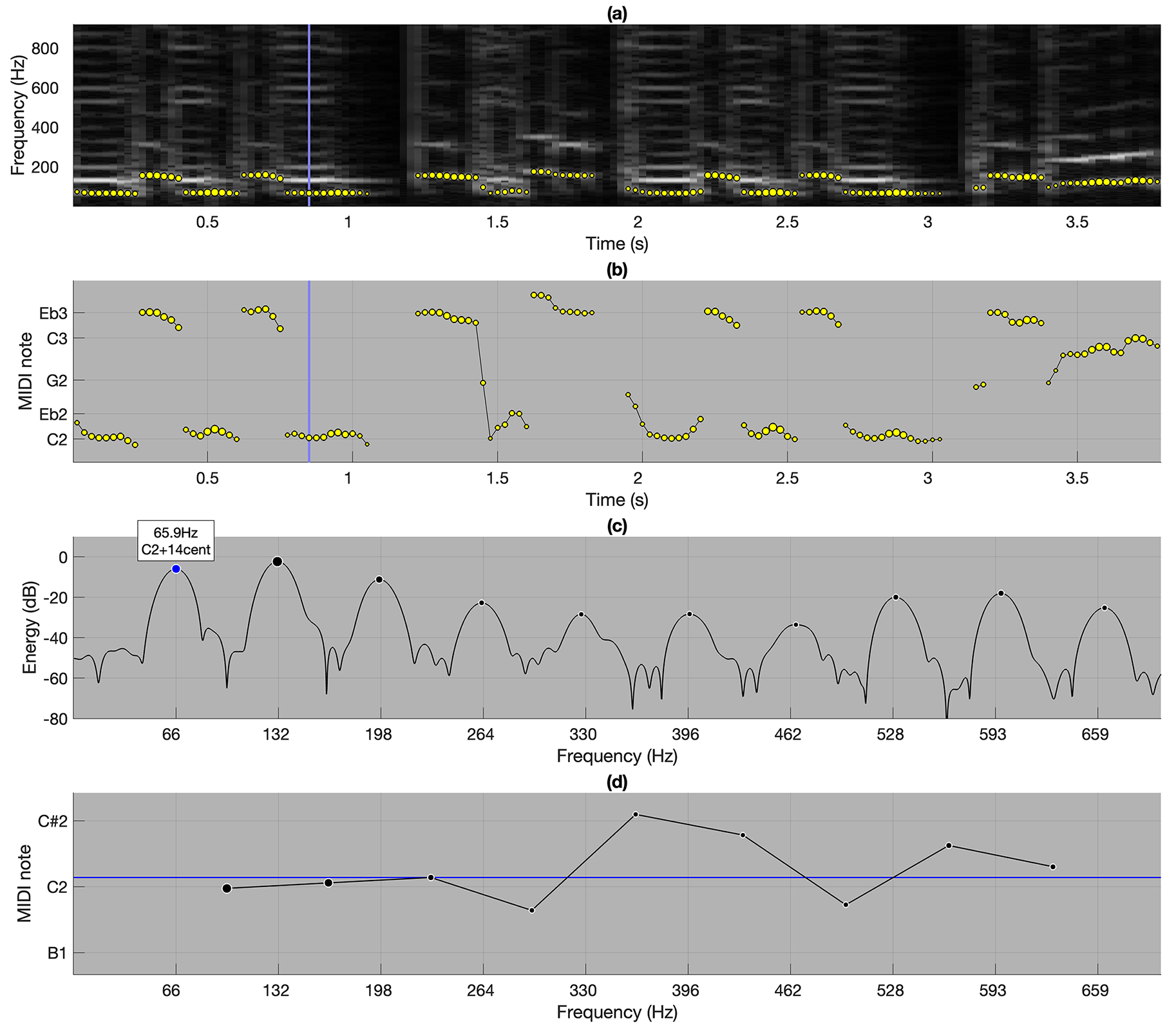}
  \caption{`R U Ready', electric bass track, 0'32 to 0'36, unweighted. (a) STFT. The yellow scatter plot tracks the $f_0$ values near the perceived pitches. (b) MIDI note corresponding to the $f_0$ value from (a). (c) FT at frame indicated by blue line in (a) and (b). The x-axis grid is set on the harmonic positions for the lowest partial. (d) Frequency difference between consecutive partials from (c). The horizontal blue line indicates the $f_0$ estimated from the lowest partial in (c).}
\label{fig:areyoureadyelecbass}
\end{framedfigure}

\clearpage
\subsection{`Silver', bass and kick drum}\label{ref:silverbass}

The bass for `Silver' was generated using Omnisphere's `808 Woofer Warfare' patch, with the `tone' setting at the highest possible value. The producers then used Omnisphere's Waveshaper and Ring Modulator to selectively boost partials, Sonic Extension's Seismic Pump to boost medium frequencies, and Seismic Verb to boost high frequencies. They used two different settings, resulting in two bass types: type 2 runs between 0'41 and 0'57, and type 1 runs during the rest of the song. 

\vspace{1cm}

\subsubsection{Bass type 1, unweighted}\label{ref:silverbasstype1unweighted}

Figure~\ref{fig:SilverBass} shows features from the `Silver' bass, type 1, unweighted. Figure~\ref{fig:SilverBass} (b) shows that the evolution of the bass's $f_0$ value is continuous. The $f_0$ value never stabilises to a fixed value. The downward glide ranges from one to three semitones. Figure~\ref{fig:SilverBass} (c) shows that the bass only involves odd harmonics. Figure~\ref{fig:SilverBass} (d) shows that the bass is inharmonic. Figure~\ref{fig:SilverBass} (e) shows that the pitch value derived from temporal modelling is ca. 50 cents below the $f_0$.

\subsubsection{Bass type 1, weighted}

Figure~\ref{fig:SilverBassWeighted} shows features from the `Silver' bass, type 1, weighted. Figure~\ref{fig:SilverBassWeighted} (a) and (b) show that some upper partials are significantly louder than the fundamental and that the loudest partial changes over time. Figure~\ref{fig:SilverBassWeighted} (c) shows the spectrum at the frame indicated in (a) and (b). The producers deliberately highlighted changing partials, creating a melody from the sequence of loudest partials.

\begin{framedfigure}[h!]
  \centering
  \includegraphics[width=1\columnwidth]{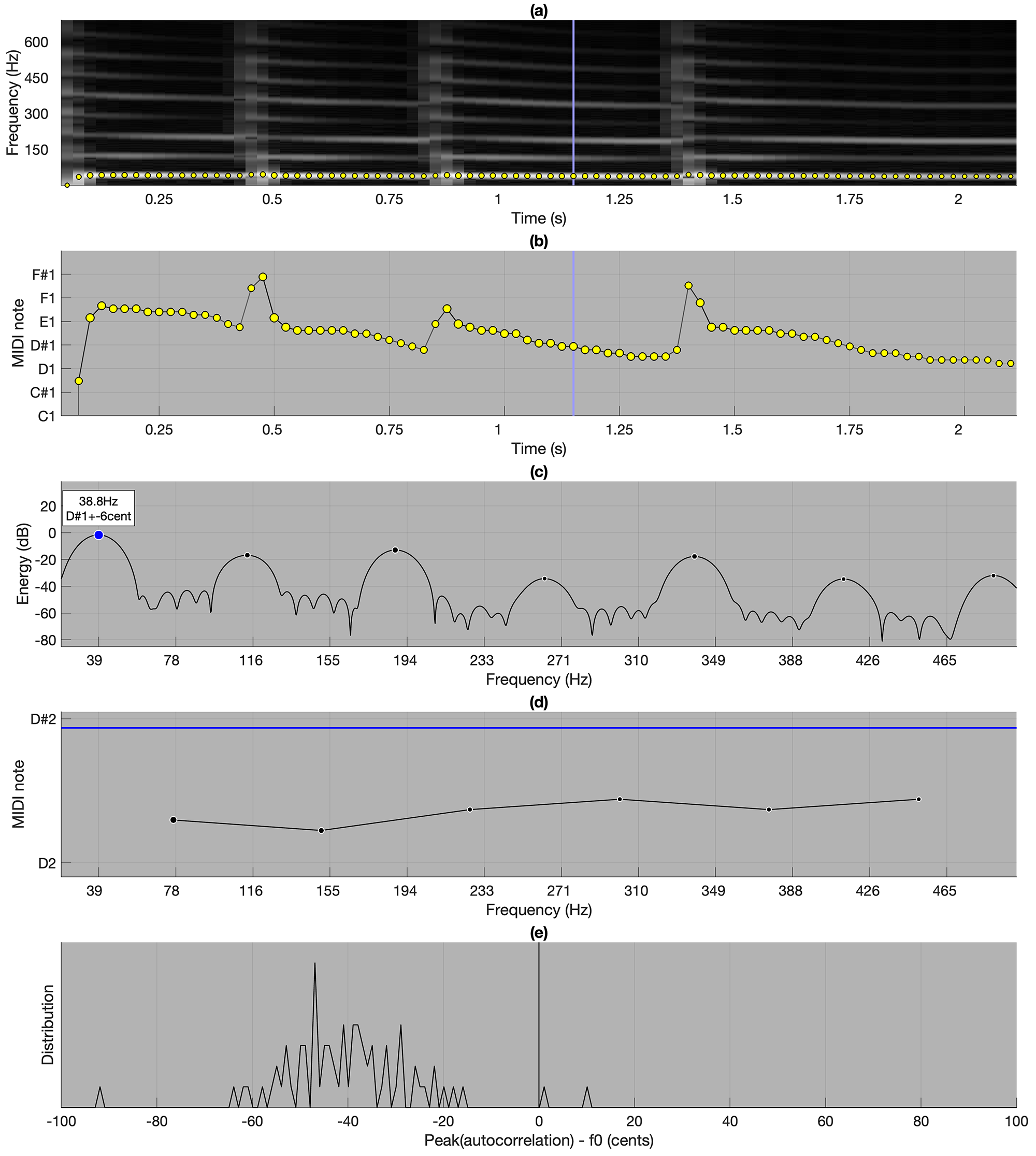}
  \caption{`Silver', bass track (type 1), 0'15 to 0'17. (a) STFT, unweighted. The yellow scatter plot tracks the lowest partial. (b) MIDI note corresponding to the lowest partial. (c) FT at the frame indicated by the blue line in (a) and (b). The x-axis grid is set on the harmonic positions for the lowest partial. (d) Frequency difference between consecutive partials from (c). The horizontal blue line indicates twice the $f_0$ value from (b) (only odd-numbered harmonics). (e) Difference between the autocorrelation peak after the first zero-crossing and the lowest partial's frequency.}
\label{fig:SilverBass}
\end{framedfigure}

\subsubsection{Bass type 2, unweighted}\label{ref:silverbasstype2}

Figure~\ref{fig:SilverBass2} shows features from the `Silver' bass, type 2, unweighted. Figure~\ref{fig:SilverBass2} (b) compares the evolution of the $f_0$ values from type 1 and 2 basses. Both $f_0$ values are continuous and never stabilise. The $f_0$ value corresponding to type 2 is generally one semitone lower than the value corresponding to type 1. Figure~\ref{fig:SilverBass2} (c) shows that type 2 bass differs from type 1 bass. Whereas the type 1 bass used odd harmonics, the type 2 bass relies on the first few harmonics, then involves harmonics 6, 9, 12, etc. Figure~\ref{fig:SilverBass2} (d) shows the distance between the partials from (c). Unlike the type 1 bass, this distance corresponds to a higher pitch than the fundamental. Figure~\ref{fig:SilverBass2} (e) confirms this observation by showing that the pitch value derived from temporal modelling is ca. 35 cents above the $f_0$.

\begin{framedfigure}[h!]
  \centering
  \includegraphics[width=1\columnwidth]{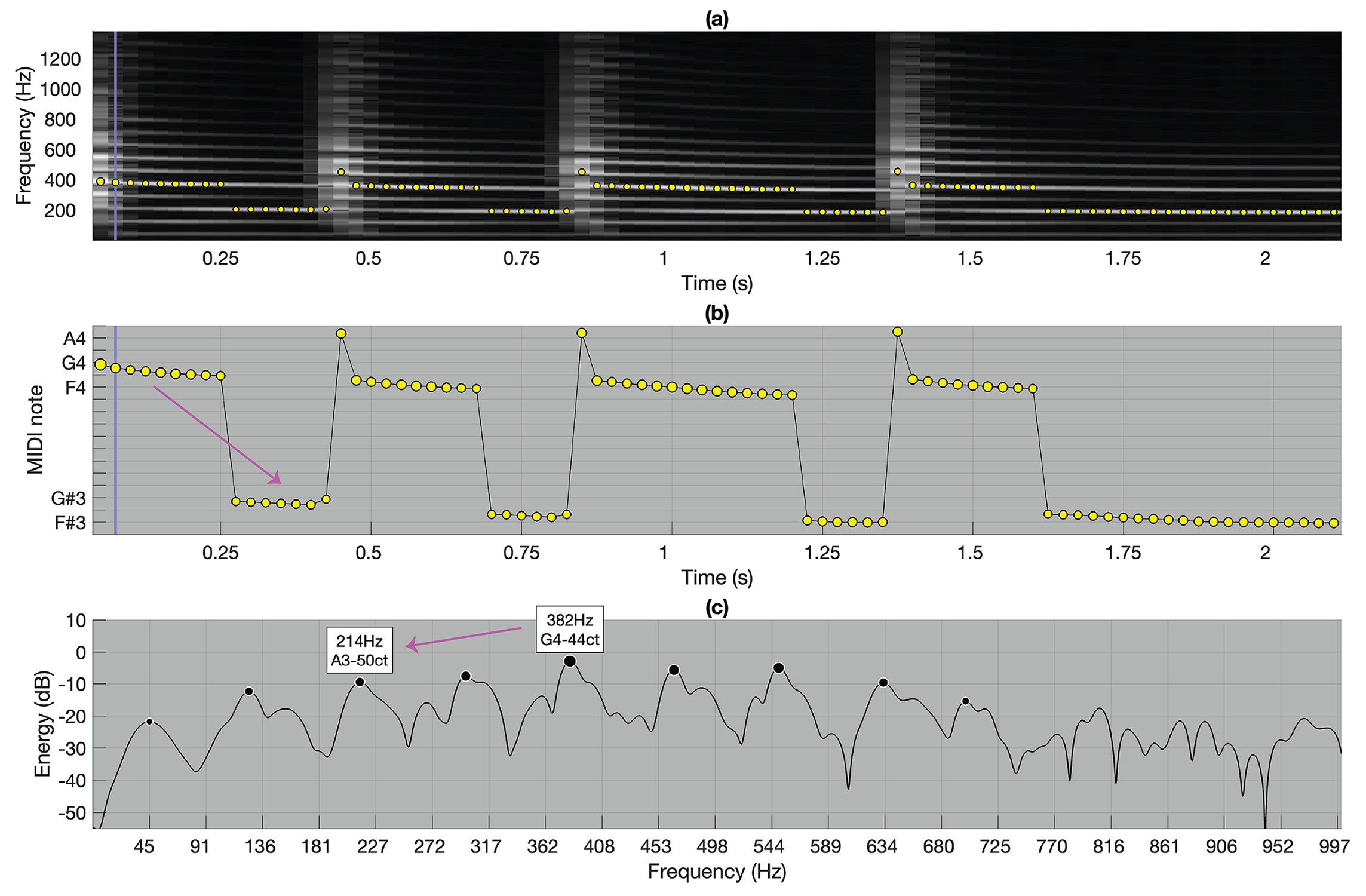}
  \caption{`Silver', bass track (type 1), weighted. (a) STFT, weighted. The yellow scatter plot tracks the loudest partial. (b) Frequency of the loudest partial, expressed as MIDI note. (c) FT at the frame indicated by the blue line in (a) and (b). The blue arrows identify the changing loudest partial in the temporal and spectral representations.}
\label{fig:SilverBassWeighted}
\end{framedfigure}

\clearpage

\begin{framedfigure}[h!]
  \centering
  \includegraphics[width=1\columnwidth]{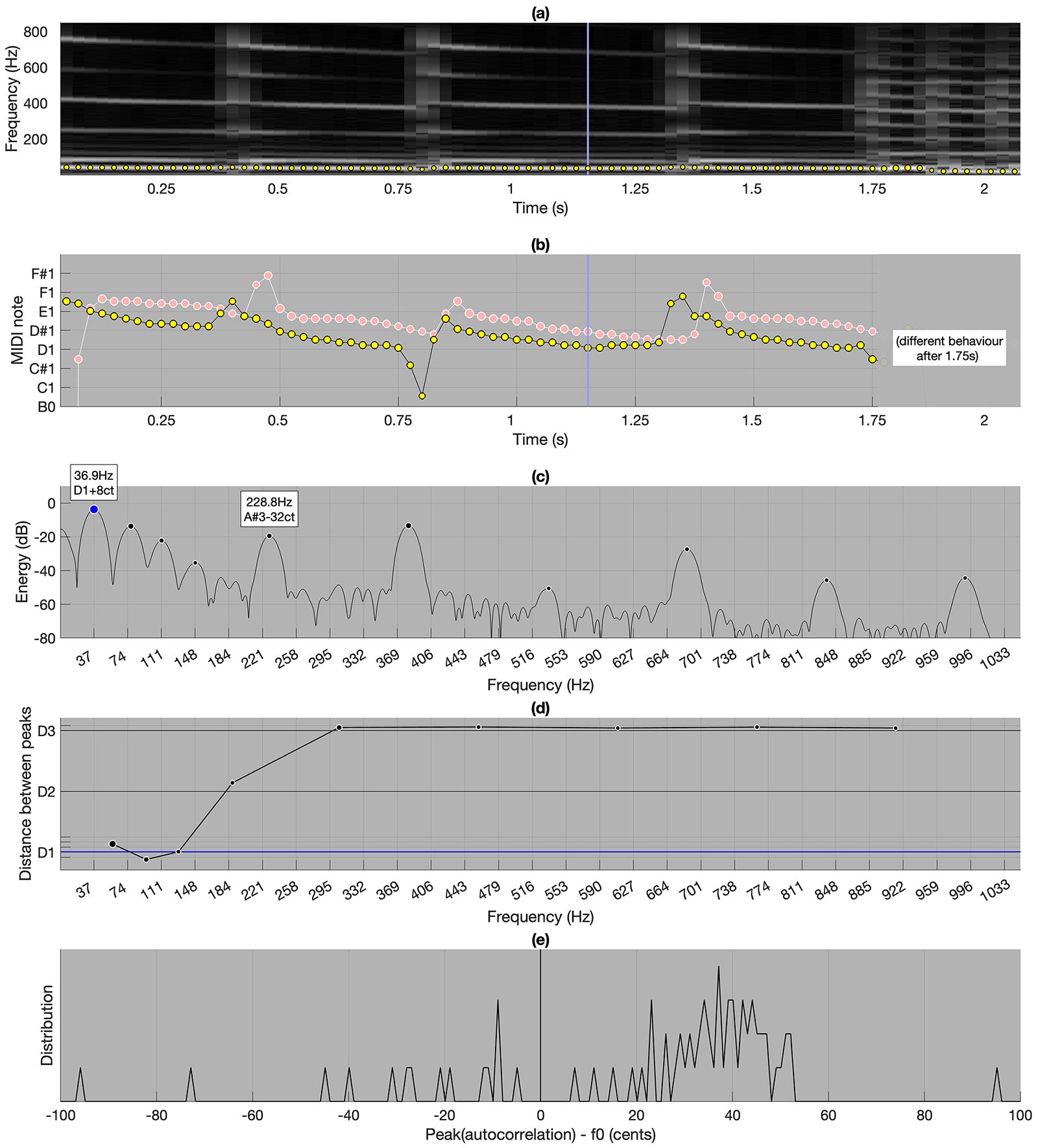}
  \caption{`Silver', bass track, 0'47 to 0'49 (type 2), unweighted. (a) STFT. The yellow scatter plot tracks the lowest partial. (b) Yellow: note corresponding to the $f_0$ from (a). Pale red: lowest partial from the `Silver' bass extract in Figure~\ref{fig:SilverBass} (type 1). (c) FT at the frame indicated by the blue line in (a) and (b). (d) Black line: interval ratio corresponding to the frequency difference between the partials highlighted by the scatter plot in (c). (e) Difference between the lowest partial and the autocorrelation peak after the first zero-crossing.}
\label{fig:SilverBass2}
\end{framedfigure}

\clearpage

\subsubsection{Kick drum}\label{ref:silverkickdrum}

This section concerns the unfiltered kick drum sample in `Silver', present from 0'07 to 0'41 and from 0'58 to the end. The kick drum occurs simultaneously with bass type 1. It contains one harmonic complex whose $f_0$ is near D2. It is tuned to the bass pitch derived from temporal modelling of pitch perception, rather than to the value derived from spectral modelling (see Section~\ref{ref:silverbasstype1unweighted}, Figure~\ref{fig:SilverBass}).

\begin{framedfigure}[htbp]
  \centering
  \includegraphics[width=1\columnwidth]{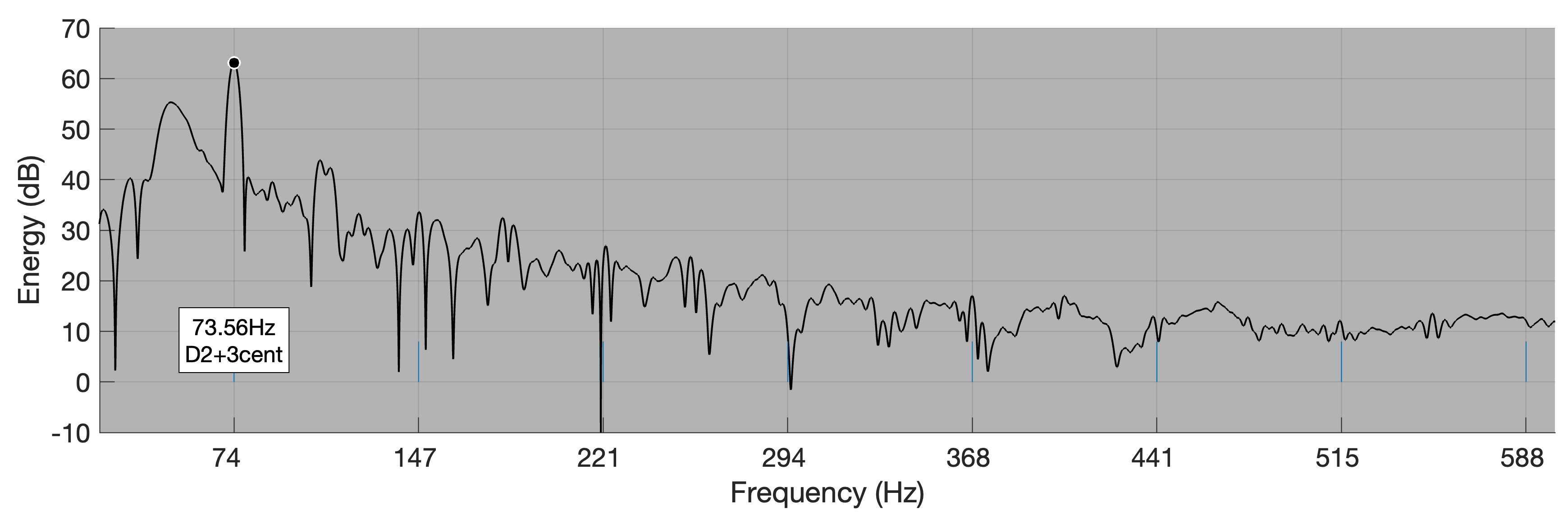}
  \caption{`Silver', kick sample, unweighted, FT. The x-axis grid is set on the multiples of the strongest partial.}
\label{fig:SilverKick}
\end{framedfigure}

\subsection{`Sweet Money', bass}\label{ref:sweetmoneybass}

The bass for `Sweet Money' was generated using Omnisphere's `808 Woofer Warfare' patch, in combination with Omnisphere's Waveshaper and Unison effects. It was then further processed with Decapitator. The glides were produced using Omnisphere's Glide feature in monophonic mode\footnote{\url{https://support.spectrasonics.net/manual/Omnisphere2/25/en/topic/main-page10}}. The producers played a sequence of two notes, and the Glide generated a portamento. The Glide rate was set so that the portamento could reach two octaves during the allotted time span. The glides at the end of the segment form a second voice. Their start and end pitches are random.

\subsubsection{Unweighted}

Figure~\ref{fig:SweetMoneyBass} (a) and (b) show only an extract of a bass pattern. The entire pattern consists of three repetitions of the first subpattern (the material before 0.6~s), followed by the second subpattern (the material after 1.6~s). Figure~\ref{fig:SweetMoneyBass} (a) shows that the bass pattern alternates between, and then combines, relatively stable pitch values and glides. Figure~\ref{fig:SweetMoneyBass} (b) shows that the pitch in relatively stable sections evolves within an ambitus of two to six semitones, and that the glides reach almost two octaves. Figure~\ref{fig:SweetMoneyBass} (c) shows that some sections of the bass track involve sixteen regularly spaced inharmonic partials. Figure~\ref{fig:SweetMoneyBass} (d) shows that the distance between partials is irregular. Figure~\ref{fig:SweetMoneyBass} (e) shows that some other sections involve three regularly spaced inharmonic partials. The bass sound is not static. The relatively stable section of the first pattern (0'00--0'06) converges to C1 after the initial 808-like bend. The note corresponds to the final of the song's mode. 

\clearpage

\begin{framedfigure}[h!]
  \centering
  \includegraphics[width=1\columnwidth]{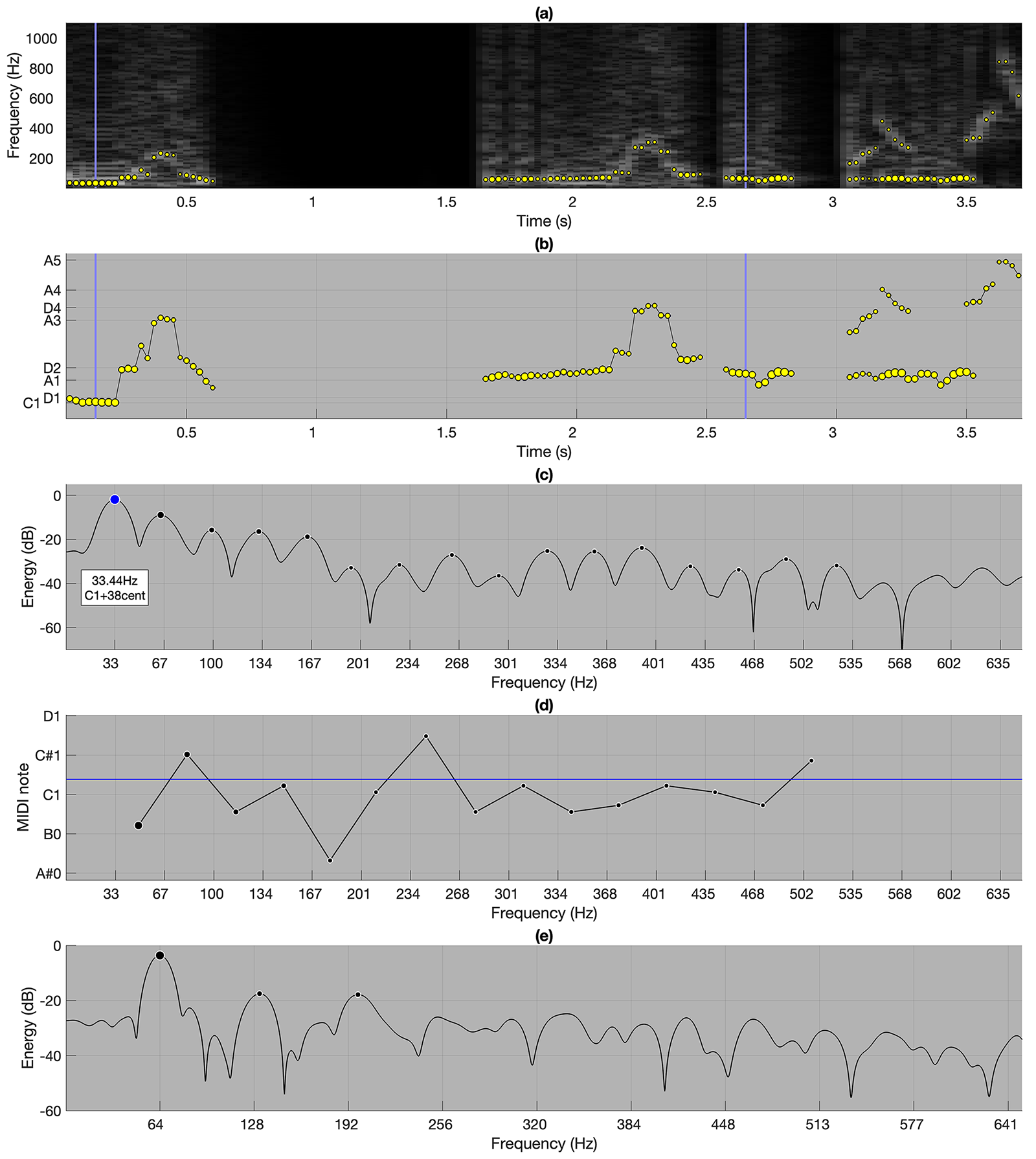}
  \caption{`Sweet Money', bass track, unweighted. (a) STFT. The yellow scatter plot tracks salient partials. (b) Pitch evolution of the peaks in (a). (c) FT at frame indicated by leftmost blue line in (a) and (b). The x-axis grid is set on the harmonic positions for the lowest partial. (d) Frequency differences between each partial and the lowest partial. (e) FT at frame indicated by rightmost blue line in (a) and (b).}
\label{fig:SweetMoneyBass}
\end{framedfigure}

\clearpage

\subsubsection{Weighted}

Figure~\ref{fig:SweetMoneyBassWeighted} (a) suggests that the high frequencies of most bass notes in `Sweet Money' are noisy. Figure~\ref{fig:SweetMoneyBassWeighted} (b) shows that during the first glide, the sound only features two clear upper partials, harmonics 3 and 5. Figure~\ref{fig:SweetMoneyBassWeighted} (c) shows that the end glides feature the same upper partials, only more clearly. The comparison between Figure~\ref{fig:SweetMoneyBass} (c) and (e) and Figure~\ref{fig:SweetMoneyBassWeighted} (b) and (c) confirms that the organisation of partials in this bass track is constantly evolving.

\vspace{1cm}

\begin{framedfigure}[h!]
  \centering
  \includegraphics[width=1\columnwidth]{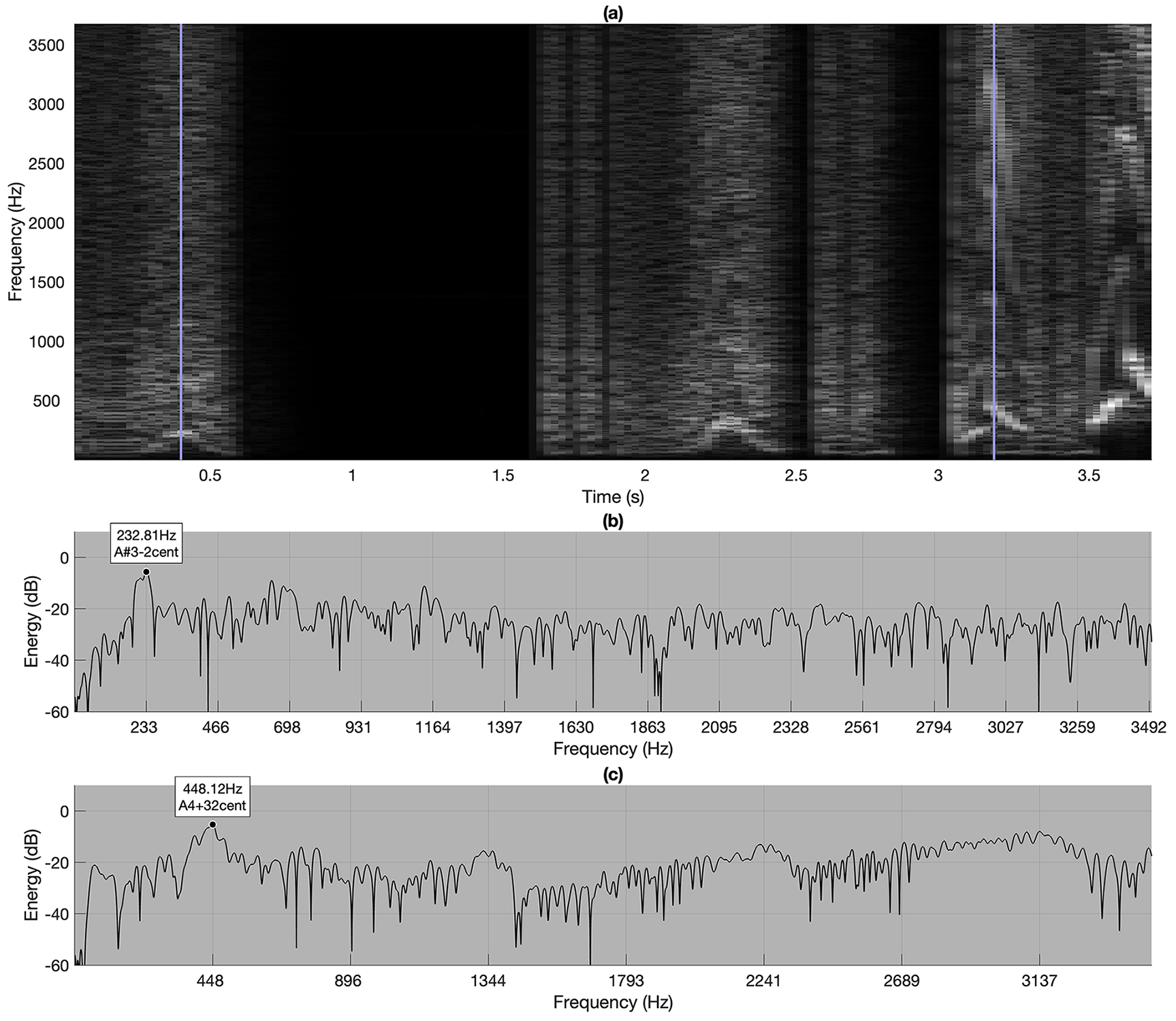}
  \caption{`Sweet Money', bass track, weighted. (a) STFT. (b) and (c), FT for the frames indicated by the blue lines. The x-axis grid is set on the harmonic positions for the lowest partial.}
\label{fig:SweetMoneyBassWeighted}
\end{framedfigure}

\clearpage

\subsection{`Whomp', gimmick}\label{ref:whompgimmick}

`Whomp' includes a `gimmick' track, a simple melody involving a sample from an unspecified keyboard. From 0'48 to 1'05, 1'53 to 2'09, and 2'35 to 2'57, the gimmick is processed with Sony CSL Resonance EQ. The process boosts the partials. It is possible to hear two patterns simultaneously, `G$\flat$2 F2 F2 F2 F0' and `B$\flat$4 A4 A4 A4 A4'. The interval between the two patterns is two octaves plus a major third. The interval is consistent with a remark from the producers, according to which the sample was chosen on the basis that harmonic 5 was particularly audible. Figure~\ref{fig:WhompGimmick}~(b) shows that the pitch value heard as F2 comes from the fundamental. Figure~\ref{fig:WhompGimmick}~(d) shows that the pitch value heard as A4 comes from harmonic 5.

\vspace{.2cm}

\begin{framedfigure}[h!]
  \centering
  \includegraphics[width=1\columnwidth]{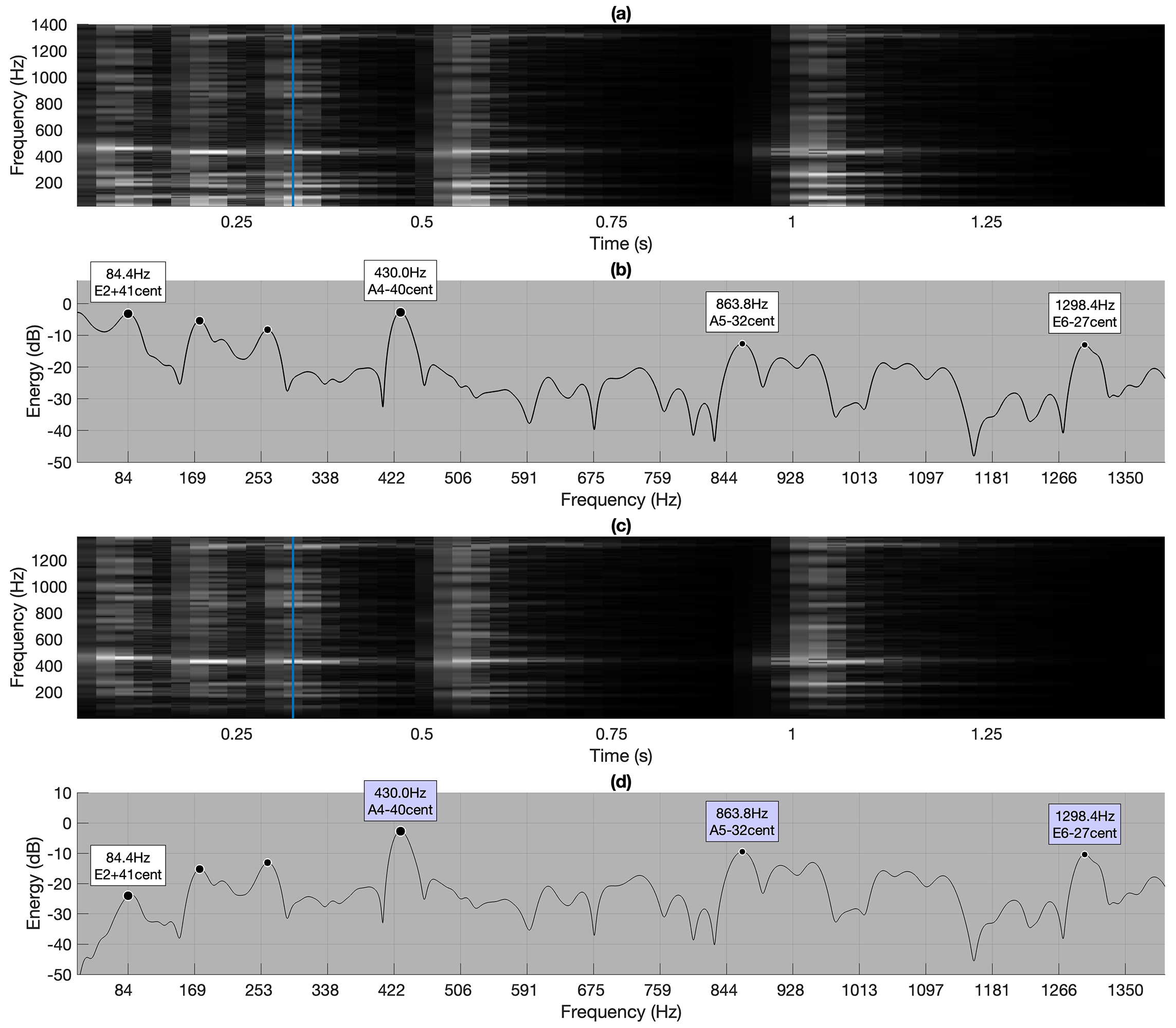}
  \caption{`Whomp', gimmick track. (a) STFT, unweighted. (b) FT at frame indicated by blue line in (a). (c) STFT, weighted. (d) FT at frame indicated by blue line in (c). In (b) and (d), the x-axis grid is set on the harmonic positions of the lowest partial. The blue frames highlight the upper audible pitch.}
\label{fig:WhompGimmick}
\end{framedfigure}

\clearpage
\subsection{`Whomp', kick and snare drums}\label{ref:whompkicksnare}

\vspace{.2cm}

Figure~\ref{fig:WhompKick} (a) shows that although it is possible to identify partials in the kick drum sample from `Whomp', they are not as salient as in the non-percussive tracks. Figure~\ref{fig:WhompKick} (c) suggests that the first three partials are harmonics 1, 3, and 5 of a harmonic complex tone. Figure~\ref{fig:WhompKick} (b) shows that the first partial's frequency falls by approximately an octave and a fifth over the course of the sample. 

\vspace{.4cm}

\begin{framedfigure}[h!]
  \centering
  \includegraphics[width=1\columnwidth]{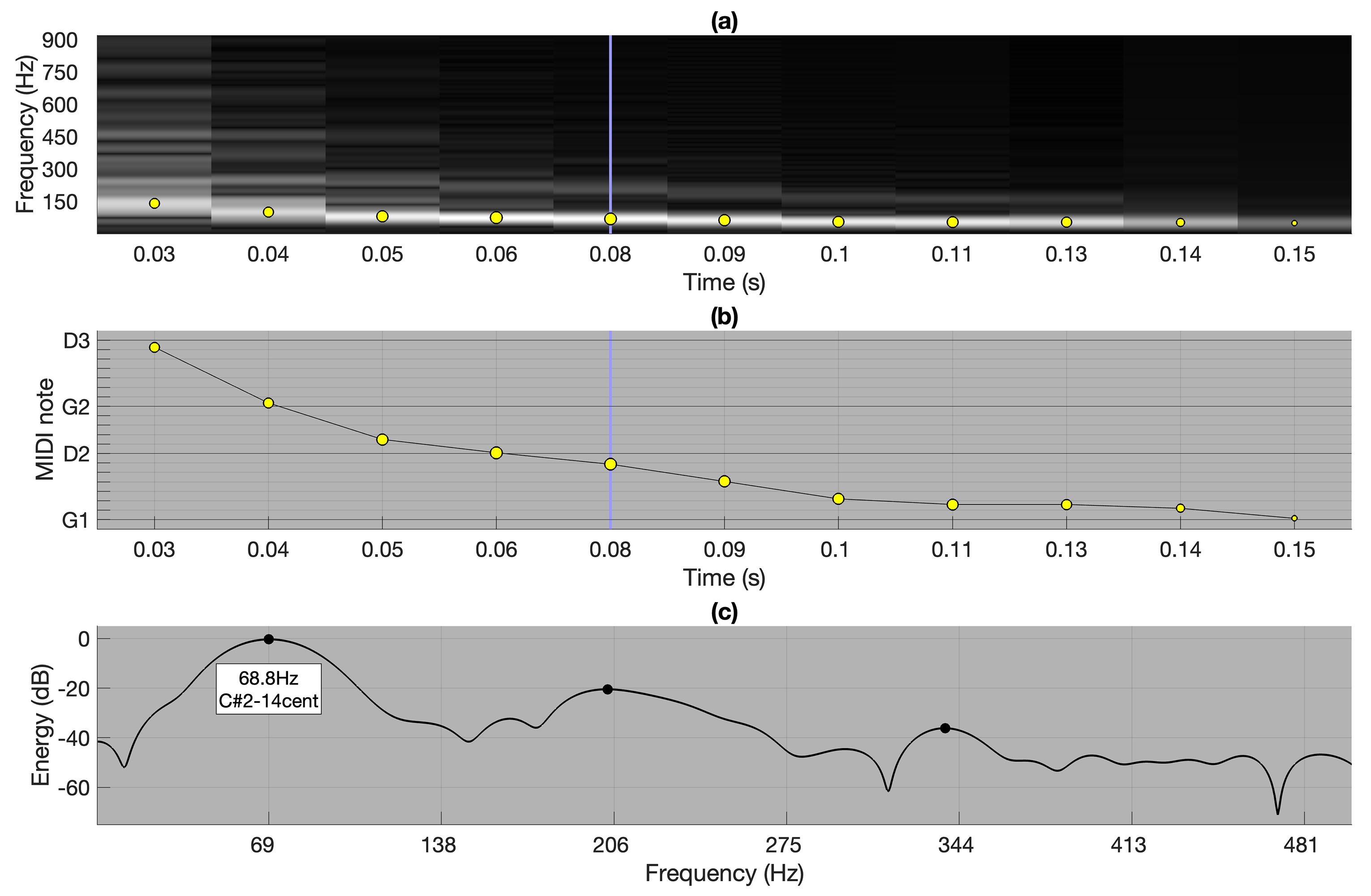}
  \caption{`Whomp', kick drum, unweighted. (a) STFT. The yellow scatter plot follows the lowest partial. (b) Pitch corresponding to the lowest partial in (a). (c) FT at frame indicated by blue lines from (a) and (b). The x-axis grid is set on the harmonic positions for the lowest partial.} 
\label{fig:WhompKick}
\end{framedfigure}

\vspace{.4cm}

The partial frequency values for the snare drum are stable over time. It is therefore possible to compute the FT and autocorrelation over the entire sample. Joint examination of Figure~\ref{fig:WhompSnare} (a) and (b) suggests that it may be possible to identify pitch values around F1/G1 and B3 in the snare sample.

\newpage

\begin{framedfigure}[h!]
  \centering
  \includegraphics[width=.8\columnwidth]{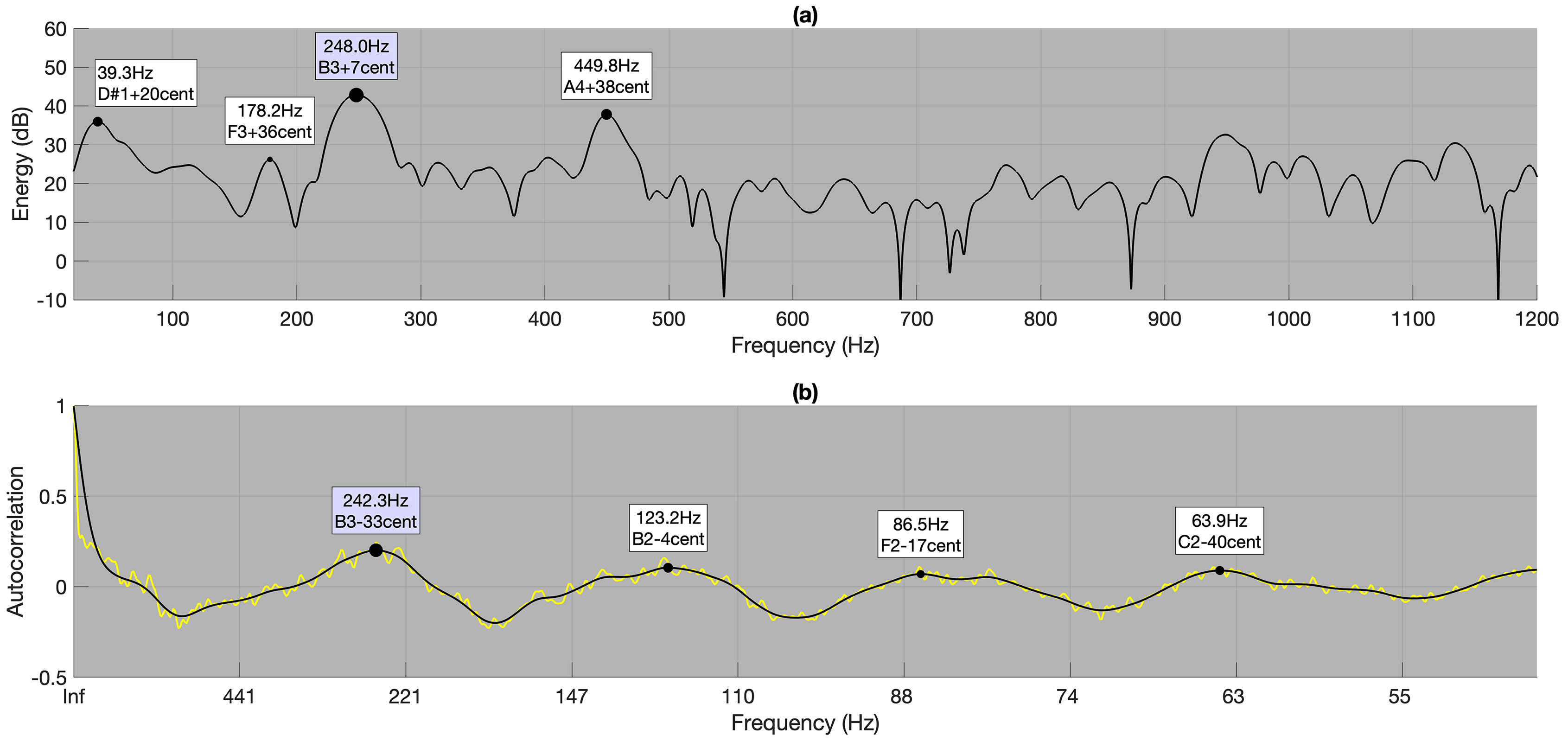}
  \caption{`Whomp', snare drum, unweighted. (a) FT. (b) Autocorrelation. Yellow line, original values. Black line, smoothed values.} 
\label{fig:WhompSnare}
\end{framedfigure}

\subsection{`Yada Yada', flute}\label{ref:yadayadaflute}

The `Yada Yada' flute track was edited from a transverse flute recording, retaining only the portions corresponding to the note `F' (Figure~\ref{fig:YadaYadaFlutes} (a)). The producers specifically kept the segments of the flute recording near the attack of each note so that the pitch values are less stable (Figure~\ref{fig:YadaYadaFlutes} (b)). The flute part around F is interrupted by glides (Figure~\ref{fig:YadaYadaFlutes} (c)).

\begin{framedfigure}[h!]
  \centering
  \includegraphics[width=.8\columnwidth]{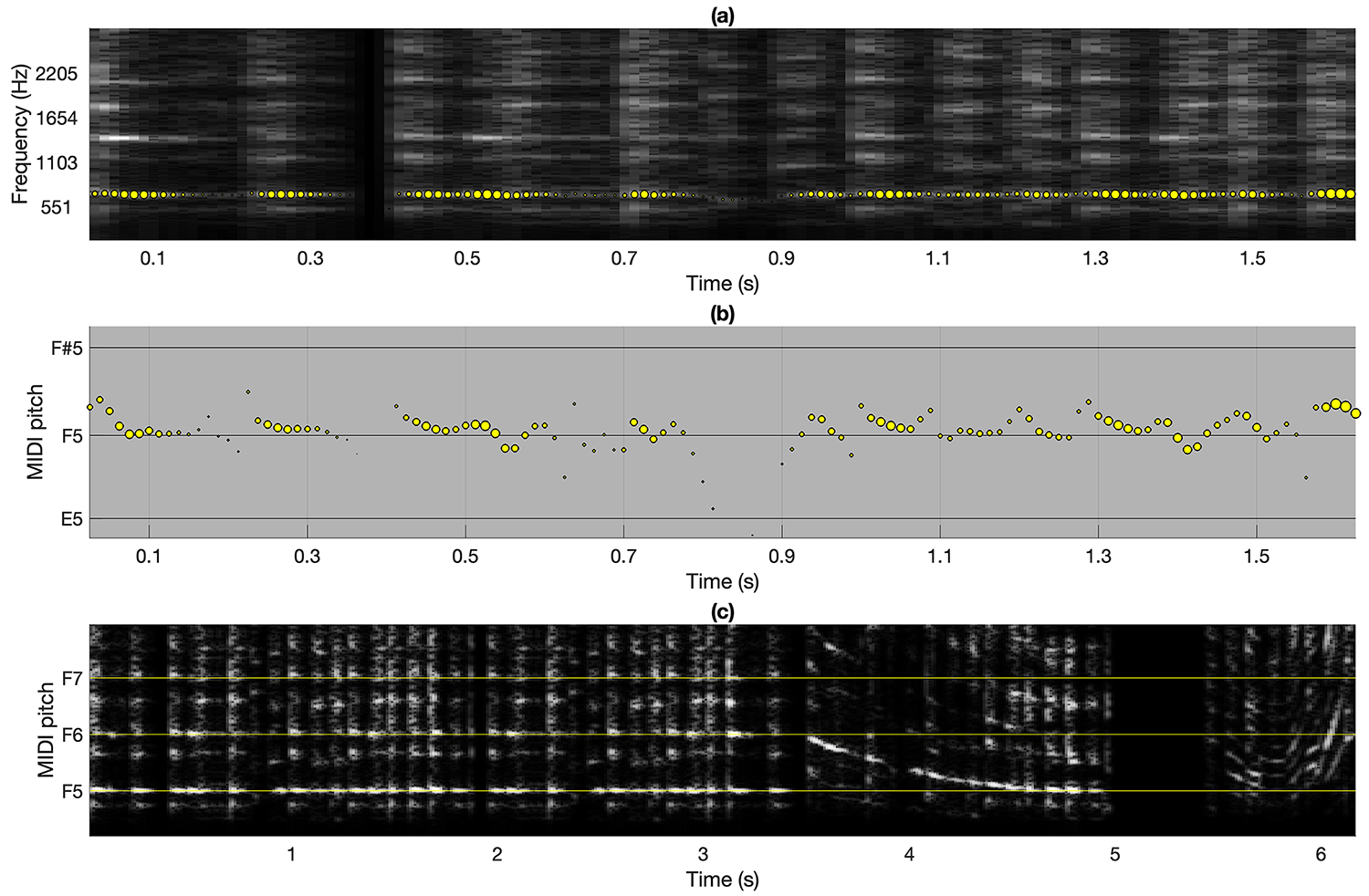}
  \caption{(a) `Yada Yada', flute track, 14 to 16~s, unweighted, STFT. (b) Pitch of the lower partial. (c) Flute track, 14 to 20~s, unweighted, STFT. Contrast is accentuated for better readability.} 
\label{fig:YadaYadaFlutes}
\end{framedfigure}

\clearpage

\section{Pitch-related musical parameters in Primaal's music}\label{sec:parameters}

\vspace{.5cm}

\begin{figure}[h!]
  \centering
  \includegraphics[width=\columnwidth]{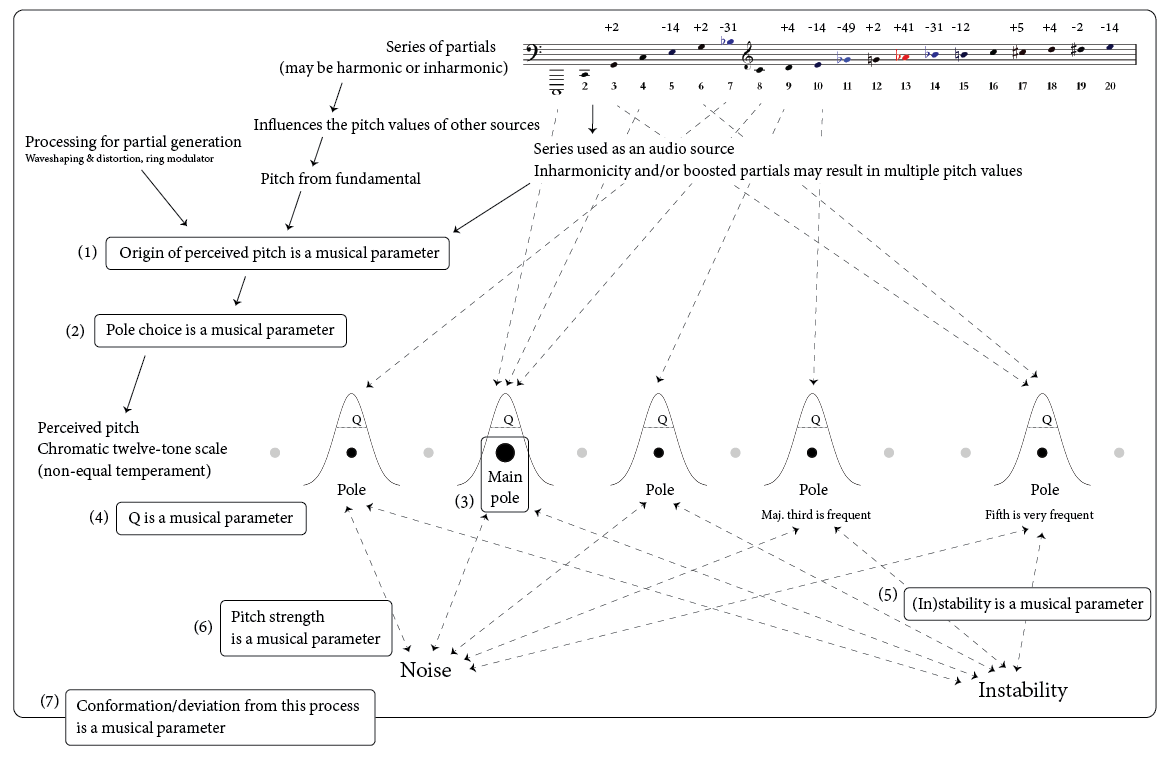}
  \caption{Summary of pitch-related processes in the production of Primaal's music. $Q$ denotes the distribution's width. Musical parameters are framed and numbered.}
\label{fig:S_summary}
\end{figure}

This section synthesises the pitch-related processes observed in Primaal's music, drawing on the signal analyses presented in Section~\ref{ref:signalanalysis}. Figure~\ref{fig:S_summary} provides an overview. The discussion identifies \textit{musical parameters} in \citepos{mcadams1999perspectives} sense, many of which relate to \citepos{schneider2009perception} dimensions for the sounds of musical instruments. Table~\ref{tab:parameters} lists the seven identified parameters, numbered as in Figure~\ref{fig:S_summary}; the remainder of this section discusses each of them in turn.

\vspace{1cm}

\clearpage

\begin{table}[htbp]
\caption{Pitch-related musical parameters in Primaal's music. Numbering follows Figure~\ref{fig:S_summary}.}
\label{tab:parameters}
\setlength{\tabcolsep}{5pt}
{\small
\begin{tabularx}{\columnwidth}{@{}lX@{}}
\toprule
Parameter & Description \\
\midrule
(1) Origin of perceived pitch & Basis of a pitch-evoking sound: fundamental of a harmonic complex tone, boosted partial, or generated partial \\
(2) `Pole' choice & Realisation of each scale step as a probability distribution centred on a most probable `pole' frequency, which may diverge from equal temperament \\
(3) Main pole & Prominent scale degree, comparable to the final in early Western modes (Sections~\ref{sec:modalitybackground} and~\ref{sec:modality}) \\
(4) $Q$ & Width of the distribution of pitches around the scale degrees (Section~\ref{sec:continousfreqs}) \\
(5) (In)stability & Frequency stability or instability and continuous frequency trajectories shaping the distribution (Section~\ref{sec:continousfreqs}) \\
(6) Pitch strength & Clarity with which a pitch is perceived \citep{zwicker1990pitchstrength,yost2009pitch} \\
(7) Conformation/deviation & Conformation to or deviation from the above processes, used to articulate form and structure (Sections~\ref{sec:largescalestructure} and~\ref{sec:smallscalestructure}) \\
\bottomrule
\end{tabularx}}
\end{table}

\vspace{.25cm}

The origin of perceived pitch---parameter~(1) in Table~\ref{tab:parameters}---refers to three techniques by which the producers create pitch-evoking sounds. First, they may use harmonic complex tones whose perceived pitches correspond to their fundamental frequencies, as in the most common way of generating perceived pitches in Western classical music. Second, they may boost specific partials in complex tones so that these partials are heard as separate `tones' with independent pitches (Section~\ref{sec:loudharmonics}); in particular, the producers commonly boost the third and fifth harmonics, and bass sounds often contain such partials. Third, they may generate multiple perceived pitches using partials produced through distortion, waveshaping (Appendix~\ref{ref:distortion}), or ring modulation (Appendix~\ref{ref:ringmodulation}). The choice of the origin of a perceived pitch (i.e., fundamental, boosted partial, or generated partial) serves as a musical parameter. Each origin has its own affordances and `idiom', much as a musical instrument does, in the sense of the techniques and patterns that feel most natural and are easiest to perform with that instrument \citep[p.~103]{huron2009characterizing}.

Choosing a particular method of generating pitches (origin of perceived pitch) may constrain the producer to a finite set of pitches---in effect, a `scale'---from which pitches can then be selected. For instance, if pitches are generated by boosting selected partials, then the choice of complex tone used as a source also fixes the set of available partials from which perceived pitches may derive. This can lead to passages being based on \emph{resultant scales}---determined by the tone itself---rather than on \emph{active scales} from which pitches are deliberately chosen. As noted in \citet{deruty2025multiple}, this distinction is reminiscent of the shift from resultant harmony to active harmony described by \citet[p.~82]{coeurdevey1998histoire} in connection with the transition from modal polyphonic writing to harmonic and tonal writing.

Perceived pitches are organised into modal scales (see Section~\ref{sec:modality}). Each scale step is realised according to a probability distribution centred on a most probable `pole' frequency---parameter~(2)---and the pole frequencies may diverge from equal temperament. Although some of the tools used by the Primaal producers, such as the Spectrasonics Omnisphere synthesiser, include temperament settings,\footnote{\url{https://support.spectrasonics.net/manual/Omnisphere/main/controls/page02.html/}} the producers choose not to use them. One scale degree---the main pole, parameter~(3)---is always prominent and can be compared to the final in early Western modes (Sections~\ref{sec:modalitybackground} and~\ref{sec:modality}).

\newpage

Parameter~(4), denoted by $Q$, is the width of the pitch distribution around each scale degree (Section~\ref{sec:continousfreqs}). The producers set this width and state that much of their music is about `hovering around a mode'.\footnote{Literal statements from the producers are between single quotes. The producers ask that we do not disclose the original transcripts of the interviews.} Frequency stability or instability and continuous frequency trajectories---parameter~(5), (un)stability---shape this distribution (Section~\ref{sec:continousfreqs}).

Pitch strength---parameter~(6)---refers to the clarity with which a pitch is perceived \citep{zwicker1990pitchstrength,yost2009pitch}. In particular, drums in Primaal's music may have variable pitch strength (see Sections~\ref{sec:pitchindrums} and~\ref{sec:tuneddrums}). Finally, conformation to or deviation from the above processes---parameter~(7), conformation/deviation---is considered a musical parameter: shifts between conformation and deviation can articulate a song's form and structure (Sections~\ref{sec:largescalestructure} and~\ref{sec:smallscalestructure}).

\vspace{1cm}

\section{Pitch uncertainty}\label{sec:pitchuncertainty}

This section outlines the musical features through which pitch uncertainty is produced in Primaal's music. This uncertainty builds on the processes described in the previous section, but does not derive from them mechanically---it arises from particular settings of the corresponding musical parameters. Most conclusions are grounded in signal analyses that substantiate uncertainty first identified through listening and subsequently confirmed by the producers. The present section identifies these features and the roles they play. Several are then placed in historical perspective in Section~\ref{sec:historical-context} and documented in detail in Section~\ref{sec:analyses-of-works}, on the basis of the analyses compiled in Section~\ref{ref:signalanalysis}.

\subsection{Inharmonicity as a source of non-standard tunings}\label{sec:inharmonicity}

Inharmonicity in Primaal's music may contribute to pitch uncertainty by yielding pitches organised according to non-standard tunings. Inharmonicity is defined as `the deviation of frequencies from an exact harmonic series' \citep{campbell2001inharmonicity}. A distinction can be drawn between noise-like inharmonic sounds and `coherent' inharmonic signals, which, like harmonic signals, sound stable \citep{deboer1956pitch}. We focus on coherent tones whose partials depart from exact harmonic positions. In Primaal's music, tones are almost always inharmonic; however, it is generally possible to associate each partial with an approximate harmonic position, making the tones \emph{quasi-harmonic}. The partials' deviations from these harmonic positions do not resemble those produced by piano strings, as studied by \citet{young1952inharmonicity}, but instead exhibit a wider variety of configurations.

\citet{jarvelainen2000effect} and \citet[p.~41]{bregman1996demonstrations} note that a single partial may become audible as a distinct pitch separate from the fundamental as its deviation from its expected harmonic position increases. In the studied extracts, we generally did not observe this phenomenon---except for one example discussed in Section~\ref{sec:lowinharmonic}. Instead, the occurrence of multiple perceived pitches from one tone appears to be driven by the high energy of specific partials (see Section~\ref{sec:loudharmonics}), rather than by their frequency positions. As a result, the primary role of inharmonicity in shaping pitch in Primaal's music appears to lie in the fine control of the \emph{resultant pitch}---see Section~\ref{sec:parameters}, parameter~(1). The producers use this control to generate pitches with non-standard tunings.

\newpage

\subsection{Very low fundamental frequencies: octave and pitch-class ambiguity}\label{sec:lowinharmonic}

Tones with very low fundamental frequencies may contribute to (a) octave ambiguity and, in one observed case with stronger inharmonicity, to (b) ambiguity in the perceived pitch classes. With respect to (a), the $f_0$ values in Primaal's bass parts often lie between 35 and 45~Hz (Section~\ref{ref:signalanalysis}). Comments from participants in the listening test reported by \citet{deruty2025multiple} indicate that octave ambiguities in such a low-frequency range are common; furthermore, very low $f_0$ values may be attenuated or not reproduced \citep{newell2001yamaha,deruty2024storch}, potentially reinforcing this phenomenon. With respect to (b), the exception discussed in Section~\ref{sec:inharmonicity} occurs in the bass part of `Silver' (Section~\ref{ref:silverbass}): its upper partials do not form exact harmonic relations with the first partial ($f_0 = 39$~Hz). On some playback systems, the pitch corresponding to this partial can be clearly heard alongside the pitch derived from temporal modelling of the upper partials; on other systems, the first partial may be greatly attenuated.

\subsection{One tone, several pitches: the amplification of upper partials}\label{sec:loudharmonics}

Quasi-harmonic tones that can give rise to the perception of two or more pitches have been linked to differences in pitch judgements across listeners, uncertainty about which pitches are being heard, and pitch--timbre ambiguity \citep{deruty2025multiple}. As documented in \citet{deruty2025multiple,deruty2025vitalictemperament,deruty2022melatonin}, the Primaal producers consistently amplify upper partials, making subsets of these partials audible as carriers of perceived pitches beyond the $f_0$. Boosting upper partials is particularly well suited to bass parts, which producers often aim to shape into a rich `harmonic entity' with `presence' and `personality'. This approach is reminiscent of the use of multiband compression to optimise bass tracks across frequency bands \citep{bloomfield2014split}, a common practice in genres such as electro house, dubstep, and drum and bass \citep{productionmusic2015split}. Primaal, however, goes further by treating the bass tone's upper register as a source of pitches: rather than adding instruments, the producers extract additional pitches from the bass itself, so that sparse arrangements can foreground the bass's harmonic richness. Particularly relevant examples of this broader treatment of upper partials include `\textexclamdown{}Fire!' (Section~\ref{ref:firebass}), `Cardinal' (Section~\ref{ref:cardinalsynthbreak}), and `Elevate' (Section~\ref{ref:elevatekey}).

Tones with prominent upper partials can be produced directly---for instance, using patches such as `808 Woofer Warfare' (Appendix~\ref{ref:bassgeneration}). Upper partials can also be amplified through equalisation or created using various processes; newly generated partials may then reinforce or replace the original partials (Appendix~\ref{ref:interactionpartials}). Table~\ref{tab:overtone_processes} summarises the processes used across the studied songs.

\begin{table}[htbp]
\caption{Processes used to generate, amplify, or otherwise modify upper partials, by song and track. The processes themselves are documented in Appendix~\ref{ref:instruments}.}
\label{tab:overtone_processes}
\setlength{\tabcolsep}{5pt}
{\small
\begin{tabularx}{\columnwidth}{@{}llXl@{}}
\toprule
Song & Track & Process(es) & Section \\
\midrule
`Boom'               & Bass              & Distortion, Equalisation        & \ref{ref:boombass}    \\
`Cardinal'           & Synth break       & Equalisation                                  & \ref{ref:cardinalsynthbreak}    \\
`Danger'             & Lower bass        & No effects                                & \ref{ref:dangerlowbass}  \\
`Danger'             & Higher bass       & Distortion                             & \ref{ref:dangerhighbass}  \\
`Elevate'            & Synth bass        & Waveshaping, Equalisation             & \ref{ref:elevatebass}    \\
`Elevate'            & Rising keyboard            & Waveshaping                                & \ref{ref:elevatekey}    \\
`\textexclamdown{}Fire!' & Bass          & Waveshaping, Frequency modulation                           & \ref{ref:firebass}   \\
`R U Ready'          & Synth bass        & Distortion, Waveshaping, Ring modulation, Frequency modulation, Reverberation           & \ref{ref:areureadysynthbass} \\
`Silver'             & Bass type 1       & Waveshaping                                & \ref{ref:silverbass}   \\
`Silver'             & Bass type 2       & Ring modulation                                & \ref{ref:silverbass}   \\
`Sweet Money'        & Bass              & Distortion, Waveshaping                        & \ref{ref:sweetmoneybass}   \\
`Whomp'              & Gimmick       & Equalisation                               & \ref{ref:whompgimmick}   \\
\bottomrule
\end{tabularx}}
\end{table}

In addition to creating pitch ambiguity, these methods blur the distinction between pitch and timbre, producing a \textit{pitch--timbre continuum}. As noted by \citet[pp.~1708--1709]{yost2009pitch}, a partial's amplitude---traditionally considered a timbral attribute \citep{peeters2011timbre}---can influence perceived pitch when it is sufficiently high. A similar continuum can be observed in music from other artists \citep{deruty2025multiple,deruty2025vitalictemperament}. To reinforce pitch uncertainty further, the producers frequently use odd-harmonic tones, as exemplified by the bass track in `Danger' (Section~\ref{ref:dangerlowbass}). Such tones promote pitch ambiguity, particularly when $f_0$ is low: \citet[pp.~1705--1706]{yost2009pitch} notes that when a harmonic complex tone lacks its lowest partials and the lowest remaining partial is an odd harmonic, temporal analysis may yield two pitches. A similar situation can occur when a tone contains odd partials and has a low $f_0$ (e.g., 30~Hz), since low-frequency partials may be poorly transmitted or not transmitted at all by playback systems (Section~\ref{sec:lowinharmonic}), making the lowest audible partial an odd upper partial.

\subsection{Continuous frequency trajectories: the TR-808 as reference}\label{sec:continuousfreqtraj}

The producers make extensive use of continuous frequency trajectories. They state that this process largely derives from the Roland TR-808 bass drum, either through TR-808-type bass-drum samples or by treating such samples as a model. The TR-808's $f_0$ drops by about a semitone before stabilising \citep{deruty2024storch}, as also observed in the bass tracks of `Elevate' (Section~\ref{ref:elevatebass}) and `Silver' (Section~\ref{ref:silverbass}). In longer bass notes, the pitch can fall through larger intervals, as in `Boom' (Section~\ref{ref:boombass}), where it drops by roughly half an octave. The TR-808 bass drum also serves as a model for non-percussive tracks, with transient-like pitch drops applied to sustained sounds. Examples include the `square synth' stem in `R U Ready' (0'26--0'40), the `sampled vocals' in `Elevate' (2'22--2'36), the `vocals' in `Silver' (0'08--0'40), and the `synths' in `Yada Yada' (0'01--0'14). Such continuous frequency changes can be achieved using bends, glides, or screws (Appendix~\ref{ref:bends}).

The continuous frequency changes are used to confuse pitch perception. The producers often set random pitch bend limits to avoid harmonic intervals like the octave or fifth, which would provide unwanted familiar references. When possible, the producers prefer pitch contours that never fully stabilise to enhance pitch ambiguity. These continuous changes instantiate (un)stability---Section~\ref{sec:parameters}, parameter~(5)---and produce pitch distributions around scale degrees whose width, $Q$, is parameter~(4). The producers choose $Q$ on the basis of the amount of confusion they want to convey.

\subsection{Drums as carriers of weak pitch}\label{sec:pitchindrums}

Musical signals are often described in terms of `percussive' and `harmonic' components: percussive sounds are typically short and noisy, whereas harmonic sounds are longer and have energy concentrated around a pitch \citep{rump2010autoregressive}. This distinction underlies early source-separation methods \citep{fitzgerald2010harmonic,yoo2010nonnegative}. However, drums can also contain pitched content \citep{richardson2010acoustic}. While their pitch strength, in the sense of \citet{zwicker1990pitchstrength}, is usually lower than that of harmonic instruments \citep{deruty2024pitchstrength}, they can be tuned \citep{toulson2009perception,drummagazine2010}. This practice is not universal; some producers disagree about whether kick drums should carry pitch at all \citep{solstate2021kicks}.

The Primaal producers often tune the kick drum, especially for longer samples, but find the snare harder to tune. They may tune other percussion sounds or leave them out of tune to act as `punctuation', as in `Cardinal', where metallic percussion contrasts with the bass. The deliberate use of drums as pitch carriers contributes to pitch uncertainty, either because their pitch strength is inherently low or because, as noted by the Primaal producers, their perceived pitch may shift in the presence of other tracks. 

The producers emphasise caution when tuning drums, noting that timbre and register should often be prioritised over pitch---a practice also documented in the work of hip-hop producer Scott Storch \citep{deruty2024storch}. When tuning is performed, the closest `pole'---Section~\ref{sec:parameters}, parameters~(2) and~(3)---is usually chosen. If transposition alters timbre too much, however, they leave the sample out of tune---echoing Luigi Russolo's practice \citep{frisius2010search}---or select another. A shift of a semitone or whole tone is generally acceptable, but not a third. For example, the producers transposed the snare drum in `Yada Yada' by one semitone to match the closest `pole'. The producers note that out-of-tune drums may not always `clash' with a song's pitches, as the perceived pitch of drums can shift or even cease to be identifiable when they are played alongside tracks with higher pitch strength.

\vspace{1cm}

\section{Primaal's treatment of pitch in a historical context}
\label{sec:historical-context}

This section relates Primaal's treatment of pitch to three previously documented aspects of music and sound: acoustic beating and combination tones (Section~\ref{sec:acousticbeatcombtones}), the long debate over whether individual harmonics can be `heard out' (Section~\ref{sec:quasiharmtonesmultiplepitches}), and the distinction between tonality and modality (Section~\ref{sec:modalitybackground}). The first two aspects relate to pitch ambiguity; beating and combination tones arise from concomitant frequencies distributed around poles---see Section~\ref{sec:parameters}, parameter~(4)---while quasi-harmonic tones evoking multiple pitches result in uncertain pitch perception.

\subsection{Acoustic beating and combination tones}\label{sec:acousticbeatcombtones}

Acoustic beats and combination tones are two phenomena that derive from the superposition of sine waves of different frequencies. \emph{Acoustical interference beats} are periodic amplitude fluctuations caused by two pure tones with slightly different frequencies. \emph{Combination tones}, or Tartini tones, are perceived `when two musical tones of considerable intensity are sounded together[: then] the ear also perceives a third ``combination'' tone, lower in pitch than either of the generating tones' \citep[p.~12]{turner1977ohm}. Although the two phenomena are `different in kind' \citep[p.~255]{christensen2006cambridge}, they derive from the same physical process: `as the difference in frequencies [increases], then the combination tone begins to be heard. Therefore the combination tone is physically identical to the phenomenon of beats; it is a beat frequency become so rapid as to be heard as a tone in its own right' \citep[p.~12]{turner1977ohm}. \citet[pp.~254--256]{christensen2006cambridge} provides a comprehensive historical perspective on both phenomena.

\newpage

In Western classical music, minimising acoustical beating was long considered important. Impure intervals, which generate acoustic beats, have historically been deemed undesirable; such a preference has been the basis for the elaboration of temperaments, `[t]unings of the scale in which some or all of the concords are made slightly impure in order that few or none will be left distastefully so' \citep{lindley2001temperaments}. Just intonation---`the consistent use of harmonic intervals tuned so pure that they do not beat', made possible `[w]hen pitch can be intoned with a modicum of flexibility' \citep{lindley2001justintonation}---was used so that intervals were purely tuned and did not produce beats.

\begin{figure}[h!]
  \centering
  \includegraphics[width=\columnwidth]{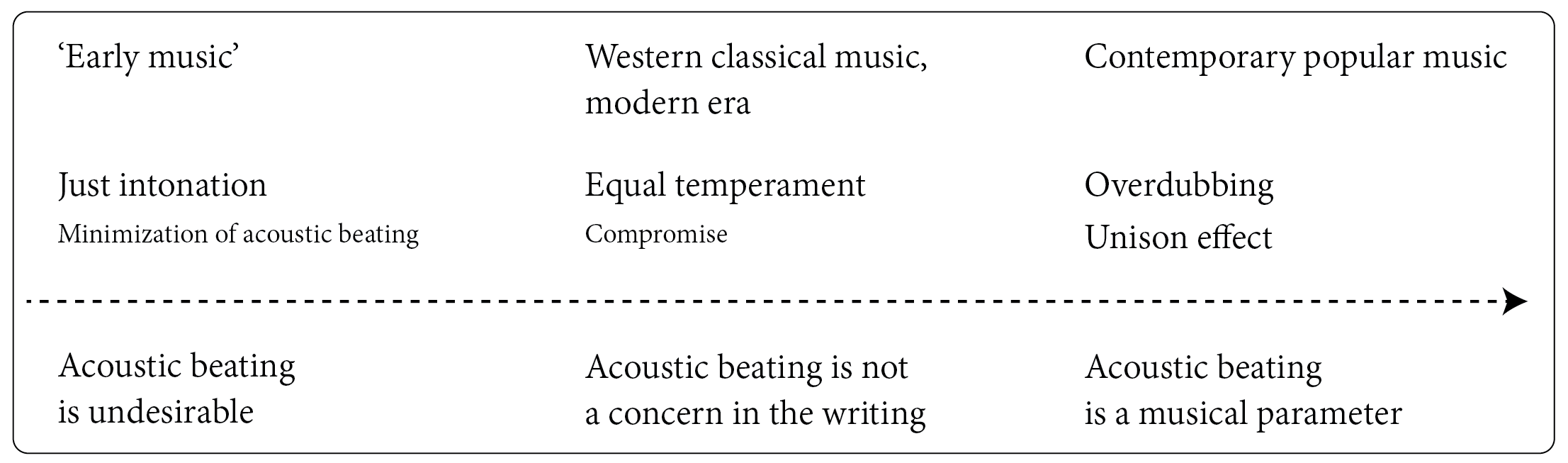}
  \caption{Development over history in attitudes towards acoustic beating in music.}
\label{fig:S_beating}
\end{figure}

The same concern motivates the tuning of instruments through the minimisation of acoustic beating: organ \citep{baggaley2023mechanisms}, harpsichord \citep[p.~255]{christensen2006cambridge}, piano \citep[p.~87]{white1917modern}, and guitar \citep{klickstein1993tuning}. Later, the adoption of equal temperament---endorsed by \citet{rameau1737generation}---prioritised the freedom to modulate in every key \citep[p.~vii]{barbour2004tuning}, and efforts towards the minimisation of acoustic beating became a concern of the past (except during the tuning of instruments). In modern popular music, techniques such as unison \citep{dailyanalog2024} and overdubbing \citep[p.~31]{huber2013modern}---both heavily used by the Primaal producers---intentionally generate beats by combining tracks with slightly different pitches; the `unison' audio effect, described in Appendix~\ref{ref:unison}, is precisely designed to \emph{produce} acoustical interference beats. In a less overtly commercial context, slow beating phenomena arising from the superposition of closely spaced frequencies are also central to certain sustained-sound practices---e.g., in the work of \'Eliane Radigue \citep{gray2024radigue}---and comparable psychoacoustic concerns can likewise be found in contemporary electronic work such as that of \citet{sheburne2023rrose}. Figure~\ref{fig:S_beating} summarises this evolution.

\vspace{.5cm}

\begin{figure}[h!]
  \centering
  \includegraphics[width=\columnwidth]{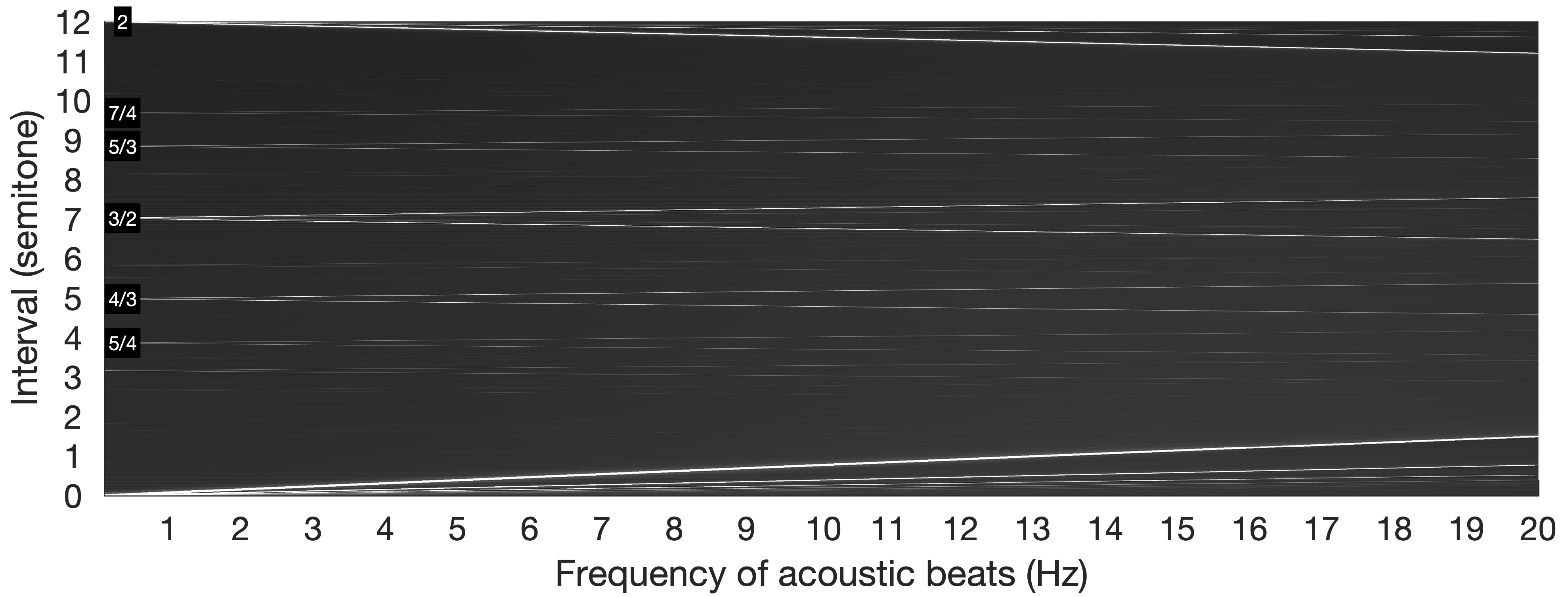}
  \caption{Frequency beating from the superposition of two harmonic tones. $Y$-axis, interval in semitones between the $f_0$s, with white-on-black ratios for pure intervals. $X$-axis, beat frequency, measured by the signal's RMS. Bright pixels denote higher RMS variation, indicating stronger beating amplitude.}
\label{fig:RMS_freq_beating}
\end{figure}

As the frequency difference increases, beating accelerates and eventually becomes audible as a distinct combination tone at around 20~Hz, as illustrated in Figure~\ref{fig:RMS_freq_beating} \citep[p.~12]{turner1977ohm}. Figure~\ref{fig:RMS_freq_beating} also shows that pure intervals (intervals expressed as small integer ratios) generate infinitely slow acoustic beats, \textit{i.e.} no acoustic beats. Helmholtz observed that the tetrads in Palestrina's \textit{Stabat Mater} minimise combination tones, suggesting that such tones were considered undesirable \citep{kursell2015third,helmoltz1885sensations}. \citet{kursell2015third} notes that, after \citet{rameau1722traite} linked triads to overtones, the distribution of notes across registers became less important, rendering combination tones largely irrelevant: `students of music have only had to understand how a triad related to its fundamental[; t]he actual distribution of notes over the registers could be disregarded'.

\clearpage

\begin{figure}[h!]
  \centering
  \includegraphics[width=\columnwidth]{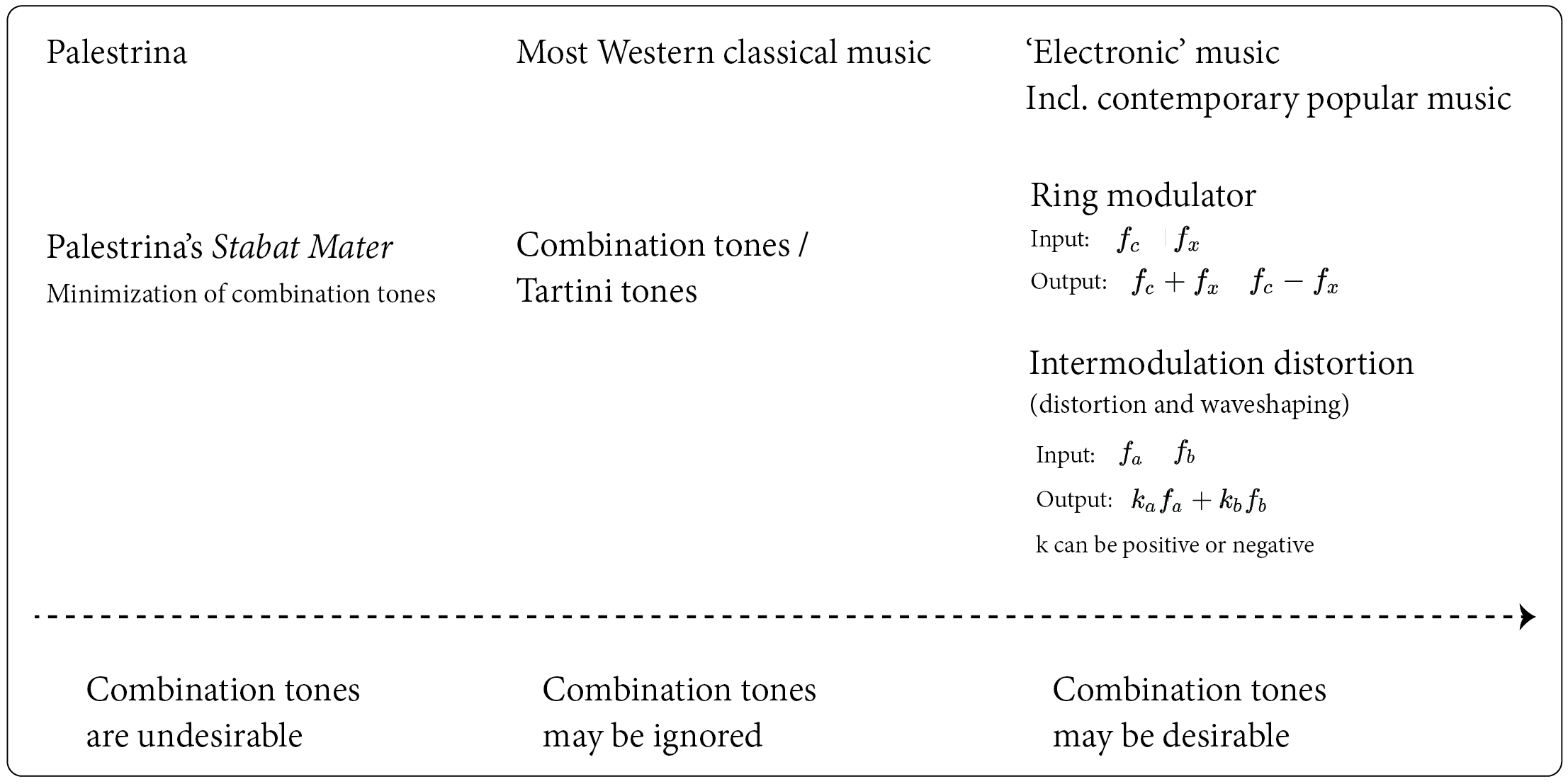}
  \caption{Development over history in attitudes towards combination tones in music.}
\label{fig:S_combination}
\end{figure}

 Later, the 20\textsuperscript{th}-century composer G\'erard Grisey, following Hindemith's \emph{ring modulation harmony} \citep{hindemith1941craft}, used combination tones to generate harmonic fields \citep[p.~52]{anderson2000provisional}. \citet{emmerson1977ring} provides examples of ring modulation harmony, and \citet{costa2019modeling} suggests relations between ring modulation harmony and temperaments. Modern popular music technologies such as ring modulators, distortion, and waveshaping also generate combination tones, and they are heavily used by the Primaal producers (Appendices~\ref{ref:distortion} and~\ref{ref:ringmodulation}). Figure~\ref{fig:S_combination} summarises this evolution.

\vspace{.3cm}

\subsection{`Hearing out' individual harmonics}\label{sec:quasiharmtonesmultiplepitches}

The debate concerning the extent to which humans can perceive multiple pitches from harmonic tones, or `hear out' individual harmonics, has a long history. Section~\ref{sec:loudharmonics} shows that the Primaal producers boost upper partials so that they carry an impression of pitch independently from the fundamental frequency; this section asks to what extent humans are able to `hear out' individual harmonics.

\subsubsection{From Mersenne to Helmholtz}

\citet{mersenne1636harmonie}, in the \textit{Livre quatriesme des Instrumens \`a chordes} (proposition~XI, pp.~208--212)\footnote{\url{https://chmtl.indiana.edu/thesauri/tfm/17th/MERHU3_4_TEXT.html}}, asserts that `the string struck and sounded freely makes at least five sounds at the same time, the first of which is the natural sound of the string and serves as the foundation for the rest, and to which one pays attention for singing and for the parts of music, since the other sounds are so weak that only the best ears can hear them easily'\footnote{In the original 17\textsuperscript{th}-century French: `Mais il faut remarquer qu'il n'a pas sceu que la chorde frapp\'{e}e, et sonn\'{e}e \`{a} vuide fait du moins cinq sons differens en mesme temps, dont le premier est le son naturel de la chorde, qui sert de fondement aux autres, et auquel on a seulement esgard pour le chant et pour les parties de la Musique, d'autant que les autres sont si foibles qu'il n'y a que les meilleures oreilles qui les entendent ays\'{e}ment.'}. According to \citet[p.~32]{euler1980rational}, this statement from Mersenne is the first documented occurrence of the hypothesis according to which the ear may behave as a frequency analyser. During the next century, \citet[pp.~13--14]{rameau1750demonstration}\footnote{\url{https://chmtl.indiana.edu/tfm/18th/RAMDEM_TEXT.html}} designated the hearing of individual harmonics as the original intuition behind his theory of harmony.

This conception of the ear as a frequency analyser became central to the Ohm--Seebeck dispute \citep{turner1977ohm} that followed the publication of \citet{ohm1843definition}. \citet{seebeck1843definition} challenged Ohm's point of view, according to which only the fundamental partial contributes to the production of a tone perceived to be of a certain pitch. On the contrary, according to Seebeck, the impression of tone is amplified (in the original German, \textit{verst\"{a}rkt wird}) by the upper harmonics. For Ohm, this amplification is an illusion: `[o]nly one element actually forms the main tone, but our ear deceives itself by drawing its overtones towards it, and thereby perceives the main tone as stronger'---to which Seebeck answered: `[h]ow can the question of what belongs to a tone be decided, if not precisely by the ear?'\footnote{\url{https://de.wikisource.org/wiki/Ueber_die_Definition_des_Tones}} \citet[pp.~58--59]{helmoltz1885sensations} elaborates on the Ohm--Seebeck dispute. Where \citet{seebeck1843definition} mentions `an irregular sequence of impressions' (original German: \textit{eine unregelm\"{a}{\ss}ige Folge von Eindr\"{u}cken}) that amplifies the tone of the lower partial, Helmholtz considers that the upper partials only prompt a `variable quality' of the tone, by which he refers to their contribution to the timbre of the sound. In terms of the contribution of the upper partials to the tone's pitch, Helmholtz stands by Ohm's hypothesis, maintaining that overtones affect timbre but not pitch, and recommends applying methods that direct the listener's attention to the upper partials. Further discussion can be found in \citet{plomp1964ear,plomp2001intelligent,turner1977ohm,kim2003theories}.

\newpage

\subsubsection{Practical experiments}

\citet{plomp1964ear,plomp1968ear,soderquist1970frequency,moore1993audibility,moore2006frequency,moore2012introduction} provide accounts of experiments that investigate the limits of the ear's ability to `hear out' harmonics. They conclude that humans seem to be able to hear the harmonics of a periodic sound individually to a limited extent: only the first 5--8 components may be identified, and the ability is limited by the width of the ear's auditory critical bands \citep[pp.~86--88]{moore2012introduction}. \citet[p.~58]{helmoltz1885sensations} argues that the difficulty of `hearing out' harmonics is not determined by their energy. \citet{plomp1964ear} disagrees, giving all partials `equal chances' to be heard out by setting all harmonics at the same loudness level.

\subsubsection{Enlarging the scope of the debate}

The hypothesis according to which the ear can behave as a frequency analyser may have been viewed negatively. \citet[p.~438]{dixonward1970} regards it as a `quarter-truth' that `has served to increase the distrust with which perceptive musicians regard scientists since it is readily apparent to them that the ear acts in this way only under very restricted conditions'. \citet{turner1977ohm} similarly speaks of individual partial identification `under certain conditions and to a very limited extent'. According to \citet[p.~2]{plomp1976aspects}, `[i]dentification of the partials may be nearly impossible in listening to tones in a musical context, but it is possible under more favourable conditions'. In light of the analysis of Primaal's music, we argue that the `musical context' mentioned by Plomp is the context of Western classical music. Contemporary popular music uses production techniques that provide favourable conditions for `hearing out' upper partials (Section~\ref{sec:examplesofboostedpartials} and Appendix~\ref{ref:instruments}). The Primaal producers' practice of boosting harmonics to make them audible, together with experimental results from \citet{deruty2025multiple}, suggests that listeners associate pitch with upper partials when these have sufficient energy---which might not be the case for most Western classical instruments, but can be routinely achieved using electronic means. The usage of individual partials and harmonics for their audible pitch values may constitute an important difference between the language of Western classical music and that of contemporary popular music.

\subsection{Tonality and modality}\label{sec:modalitybackground}

An ambiguity resides in the articulation between tonality and modality. Scholars disagree over whether modality designates a real system at all and, if so, over how sharply it should be distinguished from tonality---and consequently over whether 16\textsuperscript{th}- and 17\textsuperscript{th}-century or even medieval polyphony is better described as `modal' or `tonal'. At one end of the spectrum, the distinction is treated as largely a construct of later theory: \citet{powers1981tonal} qualifies the `modal system that is supposed to have governed medieval and Renaissance polyphony' as a `doctrine', preferring to look for `tonal relationships', and later asks whether mode is `real' at all \citep{powers1992modalite,powers2017mode}. At the other end, the two are held to be separate systems with no filiation: \citet{atcherson1973key} considers as `misleading' the conception according to which `key theory' (tonality) is a lineal descendant of `modal theory', even contesting that the major and minor scales are descendants of the church modes. \citeauthor{meeus2013modalite} occupies an intermediate position, defending the distinction while acknowledging that the two systems are related: \citet{meeus2013modalite} refutes \citepos{powers1992modalite} arguments---remarking, for instance, that church modes are essentially \emph{diatonic} whereas tonality requires a \emph{chromatic} scale---yet \citet{meeus2023theoretical} describes a `tonal center' in Renaissance polyphony, a centre which may be either `the final or the final chord'.

The concept of `mode' has itself been interpreted in various ways in the literature, for instance as a historical category in Western music or as a modern concept applied to non-Western repertoires, and as either a \emph{scale} or a \emph{melodic type}---a `particularized scale' or a `generalized tune' \citep{powers2001mode}. The definition retained in this article is the early Western one (900--1000), in which a mode is a scale organised around a prominent degree, the `final'. Such an interpretation can be found in the treatises \textit{Musica Enchiriadis} \citep{Hucbaldus900musicaenrichidis} and \textit{Scholia Enchiriadis} \citep{Hucbaldus900scholiaenrichidis}, attributed to Hucbald of St Amand (Hucbaldus Sancti Amandi, ca.~840--930)\footnote{\url{http://nicolas.meeus.free.fr/mt/0900_Hucbald.html}}. \citet{erickson1995musica} provides a translation of the treatises, in which the scale is introduced on p.~3 (\textit{Musica Enchiriadis}) and the final on p.~50 (\textit{Scholia Enchiriadis}). Similar definitions are derived from \emph{Dialogus de musica} (ca.~1000--1100, uncertain author)\footnote{\url{http://nicolas.meeus.free.fr/mt/1100_Dialogus.html}} by \citet[p.~17]{coeurdevey1998histoire} and \citet{powers2001mode}, with C\oe urdevey stating that `[a] tone or mode is the principle which distinguishes any song by its final'\footnote{In the original French: `Un ton ou mode est le principe qui distingue tout chant par sa finale'.}. The original formulation is: `[t]he tone or mode is the rule that determines the end of every song. According to the aforesaid six consonances [the permissible intervals between the final and the other notes], every principle must agree with its end. No voice can begin a song unless it is either final or consonant with the final through one of the six consonances'\footnote{In the original Latin: `Tonus vel modus est regula, quae de omni cantu in fine dijudicat. Omne principium secundum praedictas sex consonantias suo fini concordare debet. Nulla vox potest incipere cantum, nisi ipsa vel finalis sit, vel consonet finali per aliquam de sex consonantiis'.} \citep{latina1841patrologiae}.

The precise role of the final nevertheless remains ambiguous. As the passage quoted above indicates, the \emph{Dialogus} makes it both the degree on which every song must end and the one to which any starting note must be consonant \citep{latina1841patrologiae}. \citet{powers2001final} emphasises its function as the concluding degree of any melody in a mode, while also warning that `the near synonymity of ``final'' and tonic has remained a pervasive notion in Western musical culture, although many scholars working in non-Western music, folk music and even early polyphony have begun to see this notion as a cultural assumption rather than an inherent connection'. \citet{meeus2023theoretical}, for his part, calls the final the `reference' or `nominal' note.

Applying these notions to Primaal's music first requires taking a position in the debate outlined above. With \citet{meeus2013modalite} and \citet{atcherson1973key}, and against \citepos{powers1992modalite} scepticism, we treat modality and tonality as distinct systems, and we identify tonality by two properties. \citet{schenker1935satz} proposes that in tonal music, harmonic progressions ultimately reduce to I--V--I structures, and \citet[p.~82]{coeurdevey1998histoire} describes the shift from modal polyphony to tonal writing as a move from \textit{resultant} to \textit{active} harmony. Taken together, these views suggest that tonal music requires both directional harmonic progressions---deployed as part of the music's implications rather than as isolated contrasts (Sections~\ref{sec:structurebackground} and~\ref{sec:smallscalestructure})---and active harmony. Primaal's music exhibits neither. There is no directional progression because there is no chord progression at all: the producers describe their music as `mono-chordal', chord changes being seen as masking the pitch shifts around the poles on which the writing rests (Section~\ref{sec:continousfreqs}). The harmony is moreover largely \textit{resultant} rather than \textit{active}, in that pitches are often inherited from the partials of a single complex tone---frequently the bass---rather than chosen from a pre-existing scale, so that the set of available degrees is fixed by the tone itself (Sections~\ref{sec:parameters} and~\ref{sec:examplesofboostedpartials}). It may therefore not be considered tonal. It does, on the other hand, meet the early Western definition of a mode retained above: Section~\ref{sec:modality} shows that the pitch content of each song is organised around a single prominent degree, which serves as a \emph{reference note} in \citepos{meeus2023theoretical} sense rather than as a concluding degree.

\section{Detailed analysis of pitch-related features}
\label{sec:analyses-of-works}

In this section, we provide a detailed analysis of pitch-related aspects in Primaal's music, following the methods outlined in Section~\ref{sec:methods-of-analysis}. Our observations are primarily based on the signal analysis results presented in Section~\ref{ref:signalanalysis}. We address tones' inharmonicity (Section~\ref{sec:inharmonicityprimaal}), salient upper partials (Section~\ref{sec:examplesofboostedpartials}), continuous frequency distributions (Section~\ref{sec:continousfreqs}), the modal organisation of pitch (Section~\ref{sec:modality}), tuned drums (Section~\ref{sec:tuneddrums}), and the organisation of pitch parameters across large- and small-scale structure (Sections~\ref{sec:largescalestructure} and~\ref{sec:smallscalestructure}). We do not attempt to determine the perceived pitch(es). We merely describe aspects of the signal and point to potential relations with the producers' perception and ours.

\subsection{Inharmonicity in the analysed tones}\label{sec:inharmonicityprimaal}

This section documents the inharmonic constitution of the tones from which pitch originates---Section~\ref{sec:parameters}, parameter~(1). In Primaal's music, analysed tones were generally inharmonic to some degree, although most can be considered quasi-harmonic. Pitch estimates from temporal modelling often differ from $f_0$. Tones deriving from subsets of partials inherit the inharmonicity of the full partial set. Finally, some drum samples exhibit measurable partials at inharmonic positions.

Examples of quasi-harmonic tones include the vocals from `Boom' (Section~\ref{ref:boomvocals}), which deviate by $\pm 10$~cents around $f_0$. According to \citet{elvander2020harmonic}, pitch deviations in human speech are small enough for harmonic structure to be exploited in pitch estimation, which is consistent with this criterion. The electric bass from `R U Ready' (Section~\ref{ref:areureadyelecbass}) exhibits inharmonic partials only above the seventh partial, making it quasi-harmonic according to \citet[p.~1]{jarvelainen1999audibility}. A similar observation applies to the bass track of `Elevate' (Section~\ref{ref:elevatebass}). The flute track of `Yada Yada' (Section~\ref{ref:yadayadaflute}) is almost exactly harmonic.

\begin{figure}[h!]
  \centering
  \includegraphics[width=\columnwidth]{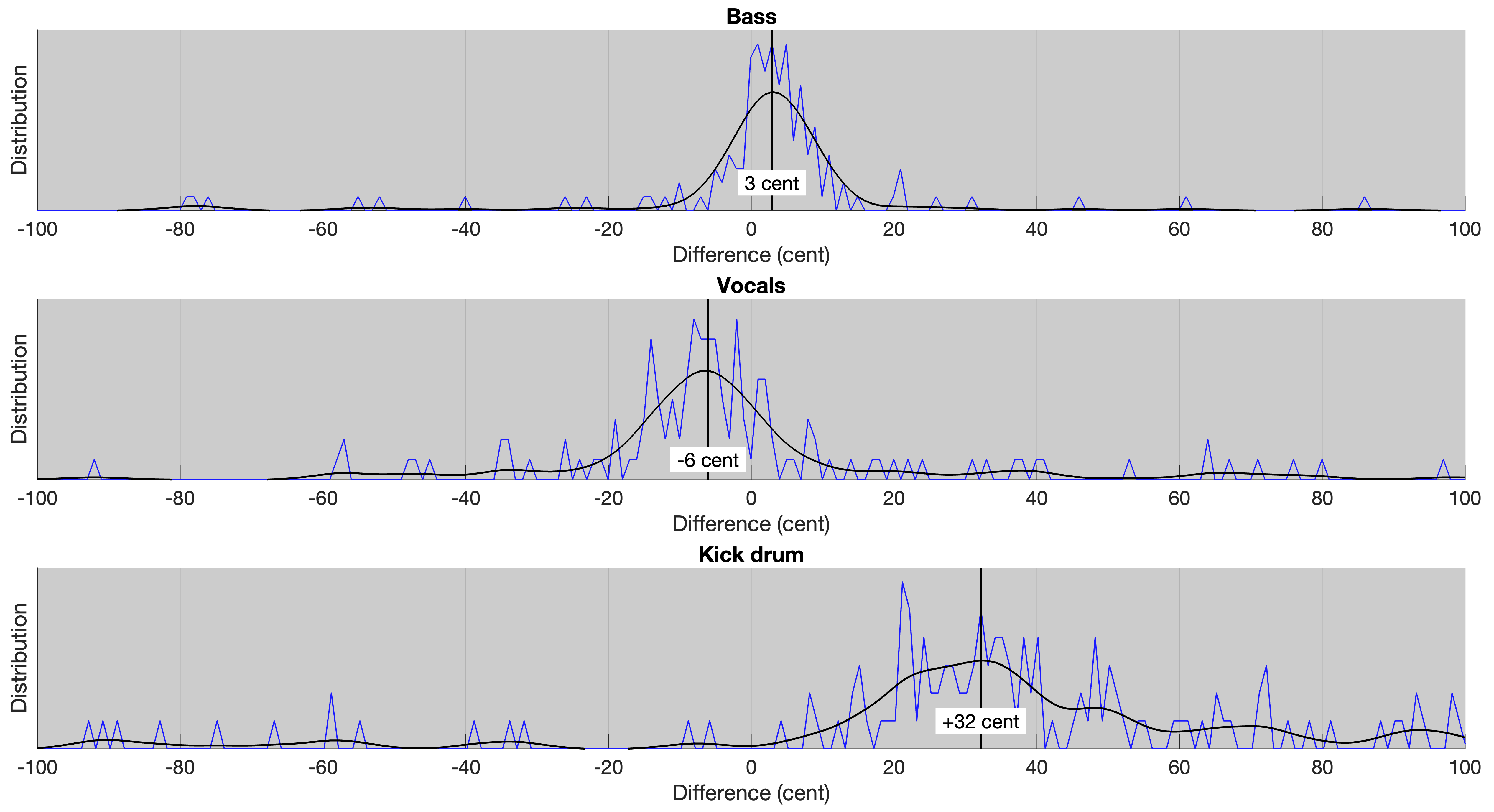}
  \caption{`Boom', bass, vocals, and kick drum track, 0'16 to 0'18. Difference between pitch derived from the lowest partial ($f_0$) and from autocorrelation.}
\label{fig:boomspeccorrcomp}
\end{figure}

The position of $f_0$ may differ from the value obtained through temporal analysis. For example, Figure~\ref{fig:boomspeccorrcomp} compares autocorrelation- and $f_0$-based pitch estimates for the bass, vocal, and kick-drum sections of `Boom' (Section~\ref{ref:boomrelations}), illustrating differences both within and across tracks. Table~\ref{tab:autocorr} summarises discrepancies observed in other tracks.

\vspace{1cm}

\begin{table}[h!]
\caption{Autocorrelation/$f_0$ discrepancy by song and track. Ranges expressed in semitones.}
\label{tab:autocorr}
\setlength{\tabcolsep}{4pt}
{\small
\begin{tabular*}{\textwidth}{@{\extracolsep{\fill}}lllll}
\toprule
Song & Track & Inter-partial difference& Distance  from $f_0$    & Section \\

\midrule
`Danger'      & Lower bass   & Regular       & $[-1,\,0]$       & \ref{ref:dangerlowbass}  \\
`Danger'      & Higher bass  & Irregular     & $[-3,\,-1]$      & \ref{ref:dangerhighbass}  \\
`R U Ready'   & Synth bass   & Irregular     & $[-2,\,0]$       & \ref{ref:areureadysynthbass} \\
`Silver'      & Bass, type 1 & Slightly  irregular     & $[-1,\,0]$       & \ref{ref:silverbasstype1unweighted}   \\
`Silver'      & Bass, type 2 & Irregular     & $[0,\,1]$      & \ref{ref:silverbasstype2}   \\
`Sweet Money' & Bass         & Irregular     & $[-2,\,1]$      & \ref{ref:sweetmoneybass}   \\
\bottomrule
\end{tabular*}}
\end{table}

\newpage

Inharmonicity may be inherited by subsets of partials. The gimmick track from `Whomp' (Section~\ref{ref:whompgimmick}) features a slightly inharmonic tone in which three partials are boosted, producing a higher inharmonic tone; both the carrier and the modulation yield an audible pitch, so the higher tone inherits inharmonicity from the lower tone's partials. A related mechanism operates in `Cardinal' (synth break, Section~\ref{ref:cardinalsynthbreak}), `Elevate' (rising keyboard, Section~\ref{ref:elevatekey}), and `\textexclamdown{}Fire!' (bass, Section~\ref{ref:firebass}), where the tone is used as a carrier: closely spaced inharmonic partials do not form an audible tone, whereas a subset of boosted partials yields a perceptible tone---see Section~\ref{sec:examplesofboostedpartials}. In these cases, the perceived tone inherits the inharmonicity of the carrier partials.

When partials can be identified in drums, their positions are generally inharmonic---see Section~\ref{sec:pitchindrums}. Table~\ref{tab:kickinharmonicity} summarises the inharmonicity observed in the four kick drum samples in which partials could be measured. These examples are relevant because they show that drums, and not only bass and melodic tracks, contribute to the music's overall inharmonicity: since the kick drum is systematically tuned in Primaal's music (Section~\ref{sec:tuneddrums}), its inharmonic partial structure participates in the resultant tuning and in the pitch uncertainty cultivated by the producers.

\begin{table}[htbp]
\caption{Inharmonicity in kick drum samples.}
\label{tab:kickinharmonicity}
\setlength{\tabcolsep}{5pt}
{\small
\begin{tabularx}{\columnwidth}{@{}lXl@{}}
\toprule
Song & Observed inharmonicity & Section \\
\midrule
`Boom'   & Frequency differences between partials up to a semitone above the $f_0$ & \ref{ref:boomrelations}    \\
`Danger' & Four slightly inharmonic partials with variable frequency differences   & \ref{ref:dangerkick}  \\
`Silver' & Harmonic partials barely louder than surrounding spectral content       & \ref{ref:silverkickdrum} \\
`Whomp'  & Three slightly inharmonic partials with variable frequency differences  & \ref{ref:whompkicksnare}   \\
\bottomrule
\end{tabularx}}
\end{table}

\subsection{Amplified partials and the pitches they carry}\label{sec:examplesofboostedpartials}

This section documents the second of the three techniques by which perceived pitches are generated---Section~\ref{sec:parameters}, parameter~(1)---namely the boosting of individual partials so that they are heard as carrying pitches of their own. The producers favour tones in which the pitch neighbouring the theoretical position of harmonic 5 (two octaves and a major third) is made audible alongside the pitch corresponding to the $f_0$; inharmonic relations may result in slightly different intervals. Examples are given in Table~\ref{tab:harmonic5}. A different approach to highlighting a major third interval can be observed in the synth bass from `R U Ready' (Section~\ref{ref:areureadysynthbass}): the $f_0$ of the first `note' in the pattern is C$\sharp$1$+33$~cents, the tone is inharmonic, and its two strongest partials lie close to B4 and D$\sharp$5, whose spacing corresponds to a major third that may be perceived as an audible interval. The producers state that they enjoy boosting the fifth harmonic for its resulting subjective colour---which they describe in terms of `timbre'---rather than for the interval \emph{per se}, and they often seek to attenuate the perception of the resulting sound as `major' while preserving the colour imparted by the fifth harmonic. They also highlight other intervals: in the `Boom' bass (Section~\ref{ref:boombass}), the 11\textsuperscript{th} harmonic (49~cents less than a tritone) was selected to give the impression of a dissonant mistake, while in the upper bass track from `Danger' (Section~\ref{ref:dangerhighbass}), an audible partial near C$\sharp$4$+16$~cents lies close to the 7\textsuperscript{th} harmonic (minor seventh), though the tone is inharmonic and slightly lower than the theoretical position.

\vspace{.5cm}

\begin{table}[h!]
\caption{Examples of highlighted harmonic 5.}
\label{tab:harmonic5}
{\small
\begin{tabular*}{\textwidth}{@{\extracolsep{\fill}}lllll}
\toprule
Song & Track & $f_0$ (approx.) & Overtones (approx.) & Section \\
\midrule
`Boom'    & Bass        & B1        & D$\sharp$4           & \ref{ref:boombass}  \\
`Silver'  & Bass type 1 & E$\flat$1 & G4                   & \ref{ref:silverbasstype1unweighted} \\
`Silver'  & Bass type 2 & D1        & F$\sharp$4           & \ref{ref:silverbasstype2} \\
`Whomp'   & Gimmick     & Between E2 and F2   & Between G$\sharp$4 and A4, plus harmonics & \ref{ref:whompgimmick} \\
\bottomrule
\end{tabular*}}
\end{table}

\vspace{.5cm}

The producers may highlight consecutive groups of partials instead of a single one or a harmonic subset. In `\textexclamdown{}Fire!' (Section~\ref{ref:firebass}), with a very low $f_0$ (ca.~37~Hz), gain is applied to groups of partials around the 36\textsuperscript{th} harmonic, resulting in an uncertain pitch around 1320~Hz. The complex tone from the first partial ($f_0$) can be seen as a \emph{carrier}, with modulation applied via selective gain. A similar approach is used in the `Cardinal' synth break (Section~\ref{ref:cardinalsynthbreak}), where groups of partials are highlighted with a filter's sliding resonance.

\subsection{Continuous frequency distributions}\label{sec:continousfreqs}

As mentioned in Section~\ref{sec:parameters}, in Primaal's music, pitch-evoking frequencies are continuously distributed around `poles'. This section documents the width of these distributions and the instability that shapes them---parameters~(4) and~(5). Such distributions may result from (1) continuous frequency evolution within a single track, often inspired by the Roland TR-808 bass drum (Section~\ref{sec:continuousfreqtraj}), or (2) small variations in discrete frequencies occurring simultaneously across different tracks. These two aspects are interlaced, as frequency evolutions in different tracks are generally not parallel.

\newpage

Continuous frequency changes may be achieved using bends, glides, or screws (Appendix~\ref{ref:bends}). These processes are used to confuse pitch perception. To increase this confusion, the producers often set random pitch bend limits to avoid harmonic intervals like the octave or fifth, which would provide familiar references. The producers may adjust glide time to the length of the musical element. In the bass tracks from `Elevate' and `Silver' (Sections~\ref{ref:elevatebass} and~\ref{ref:silverbass}), they use the time between onsets to let the glide fully develop. This contrasts with tracks like `Danger', where the glide quickly stabilises (Section~\ref{ref:dangerbassandkick}). When possible, they prefer pitch contours that never fully stabilise to enhance pitch ambiguity.

\vspace{.5cm}

\begin{figure}[h!]
  \centering
  \includegraphics[width=\columnwidth]{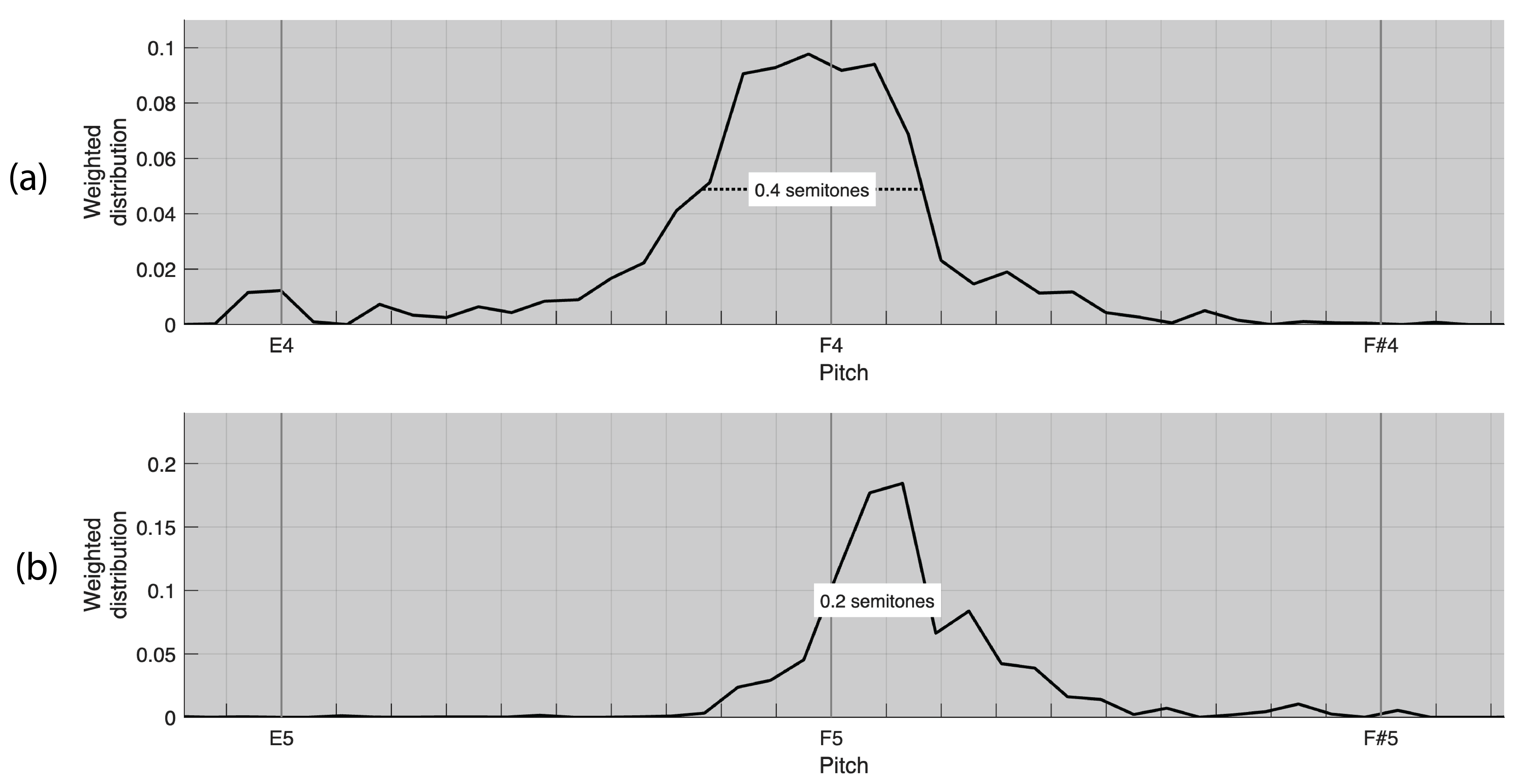}
  \caption{Distribution of $f_0$ values over time, excluding contrasts (see Section~\ref{sec:smallscalestructure}). (a) `Cardinal', vocal loop. (b) `Yada Yada', flute loop. The interval in the white box measures $Q$ from Section~\ref{sec:parameters}, evaluated as the full width at half maximum.}
\label{fig:twodistributions}
\end{figure}

Continuous frequency distributions may stem from continuous frequencies over time. Figure~\ref{fig:twodistributions} compares $f_0$ distributions in two tracks: the vocal loop in `Cardinal' (Section~\ref{ref:cardinalvocals}) and the flute loop in `Yada Yada' (Section~\ref{ref:yadayadaflute}), providing real examples of the schematic distributions in Figure~\ref{fig:S_summary}. The top distribution is wider than the bottom one, and the producers confirmed they intentionally set these widths. Regarding the `Yada Yada' flute, the producers edited the original sample to keep only content near each attack, making the pitches less stable. However, the overall result remained relatively stable, which the producers felt made it too easy for listeners to identify the pitch. To address this, they contrasted this stable part with wide glides, using the width of the distribution as a musical parameter. In Primaal's music, frequency distributions exist around a `pole', and deviations from this `pole' contribute to the music's interest. To help listeners identify the `pole', the producers introduce anchors. An example is the vocal gimmick in `Boom', where the second `note' of the loop serves as a reference for the `pole' (Section~\ref{ref:boomvocals}). However, the `pole' in one track may not match the tuning of others, as tuning is generally specific to each track.

\vspace{.5cm}

\clearpage

Continuous distributions may also stem simultaneously from the relations between tracks and from continuous frequencies over time. For example, `Boom' features a kick drum, bass, and looped vocal gimmick. The frequency trajectories for these tracks are continuously evolving. The $f_0$ values for the kick drum and bass tracks either follow similar continuous sequences or intersect (Section~\ref{ref:boomrelations}). The analysis also shows how the vocal track's $f_0$ often neighbours that of the kick drum, creating a continuous distribution through episodic proximity of the $f_0$ values. Another example is `Silver' (Section~\ref{ref:silverbass}), in which the bass's $f_0$ initially exceeds the kick drum's, creating dissonance. As the kick drum is shorter, it stops while the bass's pitch continues down. The bass `notes' 2, 3, and 4 start between D$\sharp$1 and E1 and end between C1 and D$\sharp$1. Temporal modelling suggests a perceived pitch lower than the $f_0$, while the kick drum contains a tonal component around C2, indicating initial dissonance, with the bass descending towards the kick drum's pitch (modulo the octave).

The producers use continuous frequencies and distributions to achieve several goals. They describe Primaal's music as `mono-chordal', where controlled pitch shifts from `poles' bring `perceptual richness'; in their terms, chord progressions would mask such shifts. In `Boom' (Section~\ref{ref:boomvocals}), for instance, the vocal track `hovers around stable pitches', keeping the listener wondering when the tension will be resolved---yet the tension is never fully resolved, sustaining engagement. More broadly, the TR-808 model of frequency evolution, in which partials' frequencies decrease quickly and then stabilise, creates a `tension-release' effect that holds interest. A further goal is to emphasise rhythm over pitch: by using glides to make pitches hard to pinpoint, the producers encourage listeners to turn to rhythm for familiar orientation. Finally, they use different tuning patterns for each voice to make individual parts stand out, as illustrated by the `Silver' brass theme (`D D D E$\flat$ D', 0'02, 0'25, 0'33, 0'59, 1'07), where a higher tuning was chosen for that purpose.

\vspace{.5cm}

\subsection{Modal organisation}\label{sec:modality}

This section interprets the pitch content in Primaal's music in terms of modal organisation as defined in Section~\ref{sec:modalitybackground}, and thereby documents the choice of poles and the prominence given to one of them---Section~\ref{sec:parameters}, parameters~(2) and~(3). In the previous section, we explain that the absence of chord progressions facilitates the derivation of musical parameters from continuous frequency distributions. Tonality would require chord changes; therefore, the Primaal producers prefer a modal organisation. The producers state that this approach is common in hip-hop/R\&B. Many hip-hop tracks are based on a single mode with one final (modal view) or first degree (tonal view). The producers aim for music that `hovers around one note' to emphasise other musical elements. They reference Jamie xx's `Gosh' \citep{jamiexx2016gosh} as an example. Other examples include Massive Attack's `Angel' \citep{massiveattack1998angel} and Pop Smoke's `Dior' \citep{popsmoke2019dior}.

According to the producers, all pitches in each song originate from a single chosen \emph{final}. Other pitches around the poles are then added through the processes shown in Figure~\ref{fig:S_summary}: they may originate from a scale where the final is the first degree, they may be partials of the tone for which the final is the $f_0$, or they may be partials of a tone where the final is a result of temporal modelling. The first relation links arbitrary pitches to harmonic overtones in a way similar to Western classical music, whereas the second and third directly derive pitches from upper partials (Section~\ref{sec:loudharmonics}). All degree choices are thus influenced by overtones, either indirectly or directly; in the latter cases, the determination of the scale relates to \emph{homoph\={o}noi} intervals in the sense of Ptolemy \citep{richter2001ptolemy}.

\newpage

We attempted to identify the modes for each song by ear, though pitch uncertainty complicates the task: it is straightforward for `R U Ready', difficult for `Danger' and `Sweet Money', and impossible for `\textexclamdown{}Fire!'. Table~\ref{tab:modes} summarises the transposed modes (final always set to C). While the producers state that they derive their modes from the minor and major diatonic scales, pentatonic modes, urban music, and blues without `blue notes' \citep{kubik2001bluenote}, several observations can be made. The fifth degree (G) is always present. The third (E) and seventh degrees (B) are often ambiguous, either minor or major, or both present in a song; the producers regularly balance the major third (harmonic 5) with the minor third to avoid a `too major' sound, as noted in Section~\ref{sec:examplesofboostedpartials}. The B$\flat$/B ambiguity moreover recalls the pre-13\textsuperscript{th}-century modal system, where the B did not function as a leading tone and its alteration was subject to interpretation, termed \emph{Musica Ficta} \citep{bent2001MusicaFicta}.

\citet[p.~42]{coeurdevey1998histoire} notes that, in the 16\textsuperscript{th} century, the bass note indicates the mode's final. This is also true in Primaal's music. \citet{meeus2023theoretical} introduces the notion of `reference' or `nominal' note, which may not always be the mode's final. In two of the songs, the final is the sole `reference' note, while in others, the fifth and sometimes the third (major or minor) also play a key role. The reference notes for each song are given in Table~\ref{tab:modes}. As noted in Section~\ref{sec:continousfreqs}, the presence of a minor or major third does not render the songs tonal.

\vspace{.2cm}

\begin{table}[htbp]
\caption{Modal organisation of the ten studied songs. The transposed mode column shows scale degrees with the final transposed to C (\textbf{bold}). The true final and reference note columns give the non-transposed pitch classes.}
\label{tab:modes}
\setlength{\tabcolsep}{4pt}
{\small
\begin{tabular*}{\columnwidth}{@{\extracolsep{\fill}}llll}
\toprule
Song & Transposed mode (final = C)& True final  & Ref.\ note(s)  if not final \\
\midrule
`Boom'        & \textbf{C} E$\flat$/E F G A B$\flat$          & B & F$\sharp$ (keyboard) \\
`Cardinal'    & \textbf{C} D$\flat$ E$\flat$ G A$\flat$ B$\flat$ & D & F (vocals) \\
`Danger'      & \textbf{C} E$\flat$ F G B$\flat$/B             & E$\flat$ & B$\flat$ (brass)\\
`Elevate'     & B$\flat$/B \textbf{C} D E                      & E$\flat$ & --- \\
`\textexclamdown{}Fire!' & (no chromatic scale)                & C & --- \\
`R U Ready'   & B$\flat$ \textbf{C} E$\flat$ F G               & C & G (vocals) \\
`Silver'      & \textbf{C} D$\flat$ E F G A$\flat$             & D & A (vocals)\\
`Sweet Money' & \textbf{C} D F G A$\flat$                      & B & D$\sharp$ (various) \\
`Whomp'       & \textbf{C} E$\flat$/E F G A$\flat$             & F & C (gimmick) \\
`Yada Yada'   & \textbf{C} D$\flat$ E$\flat$/E F G A$\flat$ B$\flat$ & F & A$\flat$ then C (vocals) \\
\bottomrule
\end{tabular*}}
\end{table}

\vspace{.2cm}

Exceptions to this modal organisation nevertheless occur. In Section~\ref{sec:largescalestructure}, we illustrate how one structural segment may differ drastically from others, creating long-term structure. In Section~\ref{sec:smallscalestructure}, we show how `contrasts' contradict formal implications, creating structure at short time scales. In both cases, an element can either take on exceptional values within a dimension or derive from a different system. In these sections, we provide examples of tonality and cadences being used for such contrasting structural patterns. In Section~\ref{sec:largescalestructure}, Figure~\ref{fig:twostructures}, we also show examples from `R U Ready' and `Silver', where tuning changes between structural segments, though the mode stays similar. While the modal aspect remains intact, this process differs from traditional modality.

\newpage

\subsection{Tuned drums}\label{sec:tuneddrums}

Drums are where pitch strength---Section~\ref{sec:parameters}, parameter~(6)---is most clearly exploited as a variable. The producers tuned the kick drum in all ten examples considered in this paper, and the snare drum in `Elevate', `Whomp', and `Yada Yada'. In `Danger' (Section~\ref{ref:dangerkick}), the kick drum is weakly harmonic with a pitch close to the bass's. In `Silver' (Section~\ref{ref:silverkickdrum}), tuning was based on temporal pitch modelling rather than the $f_0$, and all percussion sounds in the song were tuned. In `Whomp' (Section~\ref{ref:whompkicksnare}), the kick drum's pitch is around C, loosely tuned to the tonality's 5\textsuperscript{th} degree, while the snare was transposed from C to B to fit the mode. In `Yada Yada', the snare was transposed by one semitone to match the closest `pole'. The producers also provided insights on drums not subjected to signal analysis in this study: in `Sweet Money', pitches were derived from the opening tom strokes, providing a reference; in `Elevate', the kick drum is perceived as D (the song's final), while the snare is tuned unusually to the minor sixth (F and B$\flat$); and in `\textexclamdown{}Fire!', the kick drum was tuned even though identifiable pitches are absent, suggesting that the presence of pitch sensation without pitch identification suffices for tuning.

\vspace{.5cm}

\subsection{Musical parameters and large-scale structure}\label{sec:largescalestructure}

\begin{figure}[h!]
  \centering
  \includegraphics[width=\columnwidth]{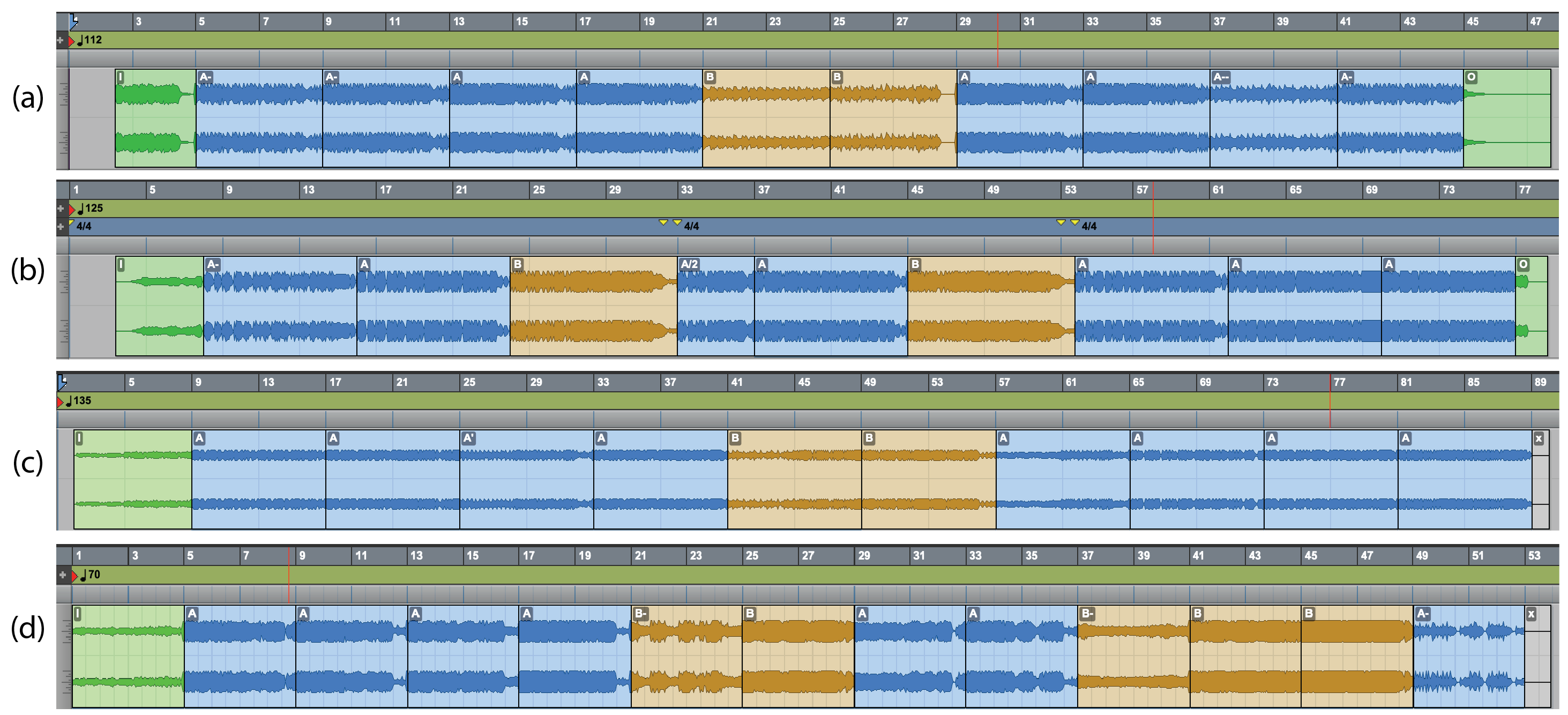}
  \caption{Large-scale structures: (a) `R U Ready'; (b) `Silver'; (c) `Elevate'; (d) `Cardinal'. Top line, bar number. Segment splitting followed \citet{bimbot2012semiotic}, with segment labelling loosely based on the same source.}
\label{fig:twostructures}
\end{figure}

In this section, we consider techniques used by the Primaal producers to articulate the large-scale structure as defined in Section~\ref{sec:structurebackground}, using the musical parameters described in Section~\ref{sec:parameters} and, in particular, conformation to or deviation from the processes those parameters describe---parameter~(7), conformation/deviation. Figure~\ref{fig:twostructures} compiles the segmentations of four songs discussed here.

\vspace{.5cm}

\newpage

In `R U Ready' and `Silver', a change in bass sound contributes to the large-scale structure (Sections~\ref{ref:areureadybass} and~\ref{ref:silverbass}). This change in sound is accompanied by a tuning shift, making the bass seem slightly detuned downwards. This shift does not correspond to a tonal modulation---i.e., `a firmly established change of key' according to \citet{saslaw2001modulation}. The mode's final remains constant, with only the tuning altered. Panels~(a) and~(b) of Figure~\ref{fig:twostructures} show the large-scale structure, where the tuning of the orange `B' segments differs from that of the blue ones. Non-Primaal examples of tuning as a determinant of large-scale structure include Diplo's `Express Yourself' \citep[1'19--1'37]{diplo2012expressyourself}, Jedi Mind Tricks' `Silence' \citep[1'00--1'11 and 2'01--2'12]{jedimind2009}, and Skrillex's `Scary Monsters and Nice Sprites' \citep{skrillex2011scary}.

In `Elevate', the producers use a mode change to build the large-scale structure. The section between 1'09 and 1'38, shown as orange segments in Figure~\ref{fig:twostructures}(c), uses G as the mode's final instead of E$\flat$ (Section~\ref{ref:elevatebass}). Additionally, the tuning changes. The simultaneous changes in mode and tuning are intended to introduce a sense of uncertainty, breaking the established mode and creating tension, which resolves when the original mode returns.

In `Cardinal', the producers introduce clearer tonality to shape large-scale structure. This reflects the use of `[c]onformation to [the composition] process [as] a musical parameter' (Section~\ref{sec:parameters}, Figure~\ref{fig:S_summary}). In Figure~\ref{fig:twostructures}(d), blue `A' segments revolve around D and F, with F as the mode's final. In the orange `B' segments, horns play sequences based on E$\flat$--D and E$\flat$--B$\flat$--D, while a synth plays arpeggios around E$\flat$ major and D minor chords. The resulting cadence is reminiscent of a flattened submediant chord \citep[p.~120]{forte1979tonal}, a typical tonal vocabulary element.

Other Primaal songs show similar processes. While `\textexclamdown{}Fire!' (Section~\ref{ref:firebass}) firmly avoids the twelve-tone chromatic scale, the producers inserted two segments (1'15--1'30 and 2'37--2'50) featuring a twelve-pitch-based chordal brass section. Non-Primaal examples in which similar structure articulation techniques are used include Britney Spears' `Work Bitch' \citep[2'49--3'19]{britney2013workbitch}, Kanye West's `On Sight' \citep[1'17--1'30]{kanyewest2013onsight}, and Skrillex's `Rumble' \citep[1'02--1'15 and 2'11--end]{skrillex2023rumble}.

\vspace{.5cm}

\subsection{Musical parameters and small-scale structure}\label{sec:smallscalestructure}

In this section, we consider techniques that articulate the small-scale structure as defined in Section~\ref{sec:structurebackground}, using the musical parameters described in Section~\ref{sec:parameters} and, again, conformation/deviation---parameter~(7)---at a shorter time scale. The notion of `determinant of form' is borrowed from \citet{caplin1998classical}.

The Primaal producers often use unexpected pitch movements as \emph{contrasts} \citep{bimbot2016system}. They explain that when they allow listeners to identify a stable pitch, they quickly disrupt it with wide-ranging pitch patterns, using bends, screws, and octaviations (transposing the audio by octaves). Examples include the vocal loop of `Cardinal' (cuts, bends, screws, and octaviations, see Section~\ref{ref:cardinalvocals}), the bass from `\textexclamdown{}Fire!' (slow upward progression followed by octaviation, Section~\ref{ref:firebass}), and the flute track from `Yada Yada' (Section~\ref{sec:continousfreqs} and Section~\ref{ref:yadayadaflute}). The producers note that creating small-scale structures using these techniques requires flexibility and precision only achievable in digital audio workstations, making these processes an \emph{idiom} of DAWs \citep{huron2009characterizing}.

\newpage

Section~\ref{sec:largescalestructure} illustrated how the producers use tonality as a deviation from their compositional process to build large-scale structure. They may implement similar deviations for short-term structure: \emph{implications} are built in a modal context, making tonality a \emph{contrast}. In `Cardinal' (1'22), a violin arpeggio on two consecutive perfect triads provides this contrast; in `Whomp', with an F final, an organ at 0'30, 0'45, 1'37, and 1'50 plays a sequence evoking `F major--G$\flat$ major--E$\flat$ major' chords; and in `Boom', with a B final, an F$\sharp$ appears at the end of three segments (0'46, 1'30, and 1'53), resolving in a V-I sequence.

\vspace{.2cm}

In `Silver', contrasts occur at 0'14, 0'22, 0'31, and 1'23, featuring a synthesiser playing B$\flat$-A-A-A, followed by a D (the final) at the next segment's start. This pattern evokes a tonal cadence with a flattened sub-mediant chord. The tuning is lower than the surrounding tracks, with the last A close to A$\flat$. The lower tuning being unexpected, it enhances the contrast. The producers note that for an unexpected element like this contrast, they would typically adjust the tuning \emph{upwards} (Section~\ref{sec:continousfreqs}). Thus, a \textit{lower} tuning adds another layer of surprise. Using terminology from \citet{conklin2010discovery}, the contrast's unexpectedness comes from both `intra-opus' relations and `inter-opus' information.

\vspace{.2cm}

Contrasts may or may not be based on a previously presented musical parameter. The producers acknowledge that every musical passage they write operates within a \emph{system}, in the sense described by \citet{bimbot2016system}. This system is based on some of the musical parameters outlined in Section~\ref{sec:parameters}. To create contrasts, the producers either (1) introduce exceptional values for these parameters or (2) introduce new, unexpected parameters. Section~\ref{sec:smallscalestructure} illustrated contrasts of type~(1), where pitch is already part of the implications, contrasts of type~(2), where the modal context introduces a tonal contrast, and the combination of both: the contrast is tonal (new parameter), but `tuning', also part of the contrast, was presented before in the song.

\vspace{1.5cm}

\section{Summary and conclusion}\label{sec:conclusion}

\vspace{.5cm}

\subsection{Study summary}

This study has used a phenomenological approach to examine the commercial music production of Primaal, a Hyper Music brand. Phenomenological reduction, used to \emph{bracket} preconceptions (Section~\ref{sec:methods-of-analysis}), has been applied to signal analyses standing in place of a score (Section~\ref{ref:signalanalysis}); these analyses form the evidence from which the pitch-related expressive parameters and their organisation have been identified (Section~\ref{sec:parameters}), and every observation reported here can be traced back to one of them. The nature of those parameters highlights how pitch manipulation in this music differs significantly from its treatment in Western classical music---exemplifying the difficulty of transcribing contemporary popular music parameters to musical scores (Sections~\ref{ref:scoreasbias} and~\ref{sec:phenomenologicalreduction}).

\newpage

A central principle in Primaal's music is the cultivation of \emph{pitch uncertainty}, achieved through tone inharmonicity (Sections~\ref{sec:inharmonicity} and~\ref{sec:inharmonicityprimaal}), tones with very low fundamental frequencies (Section~\ref{sec:lowinharmonic}), quasi-harmonic tones evoking multiple pitches including odd-harmonic tones (Sections~\ref{sec:loudharmonics}, \ref{sec:quasiharmtonesmultiplepitches} and~\ref{sec:examplesofboostedpartials}), continuous frequency trajectories (Sections~\ref{sec:continuousfreqtraj} and~\ref{sec:continousfreqs}), and tonal drums with low pitch strength (Sections~\ref{sec:pitchindrums} and~\ref{sec:tuneddrums}). Primaal's music appears primarily \emph{modal} (Sections~\ref{sec:modalitybackground} and~\ref{sec:modality}), with scale degrees defined as \emph{distributions} rather than discrete values. These distributions may stem from continuous frequency evolution over time and/or from the superimposition of slightly different frequencies at a given moment (Section~\ref{sec:continousfreqs}). Finally, the highlighted pitch-related musical parameters play a key role in the constitution of Primaal's musical structure, be it long-term (Section~\ref{sec:largescalestructure}) or short-term (Section~\ref{sec:smallscalestructure}).

\subsection{Historical perspective}

While Primaal's manipulation of pitch has little in common with late Baroque or Viennese classical music, it resonates more strongly with concerns found both before and after that tradition. On the one hand, it recalls early Western approaches to pitch organisation: their use of \emph{homoph\={o}noi} intervals (Ptolemy, 2\textsuperscript{nd} century; Section~\ref{sec:modality}), their interpretation of \emph{mode}, which resembles the early Western definitions of 900--1000 (Section~\ref{sec:modalitybackground}), and their use of \emph{combination tones} and \emph{acoustic beating}, echoing 16\textsuperscript{th}-century and Renaissance concerns (Section~\ref{sec:acousticbeatcombtones}). Related techniques also appear in 20\textsuperscript{th}-century `art' music. Maurice Ravel blurred timbre and pitch in {\em Bol\'ero\/} (1928) through parallel thirds, sixths, and fifths, producing timbral effects akin to Primaal's boosting of the fifth harmonic (Section~\ref{sec:examplesofboostedpartials}). Further instances of continuous pitch manipulation include Ligeti's {\em Atmosph\`eres\/} (1961), Penderecki's {\em Threnody\/} (1961), and Stockhausen's use of ring modulation in {\em Mixtur\/} (1964), {\em Mikrophonie II\/} (1965), and {\em Mantra\/} (1970). The producers acknowledge the influence of Ravel and Stockhausen on their work.

\subsection{Summary of contributions}

We believe this work contributes to music analysis, music information retrieval (MIR), computational creativity, music generation, and psychoacoustics.

With regard to \emph{music analysis}, we demonstrate the potential of phenomenological reduction for popular music analysis: bracketing preconceptions brings into view pitch-related dimensions that are systematically under-documented. This improves our understanding of contemporary popular music, where pitch extends beyond the notion of discrete notes. Rather than transcribing the music into scores or MIDI, we advocate starting from the signal, the listener's perception, and the production process. So that this last term is usable rather than merely invoked, we document the tools and techniques on which the observed pitch manipulations depend (Appendix~\ref{ref:instruments}), which allows each analytical observation to be related to the process that produced it. The analysis was moreover conducted with the producers themselves, their commentary being treated as empirical clarification of intentional choices rather than as an external interpretive frame; this addresses a recurring obstacle in popular music analysis, namely that analysts seldom establish whether the creators were aware of what is being described (Section~\ref{ref:producerspov}). Furthermore, the practice of boosting partials to render them audible underlines a \emph{pitch--timbre continuum}, bridging harmony-focused and production-focused approaches.

\newpage

With regard to \emph{music information retrieval (MIR)}, we show that a single complex tone does not necessarily evoke a single pitch. This bears directly on the way pitch is commonly represented in the field, where estimation tasks are generally formulated as the recovery of the $f_0$ of a harmonic complex tone (Section~\ref{ref:mircommunity}). Pitch tracking should instead estimate the probability of perceived pitches in alignment with psychoacoustics, rather than attempting to infer the `notes' that generated the sounds. More broadly, the continuous pitch distributions observed in this music suggest that representing pitch as discrete notes may be generally inappropriate for popular music.

With regard to \emph{computational creativity} and \emph{music generation}, this work highlights that expressive variations in pitch and timbre are crucial for music generation. More accurate descriptions of contemporary music, including pitch-manipulation techniques, could improve text-to-music interfaces and open new avenues for practical music-generation applications, for instance through richer forms of user interaction.

With regard to \emph{psychoacoustics}, rather than debating whether partials can be heard individually, we suggest that individual partials or subsets of partials may become audible when sufficiently intense, which motivates the study of audibility thresholds; conditions for hearing out individual partials may occur more often than previously assumed. Inharmonicity in the studied music also offers a new perspective on the assumption that `[t]he most commonly considered form of pitch-evoking sound is a harmonic complex tone' \citep{oxenham2012pitch}, and TR-808-derived bass tones---in which the first partial is often inaudible---challenge the assumption that tones sharing the same $f_0$ necessarily share the same pitch.

\subsection{Future work}

This study has relied on the producers for validation; future work should use complementary methods, such as the listening tests performed by \citet{deruty2025multiple} on tones similar to the bass tones used by Primaal. A similar analytical procedure should also be applied to other artists, as has been done for Vitalic by \citet{deruty2025vitalictemperament}: since Primaal primarily operate within a commercial context, their productions are expected to align with current trends, and studies involving other artists may confirm or refute this hypothesis.

\vspace{1cm}
\appendix

\interlinepenalty=10000
\renewcommand{\topfraction}{0.9}
\renewcommand{\bottomfraction}{0.9}
\renewcommand{\textfraction}{0.06}
\renewcommand{\floatpagefraction}{0.75}

\section{Tools and techniques}\label{ref:instruments}\label{sec:tools-and-techniques}

This appendix lists some of the tools and techniques used by the Primaal producers during the production of the analysed tracks. Each is documented for what it does to the partial structure of a tone, which is what the analyses of Section~\ref{ref:signalanalysis} measure; Table~\ref{tab:overtone_processes}, in Section~\ref{sec:loudharmonics}, indicates which of these processes is used in which song and track. One central tool, used for both generation and processing, is Spectrasonics Omnisphere\footnote{\url{https://www.spectrasonics.net/products/omnisphere/}}. Omnisphere has been considered one of the world's best synthesisers for several years \citep{nagle2015omnisphere}. This suggests that observations made in the case of Primaal's music may apply to other commercial tracks.

\subsection{The Roland TR-808 bass drum}\label{ref:bassgeneration}

The Roland TR-808 Rhythm Composer is an analogue drum machine manufactured between 1980 and 1983 \citep{hasnain2017tr}. Described as `one of the most influential and unique drum machines of its time' \citep{meyers2003tr}, it remains `a benchmark against which all other analog drum machines are measured' \citep{werner2014physically}. It is a fixture of hip-hop culture, `not only as a tool for producers but as a defining sound of the genre' \citep{hasnain2017tr}---R'n'B producer Scott Storch observes that in modern trap music, producers `live in an 808 world' \citep{storch2022}---and its presets are equally classic in techno, electro, R\&B, and house \citep{dayal2014tr}.

 The Primaal producers use 808 bass drum-style generators and sample libraries for their bass parts, not for their kick drums, for which they use samples from other sources. This usage follows a well-documented shift. The 808's `long and velvet deep, almost subsonic' bass drum \citep{carter1997tr}, originally designed as a kick, came to serve as kick and bass at once: producer Remi Kabaka Jr.\ recalls that `the kick drum would play the bass at the same time [...] so the fill was not just complex and rhythmical, but it was also tonal' \citep[1:11:11]{Dunn2015}, and 808 kicks were soon loaded into samplers and played `no longer as drum samples but as full-on booming bass lines' \citep{burke2019}. In trap and hip-hop especially, the 808 now `often carries the bassline, providing both the low-end foundation and outlining the harmonic progression of the song' \citep{lavoie2020}, to the point that many producers program their drum patterns with a tuned 808 in place of a kick drum sample \citep{burchell2022}.

 According to \citet{shier2023differentiable}, `percussive audio [...] often contains rapidly decaying signal components following an impulsive event'. Indeed, decreasing frequency values in kick drums can be observed in all the provided Primaal songs (see for instance Figure~\ref{fig:WhompKick} (a), Section~\ref{ref:whompkicksnare}). To emulate this phenomenon, the 808 bass drum features a `decaying pseudo-sinusoid with a characteristic sighing pitch' \citep{werner2014physically}. 

\vspace{.3cm}

\begin{figure}[h!]
  \centering
  \includegraphics[width=.9\columnwidth]{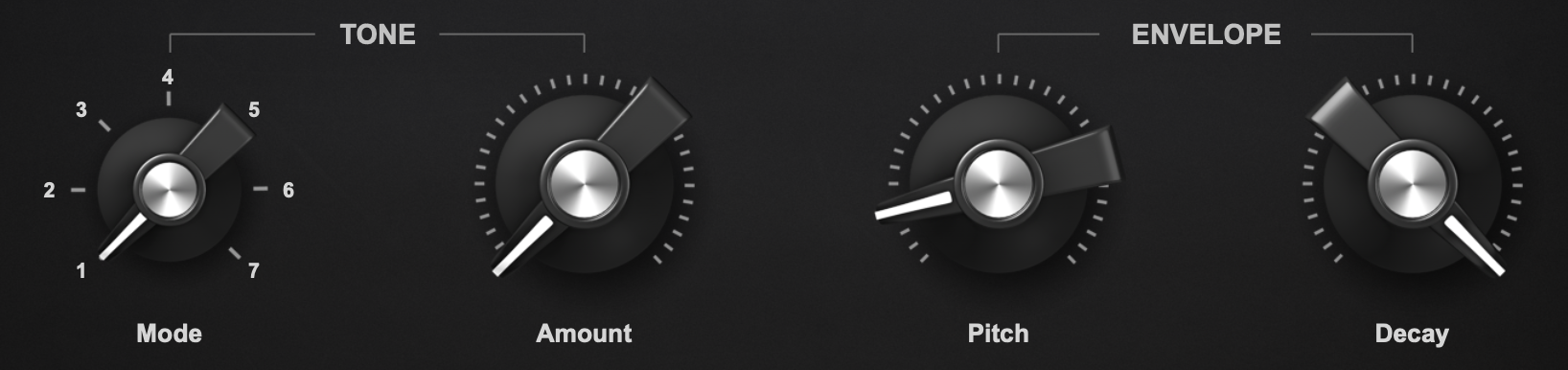}
  \caption{`808 Woofer Warfare' patch, key controls.}
\label{fig:808WooferWarfare}
\end{figure}

In Primaal's music, the main tool for 808-style bass generation is the `808 Woofer Warfare' patch in Omnisphere's Seismic Shock library\footnote{\url{https://www.sonicextensions.com/seismic-shock/}}. Figure~\ref{fig:808WooferWarfare} shows the patch's key controls. `Mode' selects the harmonics to be highlighted. `Amount' specifies to what extent the harmonics should be highlighted. The `808 Woofer Warfare' info panel states: `Use the Mode and Amount knobs in tandem to dial in different harmonics'. `Pitch' controls the amplitude of the initial downward frequency glide. `Decay' sets the length of the `note'.

\vspace{.3cm}

\clearpage

\begin{figure}[h!]
  \centering
  \includegraphics[width=.9\columnwidth]{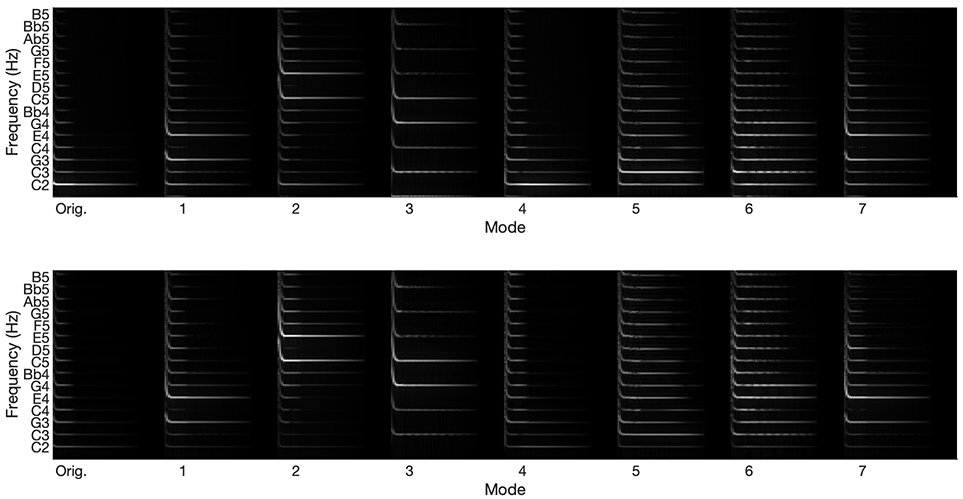}
  \caption{Omnisphere, Seismic Shock library, `808 Woofer Warfare' patch, STFT for the seven modes. (a) Unweighted audio. (b) Weighted audio.}
\label{fig:808warfare}
\end{figure}

Figure~\ref{fig:808warfare} shows the partials in the original 808 Woofer Warfare audio and in audio generated using the different modes. Mode 1 favours harmonics 3 and 5, mode 2 favours harmonics 8 and 10, mode 3 favours even harmonics, 6 and 8 in particular. The Primaal producers enjoy mode 2, which makes the major third audible. In addition to making different harmonics audible, each mode results in a different RMS envelope for the output sound: mode 2, for instance, provides a smooth envelope, whereas mode 4 results in an envelope with wide variations.

\vspace{.5cm}

The producers may, on occasion, use TR-808 audio libraries instead of the `808 Woofer Warfare' patch. An example of such libraries is Julez Jadon's Hot Sauce `The 808 Pack'\footnote{\url{https://julezjadon.com/collections/all/products/hot-sauce-the-808-pack-vol-ii}}. One set of samples from this library, which is used on `Elevate', is eloquently called `808 Drunken Pig' as a reference to the strong glides it features.

\vspace{.5cm}

\subsection{Distortion and waveshaping}\label{ref:distortion}

We consider distortion and waveshaping together, as waveshaping is a way to produce non-linear distortion \citep{roads1979tutorial}. For distortion and waveshaping, the Primaal producers use Omnisphere's Waveshaper\footnote{\url{https://support.spectrasonics.net/manual/Omnisphere2/25/en/topic/layer-page-oscillator-page20a}}, SoundToys' Decapitator\footnote{\url{https://www.soundtoys.com/product/decapitator/}} and the SansAmp PSA-1 plug-in\footnote{\url{https://www.avid.com/plugins/sansamp-psa1}}.

\clearpage

\subsubsection{Distortion and waveshaping generate partials}

Distortion and waveshaping apply a non-linear transformation to the signal, therefore prompting intermodulation distortion \citep[p. 464]{newell2017recording}. Intermodulation distortion produces partials from each input partial and their combination. If $f_a$ and $f_b$ are the input frequencies, then the output contains the frequencies $k_a f_a + k_b f_b$, where $(k_a, k_b) \in \mathbb{Z}^2$. The generated partials have both lower and higher frequencies than the input.

\subsubsection{Interaction between original and generated partials}\label{ref:interactionpartials}

\begin{figure}[h!]
  \centering
  \includegraphics[width=1\columnwidth]{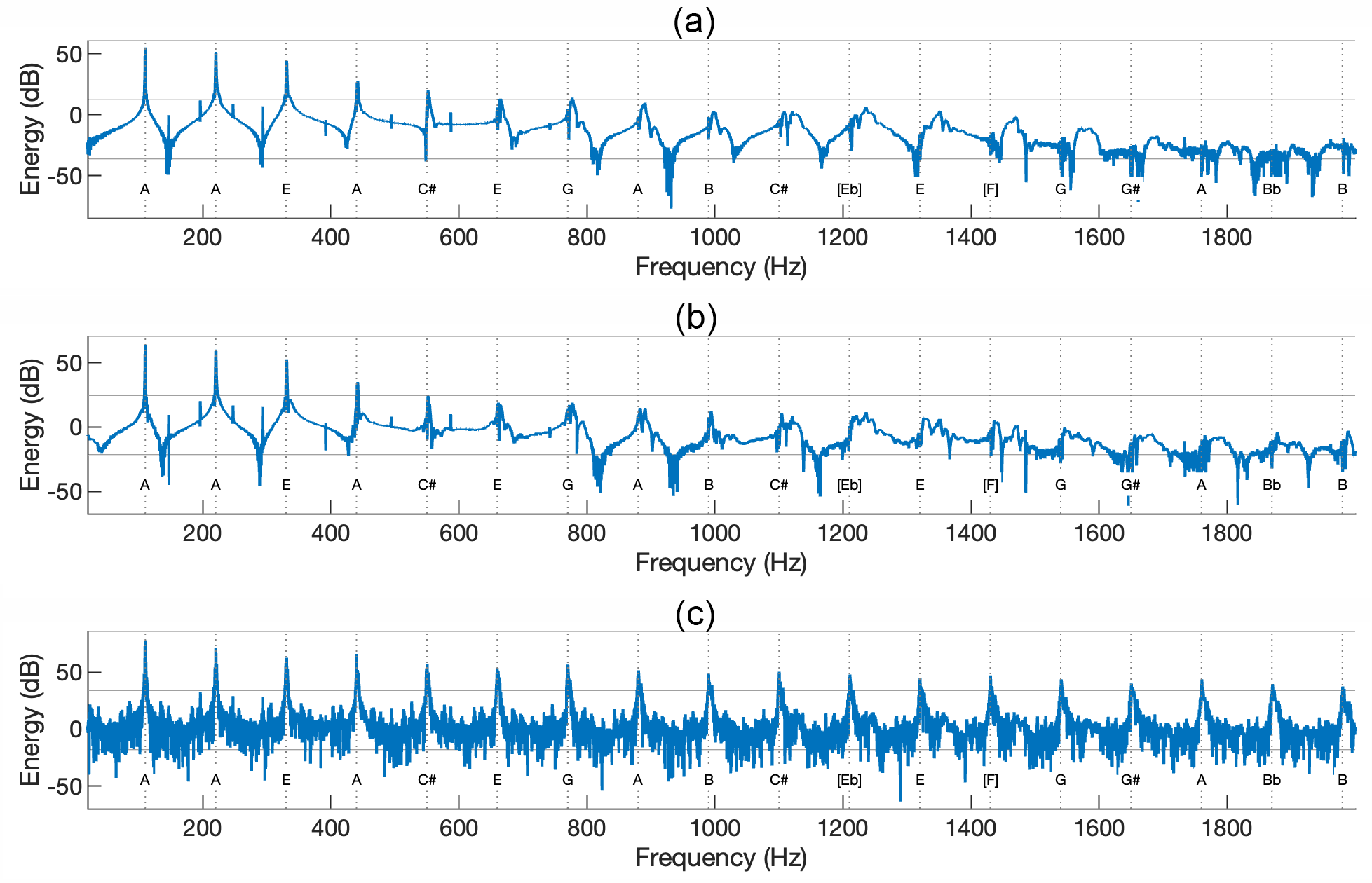}
  \caption{Epiphone Emily the Strange guitar sample processed with SoundToys Decapitator. (a) Power spectrum of the original sample. (b) Power spectrum of the processed sample using Primaal's favourite settings (`N' mode, drive to 4, low cut to 3, high cut to 7); (c) Power spectrum of the processed sample using maximum distortion.}
\label{fig:decapitator}
\end{figure}

\begin{figure}[h!]
  \centering
  \includegraphics[width=1\columnwidth]{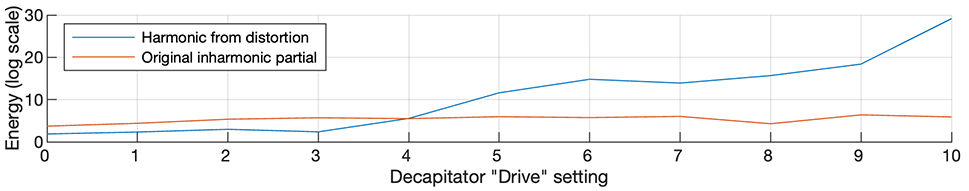}
  \caption{Amplitude of original and generated partials near 880Hz with different Decapitator `Drive' settings. The red line shows the amplitude of the original inharmonic partial. The blue line shows the amplitude of the generated 8\textsuperscript{th} harmonic.}
\label{fig:decapitator_harmonic}
\end{figure}

Primaal's distortion plug-in of choice is SoundToys' Decapitator. Figure~\ref{fig:decapitator} shows the power spectrum of an Epiphone Emily the Strange guitar sample\footnote{\url{https://www.karoryfer.com/karoryfer-samples/emilyguitar}} processed with Decapitator. Figure~\ref{fig:decapitator}~(a) shows that the sample is inharmonic, as is the case for most string instruments including the guitar \citep{jarvelainen2006perceptibility}. Figure~\ref{fig:decapitator}~(c) shows that when using maximum distortion settings, partials generated by harmonic distortion are (1) much louder than the nearby original inharmonic partials and (2) harmonic. Harmonic distortion makes the original audio harmonic if it wasn't. Figure~\ref{fig:decapitator}~(b) shows the result of processing the original sample using Primaal's favourite settings. Examination of the peaks around 1000Hz in Figure~\ref{fig:decapitator}~(b) suggests that when using these settings, the energy of the harmonics generated by the distortion process is equal to the energy of the original inharmonic partials. The finding is confirmed in Figure~\ref{fig:decapitator_harmonic}. The proximity of equal-energy partials is reminiscent of the result of the `Unison' process. It suggests that distortion, in this case, is used to modify the sound's envelope (see Appendix~\ref{ref:unison}).

\subsubsection{SansAmp PSA-1}

Besides Decapitator, Primaal may use the SansAmp PSA-1 plug-in. As shown in the example of `Boom' (Figure~\ref{fig:boombass}, Section~\ref{ref:boombass}), the SansAmp plug-in provides more radical processing than Decapitator. The generated partials may follow a frequency evolution that goes in the opposite direction from the original signal, recalling the folding of frequencies that occurs during aliasing \citep{park1993aliasing}. The producers mention that such generated partials may be a valuable element in the partial masking of pitch values, and in the reinforcement of pitch ambiguity. Finally, the producers mention that they may use the strong intermodulation distortion stemming from the SansAmp PSA-1 plug-in to add energy to the $f_0$ partial when necessary.

\subsection{Ring modulation}\label{ref:ringmodulation}

The Primaal producers use Omnisphere's Ring Modulation\footnote{\url{https://support.spectrasonics.net/manual/Omnisphere2/25/en/topic/layer-page-oscillator-page17}} to boost partials or create new ones. One difference between ring modulation and intermodulation distortion lies in the quantity of generated partials. In the case of intermodulation distortion, if $f_a$ and $f_b$ are the input frequencies, then the output contains the frequencies $k_a f_a + k_b f_b$, where $(k_a, k_b) \in \mathbb{Z}^2$. In the case of ring modulation, with the same input frequencies, the output only contains the frequencies $f_a+f_b$ and $f_a-f_b$ \citep{parker2011simple}. As a result, ring modulation is more controllable than intermodulation distortion. The Primaal producers use ring modulation when they want to target specific harmonics, including the fundamental.

\clearpage

\vspace{1cm}

\subsection{Unison}\label{ref:unison}

In Western classical music, one definition of `unison' is `[t]he simultaneous execution of one polyphonic part by more than one performer or performing group [...] either at exactly the same pitch or at the interval of an octave, double octave etc' \citep{Unison}. The `unison' audio effect, as originally found in analogue synthesisers, simulates such execution by superimposing an original part with slightly detuned versions of itself. The resulting sound is described as `much thicker'\footnote{\url{https://dailyanalog.com/glossary/unison/}}. The unison effect echoes one remark from \citet[p. 34]{roederer2008physics}, according to which `single pure tones sound very dull. Things become a little livelier as soon as we superpose two pure tones by sounding them together'. The Primaal producers make extensive use of Omnisphere's Unison effect\footnote{\url{https://support.spectrasonics.net/manual/Omnisphere2/25/en/topic/layer-page-oscillator-page22}}.

\begin{figure}[!htbp]
  \centering
  \includegraphics[width=1\columnwidth]{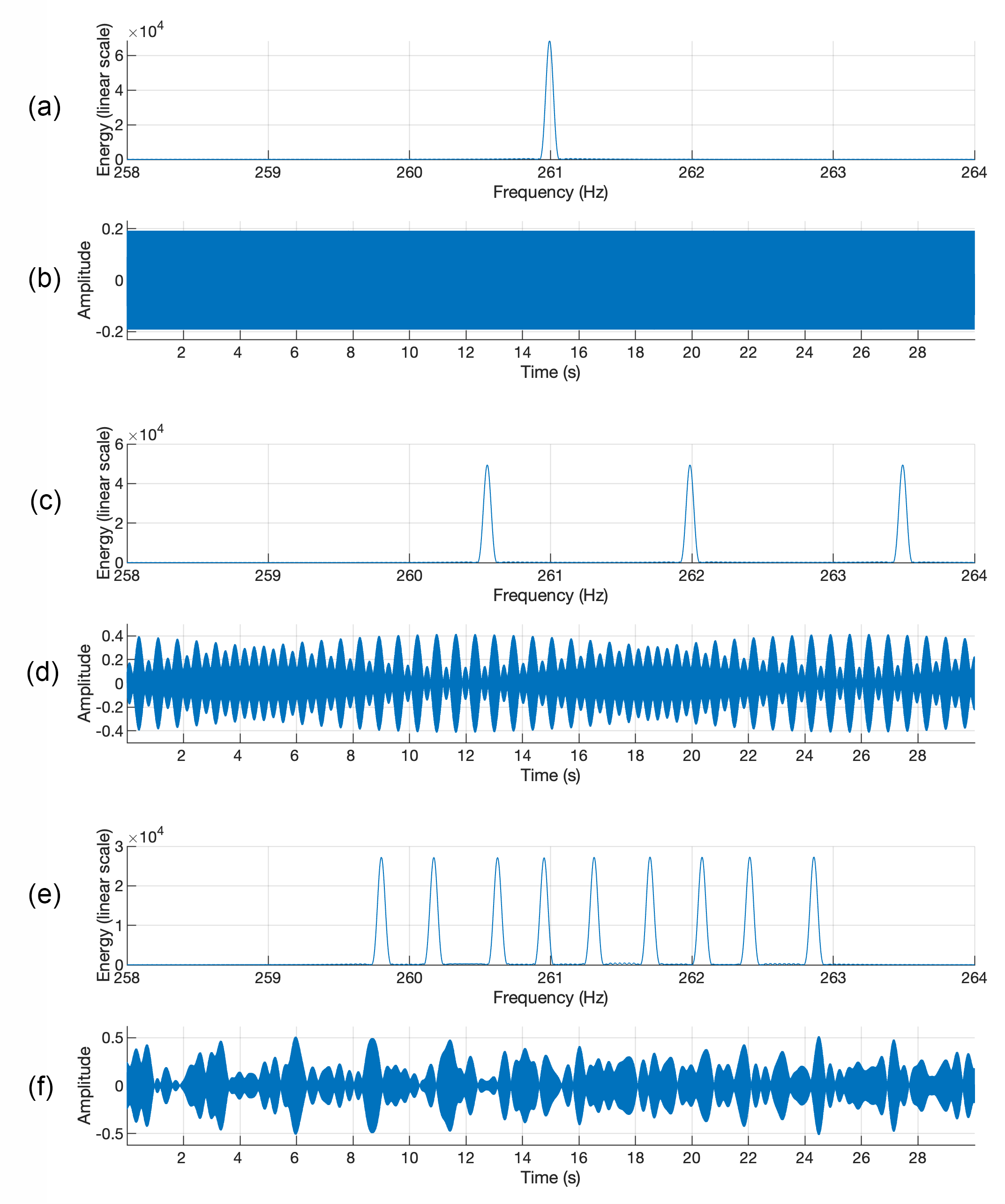}
  \caption{Omnisphere Unison's behaviour. Original sine wave, (a) power spectrum, and (b) amplitude. Unison output, minimum number of voices, (c) power spectrum, and (d) amplitude. Unison output, maximum number of voices, (e) power spectrum, and (f) amplitude.}
\label{fig:Unison_sines}
\end{figure}

One result of the unison audio effect is the emergence of acoustic beats. Figure~\ref{fig:Unison_sines}~(a) and (b) show the spectrum and waveform of a sine wave. Figure~\ref{fig:Unison_sines}~(c) and (d) show the spectrum and waveform of the sine wave processed with Omnisphere's Unison, using the minimum number of voices and a `detune' value of 0.1. Acoustic beating can be seen in Figure~\ref{fig:Unison_sines}~(d). Figure~\ref{fig:Unison_sines}~(e) and (f) show the spectrum and waveform of the same sine wave processed with the same effect, this time using the maximum number of available voices. The acoustic beating patterns are more irregular.

\begin{figure}[!htbp]
  \centering
  \includegraphics[width=1\columnwidth]{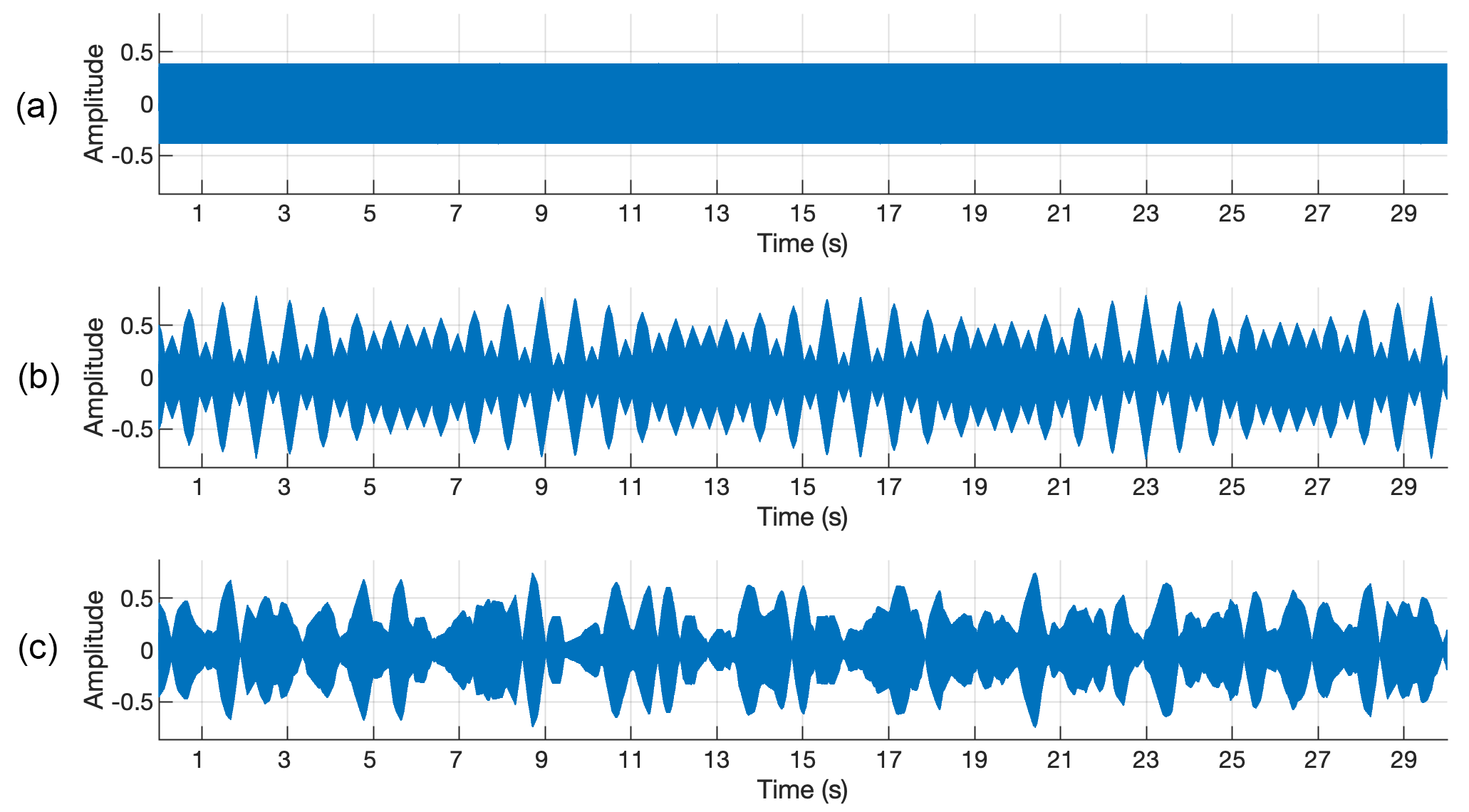}
  \caption{Omnisphere Unison's behaviour, harmonic tones, example. (a) Original triangle wave, amplitude. (b) Unison output, minimum number of voices, amplitude. (c) Unison output, maximum number of voices, amplitude.}
\label{fig:Unison_harm}
\end{figure}

Since all frequencies in a harmonic tone are multiples of the fundamental, the signal is periodic at the fundamental frequency. As a result, all periods are identical, and the signal's envelope is flat. Figure~\ref{fig:Unison_harm}~(a) shows the waveform of a triangle tone. Figure~\ref{fig:Unison_harm}~(b) shows the waveform of a triangle tone processed with Omnisphere's Unison using the minimum number of voices, and Figure~\ref{fig:Unison_harm}~(c) using the maximum number of voices. As in the case of the sine wave, using more voices results in more irregular acoustic beating patterns. One can conclude that one use of the unison audio effect is to provide more `life' and interest to a harmonic complex tone.

\subsection{Equalisation}\label{ref:equalization}

The Primaal producers use equalisers for multiple purposes. One particular purpose is to attenuate the fundamental frequency and boost upper harmonics. The equalisers the producers use come in different types. One type is the standard parametric equaliser as originally designed by \citet{massenburg1972parametric}. Another type is the `resonance' control commonly found in the lowpass filter module of analogue synthesisers \citep{moog1965voltage}. A third type is Sony CSL Resonance EQ\footnote{\url{https://cslmusicteam.sony.fr/prototypes/resonance-eq/}} \citep{grachten2019req}, which tracks the frequency of spectral formants in order to attenuate or boost the formants.

\subsection{Frequency modulation}\label{ref:frequencymodulation}

The Primaal producers use frequency modulation sparsely, and not as a direct means of audio synthesis. Section~\ref{ref:firebass} shows the example of the bass in `\textexclamdown{}Fire!', where the FM-based, Industrial Music Electronics Piston Honda\footnote{\url{http://www.industrialmusicelectronics.com/products/21}} is used to modulate a quasi-harmonic low-frequency tone using FM-generated inharmonic partials. The Primaal producers note that they involve frequency modulation when they target inharmonic relations.

\subsection{Compression}\label{ref:compression}

The Primaal producers make strong use of Omnisphere's Seismic Pump compressor\footnote{\url{https://support.spectrasonics.net/manual/Omnisphere2/25/en/topic/seismic-pump}}, for instance in `Elevate'. The producers use this compressor not only to stabilise the input signal's dynamics but also to modify the signal's spectrum. Depending on the settings, the compressor may generate partials by distorting the signal, or attenuate them by damping upper frequencies.

\subsection{Reverb}\label{ref:reverb}

The Primaal producers use Omnisphere's Seismic Verb\footnote{\url{https://support.spectrasonics.net/manual/Omnisphere2/25/en/topic/seismic-verb}} on `Danger', `R U Ready', and `Silver'. By design, Seismic Verb is not a standard reverberation algorithm. Amongst other features, it includes dynamic processing and detuning. The Primaal producers use Seismic Verb to scramble pitch values, especially in the high-medium and high ranges.

\subsection{Combined effects}

The Primaal producers use several plug-ins that combine a variety of effects such as dynamics, EQ, distortion, and reverberation. This is typically the case for amp simulators found for instance in Native Instruments Guitar Rig\footnote{\url{https://www.native-instruments.com/en/products/komplete/guitar/guitar-rig-7-pro/}}. Tape emulators such as Omnisphere Tape Slammer\footnote{\url{https://support.spectrasonics.net/manual/Omnisphere2/25/en/topic/fx-page-dynamics-page05}} also combine dynamics, EQ and distortion.

\subsection{Bends, glides, and screws}\label{ref:bends}

In this paper, we follow the Primaal producers' terminology. Other sources may provide different definitions. \emph{Bends}, or \emph{glides}, consist of applying a continuously changing pitch shift to an input sound. This can typically be achieved using a pitch wheel or pitch-bend control on a MIDI controller. \emph{Screws} are specific forms of bends. The pitch-change effect is similar to what a tape deck or turntable sounds like when speeding up or slowing down from a complete stop. The Avid Vari-fi plug-in\footnote{\url{https://www.avid.com/plugins/vari-fi-audiosuite}} can produce screws.

\vspace{1cm}

\interlinepenalty=0
\bibliography{mybib.bib}
\bibliographystyle{apalike}

\end{document}